\documentclass[letterpaper,twocolumn,10pt]{article}
\usepackage{usenix,url,amsfonts,amsmath,amsthm,multirow,amssymb,bmpsize}
\usepackage{graphicx}
\usepackage{subcaption}
\usepackage{enumitem}
\usepackage{booktabs}
\usepackage{listings}  
\usepackage{url}

\usepackage{breakurl}
\usepackage[breaklinks]{hyperref}
\usepackage{xcolor}
\setlist{noitemsep}
\graphicspath{{figs/}}
\DeclareGraphicsExtensions{.png}

\usepackage{adjustbox}
\usepackage{array}

\newcolumntype{R}[2]{%
    >{\adjustbox{angle=#1,lap=\width-(#2)}\bgroup}%
    l%
    <{\egroup}%
}
\newcommand*\rot{\multicolumn{1}{R{60}{1em}}}

\newcommand\mytt[1]{\texttt{\small{#1}}}

\begin{document}

\date{}

\title{\Large \bf Diffix Elm: Simple Diffix}

\author{
Paul Francis$^{\dag}$ \quad Sebastian Probst-Eide$^{\S}$ \quad David Wagner$^{\ddag}$ \\
Felix Bauer$^{\S}$ \quad Cristian Berneanu$^{\S}$ \quad Edon Gashi$^{\S}$ \\
$^{\dag}$Max Planck Institute for Software Systems (MPI-SWS), Germany\\
$^{\S}$Open Diffix\\
$^{\ddag}$Forschungsinstitut f\"{u}r \"{o}ffentliche Verwaltung (F\"{O}V), Germany\\
francis@mpi-sws.org, dwagner@foev-speyer.de, \{sebastian, felix, cristian, edon\}@aircloak.com
}

\maketitle



\begin{abstract}

Historically, strong data anonymization requires substantial domain
expertise and custom design for the given data set and use case.
Diffix is an anonymization framework designed to make strong data
anonymization available to non-experts. This paper describes Diffix Elm,
a version of Diffix that is very easy to use at the expense of query
features. We describe Diffix Elm, and show that it provides strong
anonymity based on the General Data Protection Regulation (GDPR) criteria.

This document is the third version of Diffix Elm. The second version added ceiling, round, and bucket\_width functions (in addition to floor). This document adds the ability to protect multiple different kinds of protected entities (a feature not found in earlier versions of Diffix). It also adds counting distinct values for any column (rather than only the AID column).

\end{abstract}

\setcounter{tocdepth}{2}
\tableofcontents

\section{Introduction}
\label{sec:intro}

Data anonymization is commonly and successfully used in a large variety
of practical settings, ranging from the public release of census and
other data~\cite{nyc-taxi} by governments, the open sale of mobility
data~\cite{mobility-data}, and the distribution of medical data to
researchers~\cite{hcup}.

In spite of occasional proclamations that data anonymity is
impossible~\cite{anon-impossible}, and somewhat more frequent
demonstrations of breaking weakly anonymized (pseudonymized)
data~\cite{sweeney2002k,aol,Narayanan:2008:RDL:1397759.1398064,taxi-break},
the track record of data anonymization in practice, as evidenced
by the lack of
reports of malicious re-identifications, is remarkably good.

The problem is not that we don't know how to effectively anonymize data. Rather, the
problem is that substantial expertise and effort is required to strongly
anonymize data while satisfying any given analytic use case. Census
bureaus employ full-time professionals to ensure that their data
releases are anonymous, and specialized companies are formed to deal
with anonymization of data in specific domains like
health~\cite{privacy-analytics} and mobility~\cite{teralytics}.

This paper describes and analyzes Diffix Elm, a strong anonymization mechanism
that is easy to use by non-experts and provides remarkably high-utility
output. Diffix Elm uses the three most common anonymization techniques,
generalization, suppression, and noise. We refer to these as the
\emph{big-three} anonymization techniques. In terms of strength of anonymization,
Diffix Elm is somewhat stronger than k-anonymity and l-diversity, but not
as strong as Differential Privacy (DP) with low epsilon and (if applicable)
low delta. Diffix Elm, however, is far easier to use and has better
utility than k-anonymity, l-diversity, or DP.

Intuitively, Diffix Elm has stronger privacy than k-anonymity
because k-anonymity use only generalization and suppression,
while Diffix Elm additionally uses noise. While Diffix Elm and DP use
all big-three techniques\footnote{
    Strictly speaking, DP and k-anonymity are measures of
    anonymity, not mechanisms per se. It would be more accurate, though
    a bit unwieldy, to say ``a mechanism that adheres to DP uses'' rather
    than ``DP uses''.
}, Diffix Elm provides weaker anonymity because for certain types of very
rare prior knowledge, Diffix Elm is less pessimistic than DP in its assumptions
about what the attacker knows. However, this difference frees Diffix Elm
from the need for a privacy budget, leading to far better utility for most
use cases compared to DP.

The source code for a reference implementation of Diffix Elm may be found
at https://github.com/diffix/reference.

This paper describes Diffix Elm and analyzes its
anonymization properties. The paper is targeted
towards Data Protection Authorities and Officers (DPA and DPO)
so that they may
evaluate the suitability of Diffix Elm in whatever legal context
applies. The paper is also targeted towards academics and other
interested privacy professionals.

Sections~\ref{sec:overview} and \ref{sec:describe} provide an overview
and detailed description of Diffix Elm respectively. The criteria for
evaluating Diffix Elm is based on the three criteria defined by the
EU~\cite{article29}, and is described in Section~\ref{sec:criteria}.
Section~\ref{sec:evaluation} presents the evaluation of Diffix Elm's
anonymization properties as a comprehensive list of attacks and
their measured effectiveness against Diffix Elm. Section~\ref{sec:dpa-guidance}
presents guidance for how a DPA or DPO may evaluate a given
Diffix Elm data release or deployment. Appendix~\ref{sec:questions}
summarizes the guidance into a list of questions.

\subsection{Differences from prior versions of Diffix}

Diffix was initially developed in a research partnership between the
startup Aircloak GmbH~\cite{aircloak-home-page} and the Max Planck
Institute for Software Systems (MPI-SWS)~\cite{mpi-sws}. Development continues
under the auspices of the Open Diffix project~\cite{open-diffix}
supported by MPI-SWS.  In
this time, Diffix has been released in a series of versions,
Aspen~\cite{diffix-aspen}, Birch~\cite{diffixBirch},
Cedar~\cite{diffix-cedar-tr}, and Dogwood~\cite{diffix-dogwood-tr}, with
each subsequent version adding new SQL features as well as new
anonymization mechanisms to defend against attacks as they are
discovered.

Although Diffix partially achieved its goal and had some success in demanding use cases, it ultimately
failed to achieve widespread use, in spite of the availability of a free
license for academic organizations and NGOs. We attribute this failure
to the overall difficulty of evaluating, deploying, and using Diffix. In releases through
Dogwood, Diffix was deployed as a software package installed
as a proxy that sits in front of a
database holding the raw data. The deployment hurdle was fairly high, requiring installation
of the proxy, and extensive and somewhat fiddly configuration with the
back end database. In addition, the anonymization mechanisms are complex
and difficult to understand, making the task of approval by Data
Protection Officers (DPO) a non-trivial effort.
We believe that these factors may have discouraged casual use of Diffix
for, for instance, occasional public releases of aggregated, low-dimensional
statistics.

At the same time, a variety of restrictions on SQL features, combined
with noise and suppression of high-dimensional data, made Diffix
challenging to use by analysts, and therefore unappealing in scenarios
where pseudonymized data could be used instead.

This paper presents the latest release, Diffix Elm. Elm represents a
massive simplification of Diffix, with a goal of extreme ease-of-use. All
but the most critical SQL features have been eliminated.
The number of anonymization mechanisms are likewise reduced, leading to
a system that is much easier to understand and evaluate. Elm is integrated
with the database rather than deployed as a proxy to a
database as with prior Diffix versions.

\begin{figure}[tp]
\begin{center}
\includegraphics[width=\linewidth]{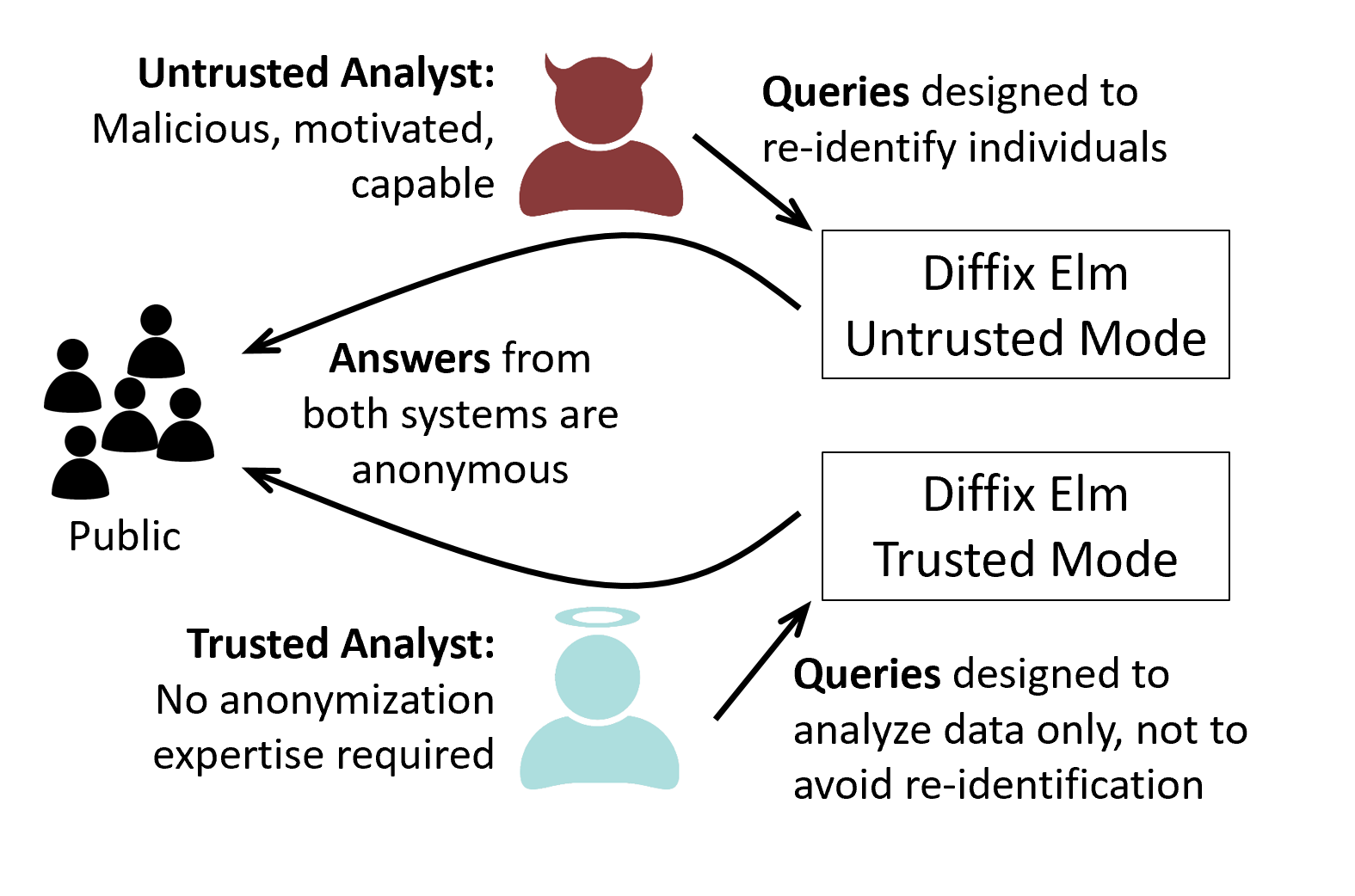}
\caption{Diffix Elm has two modes of operation, trusted and untrusted.
Untrusted Analyst Mode (UA-Mode) protects against a intentional re-identification, whereas
Trusted Analyst Mode (TA-Mode) protects against accidental re-identification. The answers
from UA-Mode are always anonymous. The answers from TA-Mode
are anonymous so long as the analyst does not intentionally try To
re-identify individuals.}
\label{fig:trust-line}
\end{center}
\end{figure}

Diffix Elm has two modes of operation, \emph{Trusted Analyst Mode} and
\emph{Untrusted Analyst Mode} (see Figure~\ref{fig:trust-line}).
In this regard, it departs from prior versions of
Diffix. Prior versions assume that an analyst is malicious, motivated,
and capable, or in other words, \emph{untrusted}. While Untrusted Analyst
Mode is of course necessary, we found that often Diffix is deployed in
environments where the analyst is on the one hand not malicious, but on
the other wants assurance that any answers received from the system
in the normal process of data analytics can
be released to the public as anonymous data. By treating trusted
analysts as untrusted, we made the job of analyzing data unnecessarily
difficult. The only technical differences in the two modes are the SQL
features that are made available: Trusted Analyst Mode has more SQL features.

The analyst in Trusted and Untrusted Modes is described as follows:

\begin{description}
\item[Untrusted Analyst Mode (UA-Mode):] The analyst is malicious, motivated, and capable.
They aim to re-identify individuals in the data. The system
protects against \textbf{intentional} re-identification of data.
\item[Trusted Analyst Mode (TA-Mode):] The analyst is trusted to not attempt to re-identify
individuals in the data. The analyst does not require any knowledge of
anonymization in order to protect the data. Rather, the analyst can 
simply go about the normal business of analyzing data, and the resulting
answers are anonymous and safe to release to the public. The system
protects against \textbf{accidental} re-identification of data.
\end{description}

In evaluating whether UA-Mode is anonymous by GDPR standards,
a DPA or DPO
only needs to evaluate the system, not the specific queries made
or the use case. This is because no known query or set of queries
violates anonymity. By contrast, in TA-Mode, a DPA or DPO must
additionally evaluate whether the queries that lead to a public
release of data may have led to re-identification.

\section{Overview of Diffix Elm}
\label{sec:overview}

This section provides a complete overview of Diffix Elm and its
anonymization properties. This section suffices for a reader interested
in only a high-level but nevertheless complete understanding of Diffix Elm.

We define a \emph{protected entity} as the entity whose privacy is being protected. Normally this is a natural person (individual), but it can be something that represents an individual (like a phone or a car), or a small group of individuals like a house or a joint bank account.

Diffix Elm allows for multiple different protected entities. For instance, a table with transactions can have both the sender and receiver protected. Likewise a table with both individuals and households and have both protected. Note that earlier versions did not have this capability.

\begin{figure}[tp]
\begin{center}
\includegraphics[width=\linewidth]{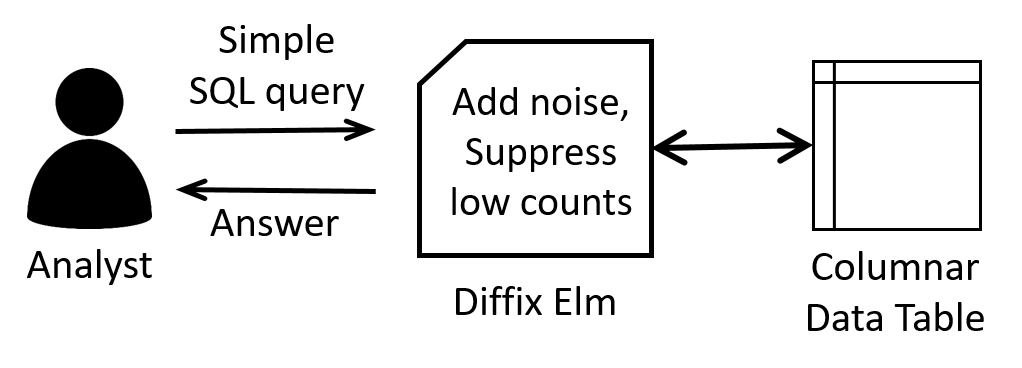}
\caption{Diffix Elm receives simple SQL queries and returns anonymized answers.
It operates on a single columnar table. If the table has more than one row per
protected entity, then one or more AID (Anonymization ID) columns must exist and
be configured as such.}
\label{fig:elm-basic}
\end{center}
\end{figure}

Figure~\ref{fig:elm-basic} illustrates the setup for Diffix Elm. Diffix
Elm offers a minimal SQL interface to an analyst or application. The
answers returned by Diffix Elm are syntactically correct SQL responses, but
are anonymized with the addition of noise and suppression of answers
that pertain to too few protected entities.

Diffix Elm operates with a single table only; it does not support table joins\footnote{Joins are planned for Diffix Fir.}. The table can have column data types of text strings, numbers, and dates and times. The table must either:
\begin{itemize}
  \item be constrained such that every protected entity in the
table occupies a single row, or
  \item have one or more AID (Anonymizing ID) columns consisting of a unique
identifier for each protected entity.
\end{itemize}
In the former case, Diffix Elm internally derives an AID value from the table index.  Diffix Elm anonymizes the data to protect protected entities as identified by the AIDs.

Diffix Elm places no limitations on the number of queries an analyst may
make.

\subsection{SQL Constraints}

Diffix Elm allows three count aggregates, \mytt{count(*)}, \mytt{count(column)}, and \mytt{count(DISTINCT column)}. The only allowed SQL keywords are \mytt{SELECT}, \mytt{FROM}, and \mytt{GROUP BY}.  The following for instance is an allowed query:

\begin{lstlisting}
   SELECT age,gender,count(*)
   FROM table
   GROUP BY age,gender
\end{lstlisting}

The selected columns may be generalized. For instance, the \mytt{age} column may be generalized as into \emph{buckets} of 10 years (i.e. \mytt{floor(age/10)*10}). Only the generalizations shown in Table~\ref{tab:col-expr} are allowed. Table~\ref{tab:col-expr} also shows which additional constraints are placed on UA-Mode.  These additional constraints are the only difference between TA-Mode and UA-Mode.

\begin{table*}
\begin{center}
\begin{tabular}{ll}
  \toprule
  Expression & Notes \\
  \midrule
  \mytt{[floor,round](numeric\_col/K)*K} & Range of width K. In UA-Mode, K must be in the set \\
    &  $\langle$... 0.1, 0.2, 0.5, 1, 2, 5, 10, 20, ...$\rangle$ \\
  \mytt{ceiling(numeric\_col/K)*K} & TA-Mode only \\
  \mytt{bucket\_width(numeric\_col,L,H,C)} & TA-Mode only \\
  \mytt{substring(text\_col from O for L)} & In UA-Mode, O (offset) must be 1 (left characters only) \\
  \mytt{date\_trunc('period',date\_col)} & Rounded datetime, where period is one of \\
  & 'year', 'quarter', 'month', 'day', 'hour', 'minute', 'second' \\
  \bottomrule
\end{tabular}
\end{center}
  \caption{Columns selected by query may be generalized, but only by the functions shown here, and with
          additional constraints for UA-Mode as described.}
  \label{tab:col-expr}
\end{table*}

\subsection{Evaluation Criteria}

The primary evaluation of the anonymization strength of Diffix Elm is based on measuring the effectiveness of an exhaustive set of attacks against Diffix Elm. The measure, called the PI/PR measure (Precision Improvement and Prediction Rate, Section~\ref{sec:measure-anon}), measures the ability of an attacker to make correct predictions about individuals in the dataset.  Specifically, it measures the \emph{improvement} in precision gained by the attack over the precision obtained prior to the attack based only on prior knowledge. This prior knowledge can include specific knowledge about individuals in the dataset as well as general statistical knowledge about the data.

The predictions we measure incorporate the three criteria for
anonymization defined by the European Data Protection Board\footnote{
  Formerly the Article 29 Data Protection Working Party.}
(EDPB)~\cite{article29}. The criteria are \emph{singling out},
\emph{inference}, and \emph{linkability}.

Singling out is a prediction that says ``There is a single individual with attributes A, B, and C.'' Singling out is problematic because it may allow an attacker to subsequently \emph{identify} the individual (e.g. associate a name, address, or some other personally identifying information to the singled-out individual). Inference is a prediction that says ``Individuals with attributes A, B, and C also have attribute D.'' Singling out as defined here also incorporates linkability because the ability to single out from the protected dataset may allow an attacker to link with a known dataset that has the same attributes (see Section~\ref{sec:three-criteria}).

We evaluate Diffix Elm by running all attacks known to us, and measuring the extent to which the attacks are effective. In each attack, we make multiple \emph{predictions} regarding the criteria. Attacks where a higher fraction of predictions are correct are relatively more effective. We have for several years been collecting attacks, both those discovered by ourselves, and those discovered by others, including through bug bounty programs~\cite{diffix-dogwood-tr}.

In addition to the PI/PR measure, we also take into consideration the \emph{prior knowledge} and \emph{data conditions} required for the attack.  We identify three classes of prior knowledge. Class A is simply knowledge of a single individual, and is the kind of external knowledge typically used to break ``anonymization'' in the Massachusetts medical data~\cite{sweeney2002k}, the an AOL search dataset~\cite{aol3}, and the Netflix prize dataset~\cite{Narayanan:2008:RDL:1397759.1398064}.  Class B requires prior knowledge of multiple individuals, and Class C requires still additional prior knowledge (see Section~\ref{sec:pk-class}).  While Class A prior knowledge is indeed easy to obtain, Classes B and C are much less likely to occur in practice.

\subsection{Diffix Elm Anonymization}

Broadly speaking, anonymization mechanisms produce one of two types of outputs:
\begin{description}
  \item[Individual records:] Each record pertains to a single
    protected entity. Implementations of k-anonymity generally produce individual
    records (even though the sensitive columns of each looks like k-1 others)~\cite{sweeney2002k}.
  \item[Statistical aggregates:] Each output is a statistical aggregate, like count or sum, that pertains to one or more (usually more) protected entities.  Differential Privacy~\cite{Dwork06} is usually used this way.
\end{description}

Diffix Elm produces statistical aggregates that always pertain to multiple protected entities.

The EDPB opinion on anonymization~\cite{article29} lists randomization and generalization as the two main anonymization mechanisms. Virtually all strong anonymization techniques exploit one or both of these mechanisms in one way or another.

With generalization, fine-grained data values are rounded or mapped into broader groups or categories. For instance, date of birth is mapped into 10-year buckets, or 5-digit zip codes are reduced to the first three digits.  Generalization is the primary mechanism for k-anonymity.

Randomization can be used either to change individual data values (i.e.  a date-of-birth is changed randomly to some other date within plus or minus one year), or can be used to change statistical aggregate values (e.g. a count of 428 is changed randomly to 436). Differential Privacy primarily depends on randomization.

Diffix Elm exploits both randomization and generalization.

\begin{figure}[tp]
\begin{center}
\includegraphics[width=0.7\linewidth]{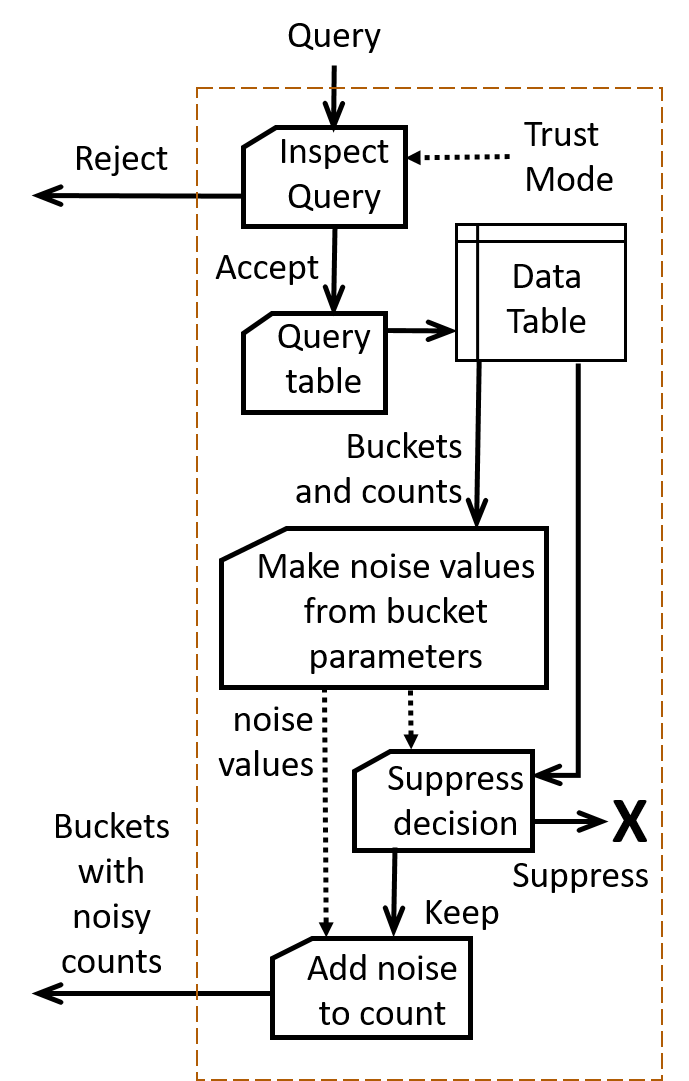}
\caption{Diffix Elm rejects queries that do not adhere to the allowed
  SQL. Per bucket, Diffix Elm generates sticky noise values from the bucket
  parameters. Buckets whose counts fall below a noisy threshold are
  silently suppressed. Noise is added to the remaining bucket counts.}
\label{fig:elm-simple-flow}
\end{center}
\end{figure}

Diffix Elm indirectly forces generalization by \textbf{suppressing} buckets that pertain to too few protected entities (Figure~\ref{fig:elm-simple-flow}): data that does not pertain to enough protected entities will be suppressed, so the analyst must generalize in order to avoid suppression.

Diffix Elm adds \textbf{noise} to counts by perturbing them
according to a zero-mean Gaussian distribution. Among other things, this
prevents analysts with prior knowledge of the data from deducing
facts about protected entities.

Noise in Diffix Elm is \emph{proportional}. In the case of \mytt{count(*)} or \mytt{count(col)} queries on tables that have multiple rows per protected entity, the amount of noise (the standard deviation) is proportional to the number of rows from \emph{heavy contributors}: protected entities that contribute the most rows. There is also a mechanism called \emph{flattening}. Flattening reduces the row contribution of \emph{extreme contributors}: the two or three protected entities that contribute substantially more than others (if any).

The threshold used to determine if a bucket should be suppressed is itself a noisy value. For any given suppression decision, a value that varies up or down from a mean value is used. This defends against an analyst using the suppression decision itself, combined with a priori knowledge of the data, from deducing facts about protected entities.

Both noise and suppression are determined by the worst-case AID. The noise standard deviation as well as the suppression decision are computed per-AID. Suppression occurs if any on the per-AID suppression decisions are positive. The noise standard deviation used is the highest of the per-AID computations.

A key feature of Diffix Elm is that it allows an analyst to make an
unlimited number of queries while still providing strong anonymity. If
Diffix Elm used a different random noise sample with each query, as most
Differential Privacy systems do, then the noise could be averaged away
with repeated queries. To defend against this, Diffix Elm uses
\textbf{sticky noise}, both for the noisy counts and noisy thresholds.
The high-level concept of sticky noise is that the \emph{same query produces
the same noise}.

Sticky noise operates by deterministically seeding a function which produces a pseudo-random but deterministic noise value from a Gaussian distribution. There are two Gaussian noise samples (called \emph{layers}) that are summed together. One of the layers is seeded from the set of AID values. The other is seeded from the bucket parameters themselves (column, value, and generalization parameters). Each of the two layers protects against different kinds of attacks (see
Section~\ref{sec:evaluation}).

Finally, Diffix Elm detects when a selected column of a given query, were it to be dropped in another query that is otherwise identical to the given query, would cause the complete contents of a suppressed bucket to appear in a single other bucket.  This condition, if gone unchecked, could sometimes allow an attacker to detect the presence of the suppressed bucket with high precision.  To prevent this, Diffix Elm merges the contents of the suppressed bucket with the other bucket.

In summary, Diffix Elm has the following mechanisms:
\begin{itemize}
  \item Support for multiple protected entities
  \item Strict SQL limitations
  \item Ability to generalize
  \item Sticky noise, proportional to heavy contributors
  \item Low-count suppression with sticky noisy threshold
  \item Flattening of extreme contributors
  \item Merging of suppressed buckets
\end{itemize}

\subsection{Evaluation Results}

Table~\ref{tab:attacks-summary} summarizes the evaluation of Diffix
Elm. For each known attack, it provides:
\begin{enumerate}
  \item The strength of anonymization according to the PI/PR measure
  ranging from Weak to Very Strong or infeasible,
  \item the class of prior knowledge required to execute the attack
  ranging from None to Class C (extremely unlikely to exist or obtain), and
  \item the data conditions required for the attack (from None to
  Very Rare).
\end{enumerate}

Table~\ref{tab:attacks-summary} shows that the PI/PR measure for
every attack except one either doesn't work at all, has
Very Strong anonymization, or can be configured to have Very
Strong anonymization.

The one attack for which this is not the case (\emph{Detect outlier bucket})
can be prevented by detecting the required data condition (which itself
is Very Rare) and modifying the data so that the condition no longer exists.

In addition to the Very Strong PI/PR measure, most of the attacks have
difficult prior knowledge requirements and/or rare data conditions, leading
to even less risk.

\section{Specification of Diffix Elm}
\label{sec:describe}

This section provides a concise and complete specification of Diffix
Elm for both Trusted Analyst and Untrusted Analyst Modes.
Section~\ref{sec:overview} is helpful but not strictly necessary to
understand this specification. Note that while this section specifies
how Diffix Elm works, it doesn't really describe why it works that way.
Section~\ref{sec:evaluation} justifies the design by describing how the design
defends against known attacks.

\subsection{Restrictions and Assumptions}

\textbf{Players and components:} 

\begin{description}
\item[Analyst:] The person or application that queries Diffix Elm and
receives anonymized answers. The analyst is trusted or untrusted according
to the mode of Diffix Elm.
\item[Public:] Any person that may receive data obtained by an analyst.
The public is untrusted. (Unless otherwise stated, any assumptions about an
untrusted analyst apply to members of the public as well.)
\item[Prior knowledge:] This refers to knowledge of values in the
table.  The untrusted analyst may have substantial knowledge of
the data in the table, including knowledge of entire columns or entire
rows. (A trusted analyst may know the entire table contents.)
\item[Admin:] The person who sets up and configures Diffix Elm. The
admin is trusted and has access to the table data.
\end{description}

\textbf{SQL restrictions:} Diffix Elm supports only the following SQL keywords: \mytt{SELECT}, \mytt{FROM},
and \mytt{GROUP BY}. Diffix Elm does not support sub-queries. As such, the
only SQL structure possible with Diffix Elm is the following:

\begin{lstlisting}
   SELECT col_expr1,...,col_exprN,count(...)
   FROM table
   GROUP BY col_expr1,...,col_exprN
\end{lstlisting}

The column expressions \mytt{col\_expr} are optional, and can consist
only of the expressions shown in Table~\ref{tab:col-expr}, including
the syntax limitations shown in the table. These syntax limitations
are the only difference between TA-Mode and UA-Mode.

The \mytt{count(...)} expression may be one of \mytt{count(*)}, \mytt{count(col)}, or \mytt{count(DISTINCT col)}, where \mytt{col} is a column name.

Strictly speaking, Diffix Elm accepts SQL without a \mytt{count()} aggregate:

\begin{lstlisting}
   SELECT col_expr1,...,col_exprN
   FROM table
\end{lstlisting}

But internally it modifies that SQL to include \mytt{count(*)} and
\mytt{GROUP BY} expressions corresponding to the selected columns. It then
modifies the resulting buckets on output to list each counted row separately.
In other words, there is always a \mytt{count()} aggregate, and if columns
are selected, then there are corresponding \mytt{GROUP BY} expressions, either
explicit or implicit. In the remainder of this description, references to
\mytt{GROUP BY} expressions include either explicit or implicit expressions.

\textbf{Table restrictions:} Diffix Elm operates on a single columnar
table only. The column types can be numeric (integer or real), text, date,
time, and datetime.

Diffix can protect multiple different entities, either of the same type (one or more persons) or of different types (a person and an office). There must be at least one column per protected entity that identifies the protected entity. We refer to these columns as AID (Anonymizing ID) columns, and an individual value in each column as an AIDV (AID Value).

In the case where a table has only a single protected entity, and there is one row per protected entity, then the table may not have an AID column. In these cases, an AID column must be created, which may be for instance nothing more than the row index.

\textbf{Secrets:} There is a single secret, the \mytt{salt}, associated with each table. Multiple tables (for instance in the same database) may have the same salt. The salt must not be known by an untrusted analyst.

\textbf{Untrusted analyst knowledge:} The untrusted analyst may have prior knowledge of substantial portions of the table data.  We assume that enough of the table data is unknown to the untrusted analyst that the analyst cannot derive the salt through a brute-force dictionary attack on the table. (In effect, the unknown portions of the table serve as a kind of secret password to derive the salt.) Both analysts knows the column names and column types.

\textbf{Table changes:} Diffix Elm supports two models for how tables may change: \emph{append} and \emph{update}. Existing rows in append tables do not change. If they are deleted, they are deleted in bulk (i.e. all rows for a given month). Existing rows may be modified in update tables.

Append tables are typically time-series data where new rows are appended
to the table, and a time or date column increases in value with added rows. The key attribute here is that multiple identical queries to windows of time in the past will always return the same result.

By contrast, update tables may have rows inserted anywhere, and the values in any row may be modified. Update tables are not normally time-series.

The distinction is important in Diffix Elm for managing the \mytt{salt}.  For append tables, the salt is created when the table is first created, and doesn't change as data is appended. If the table is replicated (for scalability or redundancy), then the salt must be replicated as well.

For update tables, the salt may be created from the contents of the table itself. If the table changes, then the salt also changes (from the point of view of Diffix Elm, a new table is created). If the table is replicated, then the salt does not need to be replicated as well: it can be re-computed from the table contents.

Both approaches have anonymity pros and cons. With append tables, since
the salt is constant, if the same query returns a different answer, then
the analyst knows that the contents of the table have changed. If the
change is small and predictable, then the analyst may learn something
about a protected entity. With update tables, the same query will always
produce a different answer then the table is updated (and the salt
changes), and so from a single query an analyst cannot tell that the
table has changed. However, if certain data has not changed over many
updates, then the analyst can average out the noise.

Either way, small incremental changes to the table weaken the anonymity properties
of Diffix Elm, and should be avoided. (Note that time-series data can be managed
as a series of update tables, each of which never changes. For instance,
each day in the time series can be a separate table. Obviously a table may
be both appended and updated. Such tables should be managed as a series of
update tables.)

\subsection{Configure constants}

Before use, the admin configures the three suppression constants
\mytt{low\_thresh}, \mytt{supp\_sd}, and \mytt{low\_mean\_gap}, the noise constant
\mytt{base\_sd}, and the flattening constants \mytt{outlier\_range} and \mytt{top\_range}
(see Table~\ref{tab:variables}). Diffix Elm enforces the minimum
values for these constants of
\mytt{low\_thresh=2}, \mytt{supp\_sd=1}, \mytt{low\_mean\_gap=2}, \mytt{base\_sd=1.5},
\mytt{outlier\_range=[1,2]}, and \mytt{top\_range=[2,3]}.
Note that the max value of \mytt{outlier\_range} and \mytt{top\_range} must be greater than the min value respectively.
The higher these values, the stronger the anonymity (and the worse the utility).

These minimum values in fact provide quite good anonymity, suitable
for most purposes (see the evaluation in Section~\ref{sec:evaluation}).
Extremely strong anonymity is achieved at values
\mytt{low\_thresh=4}, \mytt{supp\_sd=2}, \mytt{low\_mean\_gap=4}, \mytt{base\_sd=3},
\mytt{outlier\_range=[2,4]}, and \mytt{top\_range=[3,5]}.
Values higher than these have diminishing returns with respect to strength of
anonymity and only serve to unnecessarily degrade the utility of the output.

\subsection{Configure AID columns}
\label{sec:config-aid}

For each protected entity, the AID column for that protected entity must be identified and configured. The ideal AID column has exactly one distinct value (AIDV) per distinct protected entity.  (See Section~\ref{sec:dpa-guidance} for a discussion of what can be done if the best candidate AID column for a given protected entity does not perfectly achieve this ideal.)

Note that anonymity is not reduced if multiple AID columns are configured for the same protected entity. Diffix Elm does not care if multiple AID columns refer to the same protected entity or different protected entities.

\subsection{Table pre-processing}

The first time the admin configures the table into Diffix Elm,
there are two optional initial pre-processing (PP) steps.

\begin{table}
\begin{center}
\begin{small}
\begin{tabular}{p{0.24\columnwidth}p{0.62\columnwidth}}
  \toprule
    \multicolumn{2}{l}{\textbf{\normalsize \kern 2em Table-derived variables}}\\
    \mytt{salt} & The secret salt used to generate noise \\
    \mytt{AIDV} & A single value from the AID column \\
  \midrule
    \multicolumn{2}{l}{\textbf{\normalsize \kern 2em Query answer variables}}\\
    \mytt{AIDV set} & A set of distinct AIDVs (i.e. associated with a bucket) \\
    \mytt{AIDV con- tribution} & The number of rows contributed by each AIDV \\
    \mytt{bucket} & An answer row, as defined by the distinct set of columns in the GROUP BY \\
    \mytt{bucket\_count} & The row count of the bucket \\
    \mytt{bucket\_value} & One of the distinct set of column values for the bucket \\
  \midrule
    \multicolumn{2}{l}{\textbf{\normalsize \kern 2em SQL-derived variables}}\\
    \mytt{column\_name} & A GROUP BY column name\\
    \mytt{range\_param} & The range parameter(s) for a GROUP BY column \\
    \mytt{range\_type} & floor, ceiling, round, bucket\_width, substring, or date\_trunc (if any) \\
  \midrule
    \multicolumn{2}{l}{\textbf{\normalsize \kern 2em Suppression constants}}\\
    \mytt{low\_thresh} & The lower bound for the noisy threshold (minimum value 2) \\
    \mytt{supp\_sd} & The standard deviation of the suppression Gaussian noise
                   (minimum value 1.0) \\
    \mytt{low\_mean\_gap} & The number of \mytt{supp\_sd} standard deviations between 
                             \mytt{low\_thresh} and the Gaussian noise mean
                             (minimum value 2)\\
  \midrule
    \multicolumn{2}{l}{\textbf{\normalsize \kern 2em Noise/flattening constants}}\\
    \mytt{base\_sd} & The base standard deviation of the noise 
       (Minimum value: 1.5) \\
    \mytt{outlier\_range} & The minimum and maximum possible values of \mytt{outlier\_count}
       (max $>$ min, Minimum values: [1,2]) \\
    \mytt{top\_range} & The minimum and maximum possible values of \mytt{top\_count}
       (max $>$ min, Minimum values: [2,3]) \\
  \midrule
    \multicolumn{2}{l}{\textbf{\normalsize \kern 2em Other constants}}\\
    \mytt{AIDs} & The columns containing the AID values. Must be explicitly configured. \\
    \mytt{trust\_mode} & Set by the admin to TA-Mode or UA-Mode \\
  \bottomrule
\end{tabular}
\end{small}
\end{center}
  \caption{The variables and constants used in Diffix Elm}
  \label{tab:variables}
\end{table}

\textbf{PP step 1:} The \mytt{salt} may be generated as a cryptographically secure random number.

For update tables, the \mytt{salt} may alternatively be generated from the table, as follows.

If the table is made available to Diffix Elm as an SQL table,
then the salt is generated by:
\begin{enumerate}
\item initialize a variable \mytt{xor\_value} to 0
\item reading every cell of the table,
\item hashing the cell value,
\item XOR'ing the hash into \mytt{xor\_value},
\item one-way hashing the \mytt{xor\_value} to produce the \mytt{salt}.
\end{enumerate}

If the table is made available to Diffix Elm as a single file, for
instance a CSV file, then the salt is generated by:
\begin{enumerate}
\item Set the \mytt{salt} as a one-way hash of the file as a binary string.
\end{enumerate}

Either way (table or CSV file), the one-way hash must be
cryptographically secure and must produce at least a 128-bit salt.

\textbf{PP step 2:} If there is a single protected entity, and the table has one row per distinct entity, then Diffix Elm may automatically generate an AID column.
This \mytt{AID} column is populated with a distinct value per row. There are
no restrictions on the actual values used: simply assigning row index number
is sufficient.

\subsection{Query handling}
\label{sec:query-handling}

This section describes query handling for \mytt{count(*)} and \mytt{count(column)}. The aggregate \mytt{count(DISTINCT column)} is a special case, described in Section~\ref{sec:distinct}.
Query handling (QH) has the following main steps (see
Figure~\ref{fig:elm-simple-flow}:
\begin{description}
\item[\textbf{QH step 1:}] Inspect the query to ensure that it satisfies
the constraints imposed on SQL by Diffix Elm.
\item[\textbf{QH step 2:}] Determine the \mytt{bucket\_values} (i.e. column values),
\mytt{bucket\_count}, and for each AID, the AIDV set and associated AIDV contributions to the \mytt{bucket\_count}.
\item[\textbf{QH step 3:}] For each bucket, determine if the bucket should be
suppressed.
\item[\textbf{QH step 4:}] For each suppressed bucket, determine if the bucket
should be merged with its related non-suppressed bucket (relatively rare event).
\item[\textbf{QH step 5:}] For each non-suppressed bucket, compute
flattening and adjust \mytt{bucket\_count} accordingly. Adjust the noise amount
\mytt{base\_sd} to account for heavy contributors (proportional noise).
\item[\textbf{QH step 6:}] Compute
sticky noise and perturb \mytt{bucket\_count} with the noise.
\end{description}

By way of example, suppose that the query is:

\begin{lstlisting}
   SELECT date_trunc('year',birthdate),
          substring(zip from 1 for 3),
          count(*)
   FROM table GROUP BY 1,2
\end{lstlisting}

This query produces buckets with two \mytt{bucket\_values} (one for
\mytt{birthdate} and one for \mytt{zip}). The associated
\mytt{bucket\_ranges} are \mytt{'year'} and \mytt{1,3} respectively.

QH step 1 accepts the query. QH step 2 computes the buckets, for instance (here assuming a single AID):

\begin{center}
\begin{small}
\begin{tabular}{lll}
  bucket\_values & bucket\_count & AIDV set \\
  \midrule
  1983,'Q2V' & 31 & $\langle$4,9,18,...,92$\rangle$ \\
  1983,'P3B' & 2 & $\langle$3,12$\rangle$ \\
  1984,'Q2V' & 62 & $\langle$7,11,22,...,104$\rangle$ \\
  1984,'P3B' & 4 & $\langle$16,33$\rangle$ \\
  ... & ... & ... \\
\end{tabular}
\end{small}
\end{center}

In QH step 3, Diffix Elm may determine that the second and fourth
buckets need to be suppressed.

(QH step 4 is a rarely executed step, and not conveyed in this example.)

In QH step 5, Diffix Elm sorts the AIDVs in descending order of number
of rows. If necessary, it may adjust \mytt{bucket\_count} to reduce and hide
the contributions of extreme contributors, and may increase \mytt{base\_sd}
to make noise proportional to the contributions of heavy contributors.

Finally in QH step 6, noise is added to
the counts, leading to an answer as follows:

\begin{center}
\begin{small}
\begin{tabular}{ll}
  Bucket\_values & Bucket\_count \\
  \midrule
  1983,'Q2V' & 35 \\
  1984,'Q2V' & 60 \\
  ... & ... \\
\end{tabular}
\end{small}
\end{center}

The following sections specify the steps in detail.

\subsubsection{Seeding of noise layers}
\label{sec:seed-base}

Two of the query handling steps (suppression step 3 and noise step 6)
require that noise values from a Gaussian distribution are created.
Both steps have two noise values, each generated from different seed
materials.

One type of seed, the \mytt{aid\_seed}, is based on seed materials from an \mytt{AIDV set} as:

\begin{lstlisting}
aid_seed = owh(salt,XOR(h(AIDV1),...,
               h(AIDVn)))
\end{lstlisting}
where \mytt{owh()} is a cryptographically secure one-way hash function of at least 128 bits. There is one \mytt{aid\_seed} per AID.

The other type of seed, the \mytt{sql\_seed}, has seed materials from the SQL itself:

\begin{lstlisting}
sql_seed = owh(salt,XOR(gb_sql1, gb_sql2,
               ..., gb_sqlN))
\end{lstlisting}

where there is one \mytt{gb\_sql} per GROUP BY component (explicit or implicit).
Each \mytt{gb\_sql} is composed of the parameters associated
with the GROUP BY:

\begin{lstlisting}
gb_sql = h(column_name,bucket_value,
           range_type,range_param)
\end{lstlisting}

The \mytt{range\_type} and \mytt{range\_param} are excluded if no range function is used.

They are also excluded if the range function is \mytt{floor()}, \mytt{ceiling()}, or \mytt{round()}, the parameter \mytt{K} is 1, and the column is an integer type. This is done because in this case the behavior of the function is identical having no function at all. Without this special case, two queries, with and without the function, would produce identical buckets but with different sql\_seeds, thus reducing the effective amount of noise for the SQL noise layer through averaging.

If there are no columns selected (and therefore no GROUP BY),
then the default \mytt{sql\_seed = owh(salt, 0)} is used.

The \mytt{XOR} is there to make sure that \mytt{sql\_seed} is independent
of the order of \mytt{GROUP BY} expressions.

From the example above, the \mytt{gb\_sql} for the \mytt{birthdate}
column for the (1983,'Q2V') bucket would be a hash of the values
('birthdate', 1983, 'date\_trunc', 'year'). For the
\mytt{zip} column of the same bucket, it would be ('zip',
'Q2V', 'substring', 1, 3).

\subsubsection{QH step 2: determine AIDVs and contributions}

Every distinct set of \mytt{bucket\_values} defines a bucket. In this step,
Diffix Elm:

\begin{enumerate}
\item scans the table,
\item determines the buckets,
\item determines which rows are associated with each bucket,
\item computes the \mytt{bucket\_count} (number of rows) for each bucket,
\item for each AID, determines the set of AIDVs associated with each bucket, and
\item for each AID, determines the contribution (in number of rows) to the \mytt{bucket\_count} for each AIDV.
\end{enumerate}

Note that all but the last two steps constitute normal SQL query processing.

If the query aggregate is \mytt{count(DISTINCT aid)}, then each AIDV for that AID contributes one row.

\subsubsection{QH step 3: Make the suppression decision}
\label{sec:step3-supp}

The per-bucket input variables for this step are the per-AID \mytt{AIDV sets} and the \mytt{bucket\_count}.

The suppression decision has the following steps:

\begin{description}
\item[\textbf{QH step 3.1:}] Per-AID, generate the per-bucket seed.
\item[\textbf{QH step 3.2:}] Per-AID, generate a noise sample and a noisy threshold from the seed.
\item[\textbf{QH step 3.3:}] Suppress the bucket if the \mytt{bucket\_count} is below the noisy threshold for any AID.
\end{description}

In \textbf{QH step 3.1}, the suppression seed is generated as:

\begin{lstlisting}
supp_aid_seed = h(aid_seed,'suppress')
\end{lstlisting}

where \mytt{h()} is a hash function with relatively few collisions (say fewer
than 1/10000).

In \textbf{QH step 3.2}, the seed is used to produce a pseudo-random value mytt{supp\_noise} from a Gaussian distribution with mean zero. The standard deviation is \mytt{supp\_sd}.

The suppression threshold is computed as:

\begin{lstlisting}
mean = low_thresh + low_mean_gap
supp_threshold = 
    max(low_thresh, mean + supp_noise)
\end{lstlisting}

In \textbf{QH step 3.3}, the bucket is suppressed if the
\mytt{bucket\_count} is less than \mytt{supp\_threshold} for any AID.
If suppressed, none of the subsequent steps are executed.

With this procedure, any bucket with fewer than \mytt{low\_thresh} protected entities for a given AID will certainly be suppressed. By setting for instance \mytt{low\_thresh=2}, we can guarantee that no bucket, and therefore no column values, pertaining to a single protected entity will be released. This in and of itself does not mean that protected entities cannot be singled-out through other means, for instance exploiting the results of multiple queries.

The suppress decision is sticky because the same query generates buckets with the same count and seed, which in turn produces the same noisy threshold.

\subsubsection{QH step 4: Possibly merge suppressed bucket with non-suppressed
bucket}
\label{sec:qh-step-4}

This step takes place to handle a relatively rare scenario, whereby if a selected column of a given query were to be dropped in another query that is otherwise identical to the given query, this would cause the complete contents of a suppressed bucket to appear in a single other bucket.  This condition, if gone unchecked, could allow an attacker to often infer an unknown value associated with the suppressed bucket with high confidence.  To prevent this, Diffix Elm merges the contents of the suppressed bucket with the other bucket.

By way of example, suppose that there is query on a dataset for a university with selected columns \mytt{dept} (department), \mytt{sex}, and \mytt{title}.  Suppose that the CS dept has only two women, and that they both have the same title. Further suppose that the bucket with \mytt{dept=CS,sex=F,title=Prof} is suppressed, and the bucket with \mytt{dept=CS,sex=M,title=Prof} is not suppressed.

In this case, Diffix Elm can detect that, in a hypothetical different query with selected columns \mytt{dept} and \mytt{title} only, the contents of the suppressed bucket \mytt{sex=F} would be included in the \mytt{dept=CS,sex=M,title=Prof} bucket. This would in turn allow the analyst to detect that the suppressed bucket has value \mytt{title=Prof}.

\paragraph{Anonymized total suppression count:}
Diffix Elm generates an anonymized count which is derived from all the rows that have been suppressed in a given output (not including merged rows). The purpose of the suppression count is to inform the analyst as to how much suppression has taken place overall.

The anonymized total suppression count must be comprised of the rows of at least two suppressed buckets. Other than this, however, it is treated as a normal bucket: it is itself subject to suppression if there are too few AIDVs, noise is added, flattening occurs, etc. To seed the total suppression count, the symbol \mytt{*} is used as column values.

\subsubsection{QH step 5: Flatten and adjust base\_sd}
\label{sec:qh-step-5}

This step is executed for each AID for each bucket. The per-bucket flattening amount is taken as the max per-AID flattening amounts, and the \mytt{base\_sd} is taken as the max per-AID \mytt{base\_sd}.

This step has no effect if all protected entities contribute one row.

The amount of noise added to counts by Diffix Elm is proportional to
the amount contributed to counts by heavy contributors. In this way,
the presence or absence of any protected entity is hidden. If there is a
single extreme contributor (a protected entity contributing far more rows than
the next biggest contributor), however, then the amount of noise alone
can reveal the presence or absence of that protected entity. Therefore,
Diffix Elm \emph{flattens} the contribution of extreme contributors to make
them similar to those of heavy contributors, thus hiding extreme
contributors.

Flattening requires the following information associated with each bucket:
a \mytt{bucket\_count}, and per-AID, the \mytt{AIDV set} and the \mytt{AIDV contributions}.

Flattening has the following steps (executed per-AID):

\begin{description}
\item[\textbf{QH step 5.1:}] Adjust \mytt{top\_range} and \mytt{outlier\_range} if
needed based on the number of AIDVs. The algorithm for the given AID can terminate here if there are not enough AIDVs.
\item[\textbf{QH step 5.2:}] Sort the AIDVs by contribution amount, and by AIDV
    within a given contribution amount.
\item[\textbf{QH step 5.3:}] Generate a seed to randomly select \mytt{outlier\_count} and
    identify \mytt{outlier\_count} highest contributing AIDVs (\mytt{outlier\_group}).
\item[\textbf{QH step 5.4:}] Generate a seed to randomly select \mytt{top\_count} and
    identify \mytt{top\_count} next highest contributing AIDVs (\mytt{top\_group}).
\item[\textbf{QH step 5.5:}] Compute \mytt{top\_avg}, the average contribution of the AIDVs in the \mytt{top\_group}.
\item[\textbf{QH step 5.6:}] Compute the \mytt{bucket\_count\_adjustment} from the flattening computation from the \mytt{outlier\_group} contributions to \mytt{top\_avg} and adjust \mytt{bucket\_count} accordingly.
\item[\textbf{QH step 5.7:}] Increase \mytt{base\_sd} to account for heavy contributors.
\end{description}

The above steps produce per-AID \mytt{bucket\_count\_adjustment} and \mytt{base\_sd} values. Select the \mytt{bucket\_count\_adjustment} as that with the largest absolute value per-AID. Select the \mytt{base\_sd} as the largest per-AID \mytt{base\_sd}. Note that the AIDV set used in QH Step 6 is taken from the AID used to select the \mytt{base\_sd}. Note that the max \mytt{bucket\_count\_adjustment} and max \mytt{base\_sd} for the same bucket may come from different AIDs.

Modify the \mytt{bucket\_count} by adding the selected \mytt{bucket\_count\_adjustment}.

Each step is described in detail as follows.

In \textbf{QH step 5.1}, if there are fewer than 
\mytt{min(outlier\_range) + min(top\_range)} AIDVs, then
the reported count is set to
\mytt{low\_thresh} and the remaining steps are skipped for this AID.

If there are fewer than \mytt{max(outlier\_range) + max(top\_range)} AIDVs,
then the maximum
\mytt{max(outlier\_range)} and/or \mytt{max(top\_range)} must be temporarily
adjusted downwards so that the \mytt{outlier\_group} and \mytt{top\_group}
can be formed. The adjustment is made such that 
\mytt{max(outlier\_range) + max(top\_range)} is equal to the number of AIDVs.
Neither max value should be set lower than the corresponding min value.
Both values should be reduced at the same rate, starting with
\mytt{max(top\_range)}, until
\mytt{max(outlier\_range) == min(outlier\_range)}, after which only
\mytt{max(top\_range)} is reduced.

In \textbf{QH step 5.2}, the AIDVs and their contributions are sorted. This is necessary
both to determine the extreme and heavy contributions, but also to derive the seed
materials from the AIDVs to determine the \mytt{outlier\_group} and \mytt{top\_group}.
They are first
sorted by contribution descending. Within each group of AIDVs with the same contribution,
the AIDVs are sorted by \mytt{h(salt,AIDV)}.
Note that the sorting of AIDVs are to ensure that the seed is always derived
from the same AIDV set and is therefore sticky.

Set:
\begin{lstlisting}
max_count = max(outlier_range)+max(top_range)
\end{lstlisting}

Note that at most \mytt{max\_count} AIDVs are used for flattening, so once
the top contributing
\mytt{max\_count} AIDVs have been sorted no more sorting is needed.

In \textbf{QH step 5.3},
assign \mytt{max\_group} as the first \mytt{max\_count} AIDVs in the
sorted list. If there are not
\mytt{max\_count} AIDVs, then assign \mytt{max\_group} as all AIDVs.

Generate \mytt{flat\_seed} as:
\begin{lstlisting}
flat_seed = owh(salt,XOR(h(AIDV1),...,
               h(AIDVn)))
\end{lstlisting}
where AIDV1 through AIDVn are the AIDVs in \mytt{max\_group}.

Generate \mytt{out\_seed} as:
\begin{lstlisting}
out_seed = h(flat_seed,'outlier')
\end{lstlisting}

Set \mytt{outlier\_count} as
a pseudo-random integer distributed uniformly from \mytt{outlier\_range} inclusive,
using \mytt{out\_seed} as the seed.
The highest
\mytt{outlier\_count} AIDVs from the sorted list are selected as the \mytt{outlier\_group}.

In \textbf{QH step 5.4}, the top seed is generated as:
\begin{lstlisting}
top_seed = h(flat_seed,'top')
\end{lstlisting}

Set \mytt{top\_count} as
a pseudo-random integer distributed uniformly from \mytt{top\_range} inclusive,
using \mytt{top\_seed} as the seed.
The \emph{next highest}
\mytt{top\_count} AIDVs from the sorted list are selected as the \mytt{top\_group}.

In \textbf{QH step 5.5}, compute \mytt{top\_avg}, the average contribution of the AIDVs in the \mytt{top\_group}.

In \textbf{QH step 5.6}, for each AIDV in \mytt{outlier\_group}, compute the difference between the AIDV's contribution and \mytt{top\_avg}. Compute \mytt{bucket\_count\_adjustment} as the sum of the differences.

In pseudo-code:
\begin{lstlisting}
Initialize bucket_count_adjustment to 0
For each contribution in outlier_group:
    bucket_count_adjustment -=
        (contribution - top_avg)
\end{lstlisting}

In \textbf{QH step 5.7}, possibly increase the value of \mytt{base\_sd} to protect the
presence or absence of AIDVs in \mytt{top\_group} and the now-flattened \mytt{outlier\_group},
as follows:

\begin{lstlisting}
base_sd *= max(flattened_avg, (0.5*top_avg))
\end{lstlisting}

where \mytt{flattened\_avg} is the average contribution of all users after
flattening (QH step 5.6).

\subsubsection{QH step 6: Add noise}
\label{sec:qh-step-6}

Diffix Elm adds two noise samples (called \emph{noise layers}) to each
bucket. Both noise layers
are taken from a zero-mean Gaussian distribution. As with the suppression decision,
the noise layers are sticky by virtue of seeding.
One of the noise layers is the \emph{aid-layer}, and is seeded from the AIDVs
(\mytt{aid\_seed} from Section~\ref{sec:seed-base}) (for the AID selected for the \mytt{base\_sd}).
The other is the \emph{sql-layer}.
It is seeded by components of the SQL itself and the bucket values
(\mytt{sql\_seed} from Section~\ref{sec:seed-base}).

The per-bucket input variables for \textbf{QH step 6} are the \mytt{AIDV set} determined in QH Step 5, \mytt{bucket\_values}, the information associated from the SQL \mytt{GROUP BY} columns (\mytt{column\_name}, \mytt{range\_param}, and \mytt{range\_type}), and the \mytt{bucket\_count} (adjusted in QH Step 5).

The steps for adding noise are:

\begin{description}
\item[\textbf{QH step 6.1:}] Generate the per-bucket seeds.
\item[\textbf{QH step 6.2:}] Generate noise samples from the seeds.
\item[\textbf{QH step 6.3:}] Add the noise samples to the
    \mytt{bucket\_counts}, and round to the nearest integer.
\item[\textbf{QH step 6.4:}] If the resulting noisy count is less than
\mytt{low\_thresh}, then set to \mytt{low\_thresh}.
\end{description}

For \textbf{QH step 6.1}, the seed for the aid-layer is:
\begin{lstlisting}
noise_aid_seed = h(aid_seed,'noise')
\end{lstlisting}

The seed for the sql-layer is:
\begin{lstlisting}
noise_sql_seed = h(sql_seed,'noise')
\end{lstlisting}

In \textbf{QH step 6.2}, two noise layers are generated
as a pseudo-random sample from a zero-mean Gaussian
distribution, each using the corresponding seed.
The standard deviation for each layer is:

\begin{lstlisting}
noise_sd_layer = base_sd / sqrt(2)
\end{lstlisting}

In \textbf{QH step 6.3}, the noise layer or layers associated with each bucket
are added to the \mytt{bucket\_count}. The resulting
noisy count is rounded to the nearest integer.

Finally in \textbf{QH step 6.4}, if the noisy count is less than \mytt{low\_thresh}, then
it is set to \mytt{low\_thresh}. This is done simply to ensure that the count in the
answer is not less than the suppression mechanism allows.

\subsection{Query handling for count distinct}
\label{sec:distinct}

The aggregate \mytt{count(DISTINCT column)} is handled differently than \mytt{count(*)} or \mytt{count(column)}. Distinct query handling (DQH) has the following steps:

\begin{description}
\item[\textbf{DQH step 1:}] Transform the query into its corresponding \mytt{GROUP BY} query.
\item[\textbf{DQH step 2:}] Determine which buckets of the \mytt{GROUP BY} query would be suppressed using QH steps 1 to 3.
\item[\textbf{DQH step 3:}] For each AID, compute the contribution of each protected entity to the suppressed buckets. If there are no suppressed buckets, stop here and release the true \mytt{distinct\_count}.
\item[\textbf{DQH step 4:}] Based on these contributions and using QH step 5, compute the flattening amount and adjust the true \mytt{distinct\_count}. Likewise compute \mytt{base\_sd}.
\item[\textbf{DQH step 5:}] Compute sticky noise using QH step 6, and apply the noise to the (adjusted) \mytt{distinct\_count}.
\end{description}

These steps are specified in the following sections. (The discussion motivating this algorithm can be found in Section~\ref{atk:outlier-bucket-distinct}.)

\subsubsection{DQH step 1: transform the query}

As a first step, \mytt{count(DISTINCT col)} is transformed into a histogram of \mytt{col} values: \mytt{SELECT col, count(*)}.

For example, the query:
\begin{lstlisting}
SELECT count(DISTINCT col)
FROM table
\end{lstlisting}
would be transformed to its corresponding \mytt{GROUP BY}:
\begin{lstlisting}
SELECT col, count(*)
FROM table
GROUP BY 1
\end{lstlisting}
Note that this transformation doesn't literally occur, only that the behavior of the algorithm operates as though it did. (This conceptual perspective allows us to reuse the QH steps.)

If the query is \mytt{SELECT c1, c2, count(DISTINCT col)}, then the transformed \mytt{GROUP BY} is \mytt{SELECT c1, c2, col, count(*)}, and subsequent DQH steps operate per \mytt{c1,c2} bucket.

\subsubsection{DQH step 2: Determine suppressed buckets}

Set \mytt{distinct\_count} to the true count of distinct column values.

Using the transformed query, run QH steps 1-3 to determine which buckets would be suppressed. If there are no suppressed buckets, then use \mytt{distinct\_count} as the answer (don't perturb).

\subsubsection{DQH step 3: Compute per-AIDV contributions}
\label{sec:dqh3}

For computing \mytt{count(*)} or \mytt{count(column)}, the contribution of each AIDV is the number of rows pertaining to that AIDV. Computing the contribution of each AIDV for \mytt{count(DISTINCT column)} is different.

In this step, for each AID, each distinct suppressed column value is assigned to one and only one AIDV. The contribution of each AIDV (for a given AID) is the number of distinct suppressed columns values assigned to it.

To do this, the following steps are executed per AID:

\begin{description}
\item[\textbf{DQH step 3.1:}] For each AIDV, list all of the suppressed column values for which the AIDV is a bucket member.
\item[\textbf{DQH step 3.2:}] Sort the AIDVs according to the number of suppressed column values ascending.
\item[\textbf{DQH step 3.3:}] Repeatedly traverse the sorted list until all suppressed column values are assigned to one AIDV. For each encountered AIDV X, if there is an associated suppressed column value that is not assigned to any AIDV, then assign it to AIDV X. (If there are no unassigned values, then AIDV X may be removed from the sorted list.)
\item[\textbf{DQH step 3.4:}] The contribution of each AIDV is the number of suppressed column values assigned to it.
\item[\textbf{DQH step 3.5:}] The \mytt{AIDV set} (used for seeding) is the set of AIDVs with a contribution of 1 or more.
\end{description}

\subsubsection{DQH steps 4 and 5: Compute and apply flattening and base\_sd}

Execute QH steps 5 and 6, using the per-AID contributions computed in DQH step 3, but substituting \mytt{distinct\_count} for \mytt{bucket\_count}.

\subsection{Relation to k-anonymity and Differential Privacy}
\label{sec:relate-k-dp2}

Diffix Elm has deep similarities with both k-anonymity and Differential
Privacy (DP).

Suppression in Diffix Elm and the grouping of K identical pseudo-identifiers
in k-anonymity serve the same purpose: to prevent trivial singling-out by
simply inspecting the data. Both mechanisms force column data to pertain to
at least \emph{so-many} protected entities. In k-anonymity, \emph{so-many} is
defined by K. In Diffix Elm, \emph{so-many} is bounded by
\mytt{low\_thresh}, and its statistical average behavior is determined
by the three parameters \mytt{low\_thresh}, \mytt{supp\_sd}, and
\mytt{low\_mean\_gap}. Indeed, suppression is a key mechanism in
k-anonymity (along side generalization).

DP also needs to prevent trivial singling-out by
data inspection, but in general it does so by simply having no
mechanism for displaying column values. Rather, it forces the
analyst to state what the column values may be, and then responds
with a noisy answer.

Noise in Diffix Elm and in many DP designs serve
the same purpose: to obscure counts that may otherwise lead to 
high-precision inferences. Both Diffix Elm and DP require that
the amount of noise be proportional to the contributions of
heavy contributors. DP refers to this as sensitivity.

Both DP and Diffix Elm have the concept of flattening. In DP,
an administrator may for instance configure bounds like the maximum
row count or the maximum contribution to a sum. These bounds
both determine the amount of noise, and determines how much any
given protected entity can contribute. Protected Entities that contribute more
are flattened to the bound. By contrast, Diffix Elm determines the
amount of flattening and the amount of noise based on the
contents of the data itself. This leads to more accurate results
and simpler configuration, but at the expense of less privacy
in certain rare cases.

\section{Evaluation Methodology}
\label{sec:criteria}

Attacks on anonymity mechanisms have three key aspects:
\begin{enumerate}
  \item The \emph{effectiveness} of the attack by some meaningful
      \emph{measure of anonymity}.
  \item The \emph{prior knowledge} required by the attacker if any,
  \item The \emph{data conditions} (or other conditions)
      necessary for the attack.
\end{enumerate}

In an ideal world, an anonymity mechanism should be so powerful that
no possible attack (known or unknown)
is effective regardless of the data conditions and the attacker's
prior knowledge. Differential Privacy (DP) can achieve this
ideal when its privacy measure (Epsilon) is sufficiently low and
other conditions are met (reasonable assumptions, lack of side-channel
attacks), but success in doing so comes at the cost of very poor
data utility and usability.

Diffix Elm achieves remarkably good data utility and usability, but doing so
comes at the cost of having to do a risk assessment to demonstrate anonymity.
This risk assessment requires that we
design and measure attacks, and show that each
attack is either ineffective, or that the cost of running the attack, especially
in terms of obtaining the necessary prior knowledge, is
substantially greater than the benefit of doing so (or for all practical
purposes not feasible). It must also be the
case that a serious, transparent, and open effort was made to find all
possible attacks.

The measure of anonymity we use for Diffix Elm is based on
common sense notions of privacy that are easy to relate to.
We use two measures,
\emph{Precision Improvement} (PI) and \emph{Prediction Rate} (PR).
PI is a measure of how likely a prediction made by an attacker is
correct.
PR is a measure of how likely a high PI can be made on a
randomly chosen individual in the dataset. We can define thresholds
for PI and PR, below which Diffix Elm may be regarded as anonymous
relative to that specific attack.

PI/PR are common-sense intuitive measures in three respects. First,
the more uncertain an attacker is about a prediction, 
the stronger the privacy protection (related to PI). Second, the less
likely an attacker is to make a high-precision prediction, the
less likely a given individual's privacy is compromised (related to PR).
Finally, the less likely an attacker is to get a good PI or PR, the
less incentive the attacker has to try in the first place.

If this PI/PR measure shows that a
given attack is ineffective, then it doesn't matter how easy it is
to obtain the prior knowledge, or how common the data conditions are:
the attack is still ineffective and Diffix Elm is anonymous for that attack.

If on the other hand the PI/PR measure shows that the attack is more effective
than is comfortable (above the anonymity threshold), but it is shown
that the data conditions don't exist in the dataset, then again Diffix
Elm can be regarded as anonymous \emph{for that attack and associated dataset}.

If, finally, the PI/PR measure is not below threshold, and the data conditions
exist, then we must consider how likely it is that the attacker has,
or is willing to get, the necessary prior knowledge. If it is very unlikely
that the attacker has or is willing to get the prior knowledge (i.e. Class C),
then Diffix Elm can be regarded as anonymous \emph{for that attack, associated
dataset, and prior knowledge}.

We define three classes of prior knowledge (see Section~\ref{sec:pk-class}):
\begin{description}
  \item[Class A:] Knowledge of one individual is required for the attack.
  \item[Class B:] Knowledge of specific multiple individuals
      is required for the attack.
  \item[Class C:] Knowledge of specific multiple individuals, where the attribute
      being learned is known for most but not all of the individuals.
\end{description}
Class A prior knowledge is very common (everyone knows something about
someone). Class B is far less common, and Class C is very rare.

With this framework, we can define two PI/PR thresholds, one below
which Diffix Elm is always anonymous (\emph{Very Strong}),
and another (\emph{Strong}) below which
Diffix Elm is anonymous if the prior knowledge is Class C prior knowledge
(and the data conditions exist). While these PI/PR thresholds must be
set by a DPA or DPO, in our evaluation of Section~\ref{sec:evaluation}, we
define Very Strong and Strong thresholds as shown in Table~\ref{tab:thresholds}.

\begin{table}
\begin{center}
\begin{tabular}{l|ccl}
  \toprule
  Threshold & PI & PR & PK Class \\
  \midrule
  Very Strong & 0.5 & 1/100000 & any (A, B, or C) \\
  Strong & 0.5 & 1/1000 & Class C \\
  \bottomrule
\end{tabular}
\end{center}
  \caption{The PI/PR thresholds below which Diffix Elm is anonymous
  and associated prior knowledge Class. $PI=0.5$ means that an attacker's
  prediction about an individual in the data is correct only 50\% of
  the time, given that the predicted attribute is statistically rare.
  $PR=1/100000$ means that only one in 100000 randomly selected individuals
  have a high PI ($PI>0.95$).
  }
  \label{tab:thresholds}
\end{table}

The Very Strong threshold can be read as saying ``So long as $PI<0.5$ 
\textbf{or} $PR<1/100000$, anonymity is very strong''. $PI<0.5$ means that,
if the attacker is for instance predicting a rare attribute, there is
roughly a 50\% chance that the attacker is wrong. This in turn gives the
victim strong deniability, and therefore anonymity. Depending on the attack,
it is sometimes possible to occasionally get a higher PI. 
$PR<1/100000$ means that 1 in 100K predictions may
randomly (unpredictably) yield a high-precision prediction ($PI>0.95$).
This means that the risk of any given individual in the dataset
is very low, and therefore the system is anonymous.

The Strong threshold ($PI<0.5$ and $PR>1/1000$) can itself be regarded
as anonymous in many situations, for instance relatively non-sensitive
data shared privately. Nevertheless, when combined with a requirement for
Class C prior knowledge, it can be regarded as anonymous for virtually
any scenario.

\subsection{Classes of Prior Knowledge}
\label{sec:pk-class}

The three classes of prior knowledge are listed earlier in this section.
Here we motivate the need for defining multiple classes and describe
the classes through examples.

There is a widespread belief that almost any anonymization mechanism
can be broken if the right prior knowledge can be obtained, and that 
it can be surprisingly easy to obtain the right prior knowledge.
In his highly influential paper from 2010~\cite{ohm-failure}, Paul Ohm
cites three well-known attack demonstrations:
\begin{enumerate}
  \item Re-identifying the Governor of Massachusetts from a medical 
      dataset (2002, \cite{sweeney2002k}),
  \item Re-identifying Thelma Arnold from an AOL search dataset (2006,~\cite{aol3}),
  \item Re-identifying individuals from the Netflix dataset
      (2008, \cite{Narayanan:2008:RDL:1397759.1398064}).
\end{enumerate}
From these three examples, Ohm concludes that the evaluation of anonymization
technologies should assume that all necessary prior knowledge is known by
the attacker.

It is critical to note, however, that in all of the above examples, the
necessary prior knowledge needed to re-identify each individual
is knowledge about \emph{that one individual only}.
Furthermore, \emph{none} of the big-three mechanisms used by Diffix Elm (generalization,
suppression, and noise) were used in the above three examples. Rather
each of the datasets were only pseudonymized (removal of personally identifying
information, but otherwise complete records released). Attacking these
datasets is effectively a matter of obtaining enough prior knowledge of one
individual, and checking that only one individual in the dataset has
the matching prior knowledge.

We refer to prior knowledge about a single individual as \emph{Class A}
prior knowledge. Class A prior knowledge is indeed easy to come by, and
getting easier as more and more information about individuals can be found
online. An anonymization mechanism that depends on the attacker not having
Class A prior knowledge is certainly not anonymous.

To make this concrete, let's consider the example of re-identifying the
Governor of Massachusetts (the victim). To do the re-identification, the following
four items prior knowledge was required.

\begin{itemize}
  \item Knowledge that the victim is a patient of the hospital from which
  the dataset came (obtained from a newspaper story about the victim).
  \item The birthdate, zip-code, and sex of the victim
  (taken from public voter registration records).
\end{itemize}

Only one individual in the dataset had the same birthdate, zip-code, and sex.
These four items of information (membership, birthdate, zip, and sex) are
obviously easy to obtain, especially for acquaintances, friends, and family
but also for public figures.

Anonymization techniques that use some or all of the big-three
mechanisms are generally not susceptible to attacks using Class A
prior knowledge. Let's use k-anonymity, which uses generalization
and optionally suppression, as a simple example.

Assume a k-anonymized dataset containing birth-month, zip, sex, and
vaccination status (Table~\ref{tab:k-example}).
Suppose that an attacker knows the birth-month, zip, and sex of a
given individual (the victim), and wants to know whether the victim
has been vaccinated or not. Suppose that an attacker also knows
that there are 15 individuals in the dataset with the same
birth-month, zip, and sex, and also knows that 7 are vaxxed and
7 are unvaxxed. In other words, the attacker knows the vaccination
status of all individuals with the same birth-month, zip, and sex
except for the victim.

\begin{table}
\begin{center}
\begin{tabular}{llll|l}
  \toprule
  Birth-month & Zip & Sex & Vax & Count \\
  \midrule
  11-1995 & 12345 & Male & Yes & 7 \\
  11-1995 & 12345 & Male & No & 8 \\
  \bottomrule
\end{tabular}
\end{center}
  \caption{Part of a k-anonymized dataset used to illustrated Class C
  prior knowledge. If the attacker does not know the vaccination status
  of the victim, but knows 1) that the victim has the given birth-month,
  zip, and sex, and also knows 2) that there are exactly 7 vaxxed and
  7 unvaxxed individuals with the given birth-month, zip, and sex, then
  the attacker can deduce with 100\% precision the vaccination status
  of the victim.
  }
  \label{tab:k-example}
\end{table}

In this case, the attacker can infer with 100\% precision the vaccination
status of the victim, because that is the entry with 8 individuals.
This is an example of Class C prior knowledge, and how it can be used
to infer information about an individual from k-anonymity.
Specifically, the attacker knows about a specific group of individuals
(those with the given birth-month, zip, and sex), and also knows
about the attribute being learned (vaccination status) of most but
not all of the individuals.

There are two important points to make here. First, it is clear that
Class C prior knowledge is a very high bar: not impossible but quite
improbable. Second, if k-anonymity also added noise, as Diffix Elm
does, then even with this prior knowledge, the attacker would not
be able to infer with 100\% confidence the vaccination status of the
victim.

We don't have an example of an attack against k-anonymity using
Class B prior knowledge. Section~\ref{atk:lpr-random} gives an example
of an attack against Diffix Elm requiring Class B prior knowledge.
Table~\ref{tab:attacks-summary} lists several
attacks against Diffix Elm requiring Class C prior knowledge.

\subsection{PI/PR Measure of Anonymity}
\label{sec:measure-anon}

Our evaluation
methodology is to measure the \emph{success rate of analyst predictions}. A higher
success rate implies weaker anonymity.

This section starts with a number of examples that motivate the
kinds of predictions that PI/PR uses.
Section~\ref{sec:three-criteria} describes how the predictions
satisfy the three EDPB criteria~\cite{article29}, and
Section~\ref{sec:pi-pr-detail} describes the PI/PR measure in detail.

The PI/PR measure uses the following two predictions:
\begin{description}
\item[\textbf{Singling-out:}] There is exactly one individual with attributes A,
B, and C.
\item[\textbf{Inference:}] All individuals with attributes A, B, and C
also have attribute D.
\end{description}

As an example for singling-out, the analyst
may predict that there is a single individual\footnote{
  Here we use the term ``individual'' rather than ``protected entity'' because
  GDPR concerns itself with the protect of individuals (natural persons).
} with attributes
\mytt{(gender='male', age=48, zip=48828, lastname='Wade')}.
If this is true, then the analyst has correctly singled
out that individual. The attributes don't need to be personal
attributes as in this example. If the analyst correctly
predicts that there is a single individual with the geo-location
attributes \mytt{(lon=44.4401, lat=7.7491, time='17:14:22')},
then that individual is singled out.

On the other hand, if there are no individuals or more than one
individual with the attributes, then the prediction is false, and the
analyst has failed to single out an individual.

As an example for inference, the analyst may predict that
all individuals with attributes
\mytt{(gender='male', age=48, zip=48828)} also have
attribute \mytt{lastname='Wade'}.
As with singling out, the inference may be true or false.
(Note that strictly speaking an inference could refer to a single
individual. In this case it can be regarded as either
an inference or a singling out. It doesn't matter which.)

We define \emph{precision} as the
number of correct predictions divided by the number of total predictions.
If we think of a prediction as defined above as a Positive prediction,
a correct prediction as a True Positive (TP), and an
incorrect prediction as a False Positive (FP), then
this is exactly analogous to the definition of precision in statistics
or machine learning as $precision = TP/(TP+FP)$.

Of course, for an analyst to be able to make a prediction, the analyst must
have some basis for the prediction. In other words, the analyst must have an
\emph{attack} that allows them to make a prediction. Our evaluation
methodology measures precision \emph{for
each known attack}. Obviously the quality of our evaluation depends on
our (and others') ability to come up with possible attacks. The
limitations associated with this approach are discussed in
Section~\ref{sec:eval-limitations}.

This all begs the question, ``What constitutes good precision?'' Is
50\% precision good? 90\% precision? In fact,
it depends on the situation.

By way of example, suppose that the analyst has prior knowledge of 1000
individuals that are known to be in a table, and knows the email
addresses of these individuals, where email address is a table
attribute. By merely having this prior knowledge, the 1000
individuals are effectively singled out with perfect precision.
If the analyst ran an attack that required this
prior knowledge to work, and obtaining the 1000 email addresses
as a result, this would
not constitute an effective attack per se because
nothing new was revealed.

Now suppose that the analyst wants to predict the political party of these 1000
individuals. Suppose further that roughly 40\% of all individuals in the
table are Tory, and 40\% are Labour. Indeed the analyst can learn this
by querying Diffix Elm itself (\mytt{SELECT party, count(*) FROM
table}). The analyst could then make 1000 singling-out predictions of the form
\mytt{(email, 'Tory')} without any additional queries, and get roughly
40\% precision. Clearly this precision also does not constitute an
effective attack because it does not improve on the baseline prior
knowledge that 40\% of the individuals are Tory.

In the above examples, the baseline probability was based
on the population of the entire table.  Suppose, however, that the
prior knowledge of the analyst includes zip code as well as email
address. Since some zip codes are in more conservative districts, while
others are in more liberal districts, the analyst can improve the
success rate simply by always predicting Tory in conservative districts, and
Labour in liberal districts. Nevertheless, the improvement over always
predicting Tory still does not constitute an effective attack, because
the extent to which conservative districts have more Tories is already
known.

On the other hand, suppose that instead the analyst wants to predict
whether these 1000 individuals have a PhD. Suppose further that only 1\%
of the individuals in the table have a PhD (also learn-able with a query
to Diffix Elm). Now suppose that with some clever attack the analyst is
able to achieve 40\% precision (i.e. it makes for instance 20
predictions, and 8 are correct, these 8 being 8 of the 10 individuals among
the 1000 with PhDs).  In this scenario, 40\% is a better success
rate, because it improves substantially on the baseline of 1\%.

This example illustrates that it is not the \emph{absolute} precision that
matters, but the precision \emph{relative} to some baseline.
Further, this baseline is measured with respect to the general
population of individuals selected by the attack, not with respect to
all individuals in
the table. We refer to this measure as \emph{Precision Improvement} (PI).

While PI is the primary measure of an attack's effectiveness, there is a
second measure that is sometimes important, \emph{Prediction Rate} (PR). This
is needed because sometimes an analyst can improve PI by making predictions
from fewer attacks. For instance, suppose in the clever PhD attack
described above the analyst ran the attack 20 times, each attack
produced one \emph{prediction opportunity}, and indeed the analyst made 20
predictions, leading to a PI of 40\%. Here the prediction rate (PR) is 100\%
(every prediction opportunity led to a prediction), and the PI is 40\%. 

Now suppose that there is a variant of the clever attack whereby the
analyst knows that some predictions are more likely to be correct than
others. The analyst could improve PI by making fewer predictions relative to
the prediction opportunities. So for instance the analyst might be able to
improve PI to 80\% by making only 5 predictions. In this case, PR is 25\% (5
predictions of 20 prediction opportunities).

Note that PR is similar to but not the same as recall. Recall is defined
as the number of true positives divided by the number of all positives,
or $recall = TP / (TP+FN)$, where FN is False Negative. Recall doesn't
quite make sense in this context because our predictions are all positive
predictions: making a negative inference or a non-singling-out are not defined
as criteria for anonymity by the EDPB opinion on
anonymity~\cite{article29}. We can't measure recall without negative
predictions.

An anonymization mechanism can be still regarded as anonymous even when
the PI is quite high so long as the corresponding PR is very low (see
Section~\ref{sec:relate-gdpr}).

\subsubsection{IDPB three criteria for anonymity}
\label{sec:three-criteria}

The EDPB opinion gives three distinct criteria for anonymity:
\emph{singling out}, \emph{inference}, and \emph{linkability}.
The opinion serves to both evaluate a number of well-known
anonymization techniques, and to provide guidance for evaluating
anonymization techniques not covered by the opinion. This section
examines in more detail how the EDPB criteria apply to Diffix Elm.

The EDPB opinion defines singling out as:
\begin{quote}
Singling out, which corresponds to the possibility to isolate some or all records
which identify an individual in the dataset.
\end{quote}

The prediction used in this paper for singling out reflects this
definition closely. Diffix Elm does not reveal records, but a set of
attributes revealed in a singling out attack may be interpreted as
a record. Predicting that one individual has the set of attributes
corresponds to isolating.

The EDPB opinion defines inference as:
\begin{quote}
Inference, which is the possibility to deduce, with significant probability, the
value of an attribute from the values of a set of other attributes.
\end{quote}

The prediction used for inference matches this very well. Indeed the phrase
``with significant probability'' recognizes that deductions may be incorrect,
and so a way to measure precision is needed.

The EDPB opinion defines linkability as:
\begin{quote}
Linkability, which is the ability to link, at least, two records concerning the same
data subject or a group of data subjects (either in the same database or in two
different databases). If an attacker can establish (e.g. by means of correlation
analysis) that two records are assigned to a same group of individuals but cannot
single out individuals in this group, the technique provides resistance against
``singling out'' but not against linkability.
\end{quote}

Compared to singling out and inference, the definition of linkability
is less crisp. Indeed, the definition of what constitutes linking is
quite different for different mechanisms. For pseudonymization, 
it can relate to either associating the records within a dataset that
have the same IDs, or associating
individual records with those in external datasets. For noise addition
and permutation, it can likewise refer to associating individual records
with those in external datasets.

In each of these cases, linking is possible because there is a 1-1
correlation between records in the anonymized/pseudonymized dataset
and the original dataset. The Diffix Elm equivalent to an individual
record would be a record composed of the attributes of a singled out
individual. As such, linkability by this definition is only possible
in Diffix Elm if singling out has taken place. Therefore, the
singling out prediction itself encompasses linkability.

In the case of aggregation (k-anonymity, l-diversity, or t-closeness),
the EDPB opinion considers linking to take place merely by observing
that the records comprising a group of k individuals (i.e. a group
with the same attributes) are linked by virtue of having the same
attributes. This is effectively a tautology and doesn't appear to
represent a privacy violation in any meaningful way. Nevertheless,
the same linking takes place with any bucket produced by Diffix Elm.

In the case of Differential Privacy, the EDPB opinion considers linking
to take place when answers to two queries comprise the same set of
individuals. This interpretation can apply to Diffix Elm, but either
the linking is trivial (the answers to the same queries), or there
are no known attacks. Even if there were attacks, however, there is no
reason to believe that this constitutes a privacy violation.

Given the above, either we see no good way to design predictions based
directly on linkability (aggregation or DP), or the singling out 
prediction encompasses linkability.

\subsubsection{PI and PR in detail}
\label{sec:pi-pr-detail}

With the above intuition in place, we can now specify how PI and PR are
computed in detail. Note that this evaluation methodology is defined by the
GDA Score Project~\cite{gda-score}. A software library for computing PI
and PR is available on Github~\cite{gda-github}\footnote{GDA Score uses different
terminology: Confidence instead of Precision, and Claim Rate instead of
Prediction Rate, but the concepts are the same.}.

\begin{table}
\begin{center}
\begin{tabular}{p{0.2\columnwidth}p{0.6\columnwidth}}
  \toprule
  $TP$ & The number of correct predictions \\
  $FP$ & The number of incorrect predictions \\
  $TP+FP$ & The number of predictions \\
  $PO$ & The number of prediction opportunities \\
  $N^p_{ua}$ & The number of individuals with the given unknown
               attributes for prediction $p$ \\
  $N^p_{ka}$ & The number of individuals with the given known
               attributes for prediction $p$ \\
  \bottomrule
\end{tabular}
\end{center}
  \caption{The variables used to compute PI and PR}
  \label{tab:pipr}
\end{table}

If a given attack yields $PO$ prediction opportunities, and the analyst makes
$TP+FP$ predictions, then the prediction rate PR is simply 

\begin{equation}
  \label{eq:pr}
PR = (TP+FP)/PO
\end{equation}

PI is measured as:

\begin{equation}
  \label{eq:pi}
PI = (P-B)/(1-B)
\end{equation}

where $P$ (precision) is the ratio of correct predictions to total predictions
$P = TP/(TP+FP)$, and $B$ is the baseline probability of a correct prediction.

To compute $B$, we need to understand
what is a priori known by the analyst, and what is unknown (i.e. what is
being learned). Furthermore, of the known attributes, we use only those
that lead to the best baseline probability. So for instance if
\mytt{email} and \mytt{zip} are known, and the political \mytt{party}
is unknown, then we only use \mytt{zip} as the known attribute since
it yields the best baseline prediction for \mytt{party}.

$B$ is computed as the fraction of individuals that have the predicted unknown
attributes compared to the total number of individuals that have the used
known attributes. Since different predictions may have different known and
unknown attributes (i.e. the different known \mytt{zips} and unknown
\mytt{parties} in the attack above), $B$ is computed as the average over all
predictions:

\begin{equation}
B = (\sum_p N^p_{ua}/N^p_{ka}) / (TP+FP)
\end{equation}

\subsection{Limitations}
\label{sec:eval-limitations}

The key limitation of this attack-and-measure evaluation approach is
that it requires that all attacks are known. In practice there is no
guarantee that all possible attacks have been found. For all practical
purposes, however, this limitation exists for all anonymization
mechanisms. For instance, in the years following the definition of
k-anonymity~\cite{Sweeney:2002:KAM:774544.774552}, a series of attacks
and weaknesses were discovered, leading to improvements like
l-diversity~\cite{MachanavajjhalaGKV06} and
t-closeness~\cite{t-closeness}.

Not even Differential Privacy (DP), with its mathematical guarantees of privacy,
is exempt from an informal attack-based evaluation in practice. For instance,
severe side-channel attacks~\cite{dp-under-fire,ccs21-side-channel} have been
found in several prominent query-based DP designs,
including PINQ~\cite{pinq}, Airavat~\cite{roy2010airavat},
Chorus~\cite{DBLP:journals/pvldb/JohnsonNS18} (used in-house
by Uber), and as well in an earlier version of Diffix, Diffix Birch.

Even the proposed DP release of the US Census must
effectively undergo an informal privacy evaluation. The reason is
because the US Census plans on using a budget of around 20. A budget
this high does not provide a formal guarantee of privacy. The noise of a
single query with an Epsilon of 20 is well below 0.5 with very high
probability, thus being able to definitely expose the presence or
absence of a single user under the right circumstances. As such, the US
Census release must rely on informal non-DP
mechanisms, such as generalization, to argue that the data release is
private.

Having said all that, it is far easier to reason about simple
mechanisms than complex mechanisms. Diffix Elm is far simpler than
earlier versions of Diffix, and so an attack-and-measure approach is
more tenable.

It is worth pointing out that the EU criteria are conservative. The
possibility of singling-out, for instance, does not necessarily imply
that an attack is practical. It may be, for instance, that singling-out
is possible for only certain attributes or certain individuals, and that
these are not of interest to an attacker.

It could also well be that,
even though individuals can be singled out, they can't be identified.
For example, suppose that a single individual with the geo-location
attributes \mytt{(lon=44.4401, lat=7.7491, time='17:14:22')} is singled
out. This is of little value to an attacker unless the individual can
also be identified. The criteria for the anonymity of Diffix Elm do not
rely on the ability to identify, only to single out, link, or infer.

\subsection{Relation to k-anonymity and Differential Privacy}
\label{sec:relate-k-dp}

It is customary to measure the strength of anonymity as $K$ for
k-anonymity and as $\epsilon$ and optionally $\delta$ for DP. These
measures are specific to the mechanisms of k-anonymity and DP, and
don't apply to Diffix Elm. The reverse, however, is not the case.
The PI/PR measure can also be applied to k-anonymity and DP.
In this sense, PI/PR is a more general measure.

By way of example, consider the simple attack of
Section \ref{atk:sim-know-noise}. Here the attacker knows that there
are either $N$ or $N+1$ individuals with a certain set of attribute values.
If the attacker can determine that there are $N+1$ individuals, then
the attacker knows that the victim has those values and the victim is
singled out.

For DP and Diffix Elm, different PI and PR values may be obtained
depending on how much noise is added. The values for Diffix Elm,
displayed for three different noise settings, is shown in
Figure~\ref{fig:atk:sim-know-noise}. The values for DP would depend
on a number of factors, but if ($\epsilon$,$\delta$) DP is used, then
there would similarly be data points with low-PI high-PR as well
as data points with low-PR high-PI (though likely with better
values than in Figure~\ref{fig:atk:sim-know-noise}).

K-anonymity, on the other hand, does not protect against this
particular attack, and so would have $PI=1.0$ and $PR=1.0$, the
worst possible measure.

\subsection{Relation to GDPR}
\label{sec:relate-gdpr}

GDPR recital 26 states that data is anonymous when
\emph{``the data subject is not or no longer identifiable''}.
GDPR recital 26 further states that:

\begin{quote}
\emph{To determine whether a natural person is identifiable, account should be
taken of all the means reasonably likely to be used, such as singling out ...}

\emph{To ascertain whether means are reasonably likely to be used to identify
the natural person, account should be taken of all objective factors,
such as the costs of and the amount of time required for identification ... }
\end{quote}

The evaluation of Diffix Elm in this document is designed to support
a determination as to whether identification means are reasonably
likely to be used. It measures the effectiveness of all known attacks.
The measure, PI/PR, is directly related to metrics meaningful to
an attacker: the likelihood and frequency of correct predictions.
The predictions themselves are based on the three criteria for
anonymity set forth by the EDPB opinion on anonymity~\cite{article29}.
Given the sensitivity or value of a given dataset, a DPA or DPO can
therefore make a reasonable estimate as to what PI and PR thresholds
would render the data as having no value or marginal value to the attacker.

The evaluation in this paper also describes the prior knowledge required
to execute each attack. Using this, a DPA or DPO can estimate the cost
to an attacker of obtaining the necessary prior knowledge. So long as
the cost of obtaining the prior knowledge substantially exceeds the
marginal value of the data, or exceeds the cost of identifying data
subjects by other means, Diffix Elm may be regarded as anonymous
by GDPR standards.

\section{Evaluation}
\label{sec:evaluation}

The first version of Diffix was published in 2017~\cite{diffix-aspen}.
In the four years that have elapsed, numerous attacks have been
discovered, and corresponding defenses designed. Most of those attacks
are documented in~\cite{diffix-dogwood-tr}.  This section evaluates
the effectiveness of all these attacks, plus a few additional attacks,
on both modes of Diffix Elm. This set of attacks represent all \emph{known
attacks}. The attacks have been discovered through
our own analysis, and by others from our open
publications~\cite{diffix-aspen,diffixBirch,diffix-cedar-tr,diffix-dogwood-tr}
and the bounty programs~\cite{bounty2017,diffix-bounty-2020-tr}. We believe
that the probability of attacks being known to others and not to us is
very low. The probability that there remain attacks unknown to anybody
is somewhat higher.

A substantial fraction of the known attacks for prior versions of Diffix
cannot be run on Diffix Elm
simply because the required SQL is not supported. Those attacks are
listed in Table~\ref{tab:non-attacks} along with reason the attack cannot
be executed. (Note that the name of the attack may differ slightly from those
of ~\cite{diffix-dogwood-tr}, but the correspondence should be clear.)
In particular, both published attacks discovered by external researchers
on prior versions of Diffix are in Table~\ref{tab:non-attacks}.
These include the attack by Gadotti et.al.~\cite{montjoyeAttack} on
Diffix Birch, labeled \emph{Noise exploitation: chaff conditions} in
Table~\ref{tab:non-attacks}, and the attack by Cohen and
Nissim~\cite{cohen2019linear}, labeled \emph{Linear program reconstruction:
random user groups} in Table~\ref{tab:non-attacks}.

\begin{table*}
\begin{center}
\begin{small}
\begin{tabular}{rl}
  \toprule
  Attack & Reason attack cannot execute \\
  \midrule
   Noise magnitude report & Not reported \\
   Averaging: different syntax, floating & Not enough syntax options \\
   Averaging: different syntax, no floating & Not enough syntax options \\
   Averaging: split & No negative AND \\
   Tracker & No OR \\
   Linear program reconstruction: random user groups & No math \\
   JOINs with non-personal tables & No JOIN \\
   Difference: First derivative, negative AND & No negative AND \\
   Difference: Counting NULL & No count(col) \\
   Noise exploitation: chaff conditions & No WHERE clause \\
   Noise exploitation: extreme contribution & All protected entities contribute 1 \\
   Multiple isolating negative AND & No negative AND \\
   Shadow table exploitation & No shadow table \\
   SQL backdoor & No math \\
   Side channel: Divide by zero & No divide function \\
   Side channel: Square root of a negative number & No square root \\
   Side channel: Overflow & No math \\
   NULL producing safe function: IS NOT NULL & No safe functions \\
   NULL producing safe function: NULL within aggregation & No safe functions \\
   Side channel: JOIN timing attack & No JOIN \\
  \bottomrule
\end{tabular}
\end{small}
\end{center}
  \caption{List of attacks from prior versions of Diffix~\cite{diffix-dogwood-tr}
      that cannot be executed because the required features don't exist
      in Diffix Elm}
  \label{tab:non-attacks}
\end{table*}

\subsection{Additional evaluation information}
\label{sec:add-eval-info}

Many of the attacks described here are demonstrated in software posted in 
the repo https://github.com/diffix/attacker under the directory
\mytt{diffixElmPaperAttacks}. The code for each individual attack is in 
a subdirectory, the name of which is given in each attack's evaluation.

In general, we tested three anonymization parameter settings, which we refer
to as Private (P), Extra Private (XP) and Extra Extra Private (XXP). The
P settings are the minimum values allowed by Diffix Elm. XXP represents 
extremely strong settings beyond which diminishing privacy returns accrue.
The values are shown in Table~\ref{tab:eval-params}. Note that the parameters
associated with suppression can be configured independently from those
associated with noise. We group them here for experimental convenience.

\begin{table}
\begin{center}
\begin{small}
\begin{tabular}{r|lll}
   \toprule
   Parameter & P & XP & XXP \\
   \midrule
   \mytt{low\_thresh} & 2 & 2 & 2 \\
   \mytt{low\_mean\_gap} & 2 & 3 & 4 \\
   \mytt{supp\_sd} & 1 & 1.5 & 2 \\
   \mytt{base\_sd} & 1.5 & 2.25 & 3.0 \\
   (Per layer SD) & 1.0607 & 1.5910 & 2.1213 \\
   \bottomrule
\end{tabular}
\end{small}
\end{center}
  \caption{Privacy settings tested in this evaluation.
  Note that \mytt{low\_thresh} is not modified because it does not
  influence the results of any of the attacks per se. Nevertheless it is
  an important parameter for suppression (See Section~\ref{atk:attr}).
  }
\label{tab:eval-params}
\end{table}

The parameters \mytt{outlier\_range} and \mytt{top\_range}
only apply to attack \emph{Detect outlier bucket}~\ref{atk:outlier-bucket}.
The associated settings are discussed there.

Note that in many of the attacks, we use the function \mytt{count()}.
This is to be interpreted as either \mytt{count(DISTINCT aid)} or
\mytt{count(*)}. In general, however, unless the victim is a heavy
contributor to the number of rows, using \mytt{count(DISTINCT aid)} is a
better approach for the attacker because otherwise there will be more
noise relative to the contribution of the victim.

\subsection{How to interpret the graphs}

Much of the evaluation is illustrated with scatterplots of
Precision Improvement (PI)
and Prediction Rate (PR). An example is Figure~\ref{fig:atk:sim-know-noise}.
These plots have a shaded area where PI and PR do not meet the
thresholds for Very Strong anonymity as given in Table~\ref{tab:thresholds}.
Data points outside of this shaded ``risk area'' may be regarded as
attacks where anonymity is preserved.

The shaded area is green in cases where the prior knowledge is Class C
(very unlikely), or where a malicious analyst has access to a trusted-mode
interface. The shaded area is red otherwise.
Because Class C prior knowledge is so unlikely, attacks with data points
within a green risk area may still be regarded as anonymous. Attacks
with data points in a red risk area, however, might not be regarded as
anonymous. (For color blind readers, the prior-knowledge class and trust mode
are also summarized in Table~\ref{tab:attacks-summary})

Of course, it is up to the DPA or DPO to determine the thresholds for
anonymity. We believe, however, that the thresholds we have selected
are very conservative.

Note that, unless otherwise stated, all evaluation data uses the
\mytt{count(DISTINCT aid)} form of attack.

\subsection{Multiple AIDs}

The noise amount (standard deviation) used for any given bucket comes from the worst-case per-AID noise amount. The purpose of noise is to hide the effect of any given protected entity. Because noise amount is worst-case, the noise chosen to protect the worst-case AID naturally protects entities defined by other AIDs. Because of this, with respect to noise amount, it suffices to evaluate the various attacks on the assumption that there is a single AID.

The flattening amount used for any given bucket comes from the worst-case per-AID flattening amount. The only attack that tries to exploit the flattening mechanism is the \emph{Detect outlier bucket} attack (Section~\ref{atk:outlier-bucket}). The effect of multiple AIDs is discussed in that section.

\subsection{Attribute value inspection}
\label{atk:attr}

\paragraph{Prior Knowledge: None}
This attack requires no prior knowledge.

\paragraph{Additional conditions: None}
There are no additional conditions.

\paragraph{Goal}
In this attack, the attacker wishes to single-out protected entities by
simply displaying the column values. If any set of one or more
column values pertain to a single protected entity, then the attack succeeds.

\paragraph{Attack}
A query in this attack selects one or more columns, where each resulting
set of values would isolate protected entities were it displayed (regardless
of the count).

\begin{lstlisting}
   SELECT col1, col2, count() FROM ...
\end{lstlisting}

\paragraph{Evaluation}
The suppression mechanism suppresses any output rows that pertain
to fewer than \mytt{low\_thresh} protected entities. Since the minimum value
of \mytt{low\_thresh} is 2, a set of column values for a single protected entity
will never be displayed. Therefore strictly speaking, $PI=0$.

\paragraph{Discussion}
It is important to note, however, that while a count of 2 may strictly speaking
satisfy GDPR requirements for not singling out, there may be conditions in 
the data that nevertheless lead to privacy loss. These are discussed in 
sections~\ref{sec:relate-ind-pe} and \ref{sec:small-group}.  So long as
\mytt{low\_thresh} is set carefully, anonymity is maintained (singling-out,
inference, or linkability does not occur).

\subsection{Unique inference}
\label{atk:unique-infer}

\paragraph{Prior Knowledge: Class A}
The attacker must know enough attributes about a victim to
know that the victim is in one and only one bucket.
Note that these attributes are not unique to the victim.
The attacker must also know that the victim is in the dataset.

\paragraph{Additional conditions: Common}
There are no particular conditions on the original data per se,
but the conditions required in any given output may or may
not exist. It could be that the data conditions necessary to
produce the output don't exist, or (more likely), the conditions
exist but don't manifest themselves because of the generalization
parameters chosen by the analyst.

\paragraph{Goal}
The goal is to infer an unknown attribute given a set of
known attributes.

\paragraph{Attack}
The attack can be run on the output of any given query.
Given an output where $N$ columns are selected,
the attacker inspects the output for any bucket whereby
the values for $N-k$ columns appears in only one bucket.
The attacker then infers the values for the remaining $k$
columns.

If the attacker knows a victim that matches the values of the
$N-k$ columns, and knows that the victim is in the dataset, the
attacker then infers the remaining $k$ values.

\paragraph{Evaluation}
PR for this attack is 1.0.

There are two reasons why a unique value inference can be made:
\begin{enumerate}
   \item All but one value has been suppressed.
   \item There is indeed only one unique value.
\end{enumerate}

In the first case, the attack's precision is less than 100\%,
because the true value might be one of the suppressed values,
and the prediction would be incorrect.
In the second case, the attack's precision is 100\%.

In both cases, however, PI is always zero: the attack precision is
the same as what would come from a statistical guess (were the
actual statistics known). In other words, for the second case,
even though precision is 100\%, that does not improve on a statistical
guess.

\paragraph{Discussion}
\label{sec:atk:unique-inference-discuss}
In spite of the fact that $PI=0$, a DPO or DPA might well regard
this attack as violating anonymity on the basis of the EDPB inference
criteria. Note that this attack is essentially the same as the
vulnerability in k-anonymity which is solved by l-diversity.

A DPO or DPA can examine the output of Diffix Elm to determine
if any output buckets satisfy the criteria for this attack. If any
do, the DPO or DPA can determine the precision of the resulting
inference, and the sensitivity of the inference if the precision
is high.

As a general rule, the precision of the inference is lower when
the number of distinct AIDVs (protected entities) is low. When
there are multiple distinct values and associated buckets,
but all of them have a low number of AIDVs, then it can
easily happen that all but one of the buckets is suppressed.
In this case, precision is low and privacy is maintained
both by the PI measure and the absolute precision measure.

If on the other hand the number of distinct AIDVs in the uniquely
inferred bucket is large, then most likely the absolute precision
will be high. The exception would be where there are a large number
of suppressed buckets, such that the number of distinct AIDVs in the
suppressed buckets is about the same or more than the number of
AIDVs in the non-suppressed bucket.

Assuming that there are few or no suppressed buckets, then the
absolute precision is high. In this case, an important consideration
is whether the distribution of the unique value of values (the
$k$ column values) is substantially different in the context of the
values of the $N-k$ column values, than in the context of the entire
dataset. If it is not substantially different, then the unique
inference is not surprising, and privacy is not lost (the zero PI
value is an accurate indicator of real privacy loss).

To give an example, suppose that some column $U$ has a value $V_u$
which occupies 90\% of all rows. Further, suppose that a given 
unique inference bucket has a count of 20. We would then expect that
there are two additional rows that have a value other than $V_u$
for column $U$, and most likely these would be suppressed.
This would almost certainly be an acceptable unique inference.

If on the other hand the unique inference bucket for the same
column $U$ and value $V_u$ has a count of 2000, then we would
expect there to be an additional 200 rows and it would be very
surprising if all of these rows were suppressed. In this case,
the DPA or DPO should look at the bucket and determine if 
it represents a privacy violation or not.

For example, suppose that the unique inference bucket had two
columns, \mytt{age} and \mytt{years\_married}. It would not be surprising,
nor would it be a privacy violation, if all individuals with \mytt{age=10}
also have \mytt{years\_married=0}.

\subsection{Simple knowledge-based: Noise}
\label{atk:sim-know-noise}

\paragraph{Prior Knowledge: Class C}
In this attack, the attacker has the following prior knowledge:

\begin{itemize}
    \item A given protected entity $I$ is in the database
    \item There are $N$ protected entities in the database, none of whom are $I$,
           that have a given attribute (e.g. \mytt{age=25}) or set of
           attributes.
    \item No other protected entities in the database, with the possible exception
           of $I$, have that given attribute.
\end{itemize}

\paragraph{Additional conditions: Common}
In addition, $N$ is large enough that a query for the
attribute will not be suppressed with high probability.

\paragraph{Goal}
The goal of the attacker is to determine whether $I$ has the attribute or not.

\paragraph{Attack}
The attack is to simply query for the count of the attribute:

\begin{lstlisting}
   SELECT attr_col, count(DISTINCT aid)
   FROM ...
\end{lstlisting}

If the count is greater than $N+1/2$, then $I$ is assumed to have the attribute,
otherwise $I$ is assumed not to have the attribute.

The attacker can improve Precision Improvement PI at the expense of Prediction Rate
PR by raising the threshold at which the attacker makes a prediction. For instance,
if the attacker requires that the count must be greater than $N+2$ in order to 
make a prediction, then PI will improve, but fewer protected entities will be attacked
because fewer predictions are made.

\paragraph{Evaluation}

\begin{figure}[tp]
\begin{center}
\includegraphics[width=\linewidth]{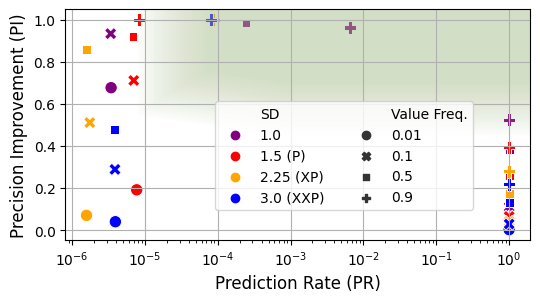}
\caption{PI and PR for attack: \emph{Simple knowledge-based attack with Noise}.
\textbf{The prior knowledge requirement is Class C.}
\emph{Value Freq.} is the frequency at which the attacked data value
appears in the column. 0.5 means that 50\% of the data has the given value.
\mytt{SD=1.0} represents the case where one noise layer can be eliminated
using the 
\emph{Averaging, different semantics same result} attack of Section\ref{atk:avg-sem}.
}
\label{fig:atk:sim-know-noise}
\end{center}
\end{figure}

Figure~\ref{fig:atk:sim-know-noise} gives the results.
The experimental parameters
are described in Section~\ref{sec:add-eval-info}.
The code for this attack is 
in the subdirectory \mytt{simpleKnowledgeBasedNoise}.

\emph{Value Freq.} is the frequency at which the given attribute appears
in the data. A value frequency of 0.5 means that 50\% of the data
has that particular attribute value.
Overall PI increases with higher value frequency. The reason for this is
that the absolute change in precision required for a given PI is smaller
for higher value frequencies. For example, if the value frequency is
90\%, a 5\% increase in absolute precision yields a PI of 0.5.
On the other hand, if the value frequency is 10\%,
the same 5\% increase in absolute precision yields a PI of only 5.3\%.

Note that Figure~\ref{fig:atk:sim-know-noise} includes data for a noise
level below the minimum allowed (\mytt{SD=1.0}). This is included to
capture the case where the 
\emph{Averaging, different semantics same result} attack of Section\ref{atk:avg-sem}
is successful in completely eliminating one noise layer, which a malicious
analyst could do for certain text columns.

Very few attack instances fall within the risk zone, and none for stronger
privacy parameters. If we assume that the analyst is non-malicious, then
only one data point falls in the risk zone, and that is for an attack
where the Value Freq. is 0.9 (90 percent of protected entities have the
same value). For this specific case, roughly 1/20K random protected entities
would have a high PI.

The cluster of measures at the lower right of the graph $PR=1$ represent
attacks where a prediction was made for every query. Here, PI is always below
roughly 0.5.

The remaining measures represent an attack whereby predictions were only made
if either the expected PI is greater than 0.95, or the PR is less
than $10^{-5}$.
Experimentally we produced these data points by increasing
the threshold until either of these conditions were met (over an
average of 100 such predictions).

\paragraph{Discussion}
The Class C prior knowledge requirements for this attack set a very high bar for
the attacker. Not only does the attacker need to have knowledge of
multiple protected entities, it would be quite unusual for an attacker 
to \emph{not} know whether the victim has a given attribute when the
attacker \emph{does} know the exact number of other protected entities with the
attribute. A plausible scenario where this could happen is where an 
analyst formerly had access to the raw data but no longer has it, and
in the interim one protected entity was added to the data set, and the analyst
subsequently has access to anonymized results.

Nevertheless, even if the prior knowledge requirement is met, the attack
is ineffective for most privacy settings and Value Frequencies.
If it is absolutely necessary to avoid the risk area, a higher
privacy setting can be set.

\subsection{Simple knowledge-based: Suppression}
\label{atk:sim-know-supp}

\paragraph{Prior Knowledge: Class C}
In this attack, the attacker has the following prior knowledge (note this is
the same prior knowledge as in the previous attack~\ref{atk:sim-know-noise}):

\begin{itemize}
    \item A given protected entity $I$ is in the database
    \item There are $N$ protected entities in the database,
           none of whom are $I$,
           that have a given attribute (e.g. \mytt{age=25}) or set of
           attributes.
    \item No other protected entities in the database, with the possible exception
           of $I$, have that given attribute.
\end{itemize}

\paragraph{Additional conditions: Common}
If the attacker wants high PI at the expense of low PR, then the number
of known protected entities $N$ must be \mytt{low\_thresh - 1}. If the attacker
wants high PR at the expense of low PI, then $N$ can be at or adjacent to
the mean suppression threshold.

\paragraph{Goal}
The goal of the attacker is to determine whether $I$ has the attribute or not.

\paragraph{Attack}
The attack is to simply query for the count of the attribute:

\begin{lstlisting}
   SELECT attr_col, count(DISTINCT aid)
   FROM ...
\end{lstlisting}

If the bucket is \emph{not} suppressed, then the
attacker knows with 100\% certainty that the victim has the attribute (at least,
given 100\% precision in the accuracy of the prior knowledge). If the bucket
is suppressed, then the attacker learns (almost) nothing new, and cannot make
a prediction.

If \mytt{N = mean suppression threshold}, then the attacker assumes that the
victim does \emph{not} have the attribute if the bucket is suppressed, and assumes that
the victim \emph{does} have the attribute if the bucket is not suppressed. In this
case, the attacker learns something in every attack, and so can make a prediction for
every attack (high PR).

\paragraph{Evaluation}

\begin{figure}[tp]
\begin{center}
\includegraphics[width=\linewidth]{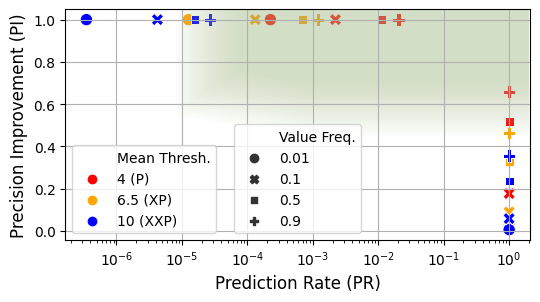}
\caption{PI and PR for attack: \emph{Simple knowledge-based attack with Suppression}.
\textbf{The prior knowledge requirement for this attack Class C.}
\emph{Value Freq.} is the frequency at which the attacked data value
appears in the column. 0.5 means that 50\% of the data has the given value.
}
\label{fig:atk:sim-know-supp}
\end{center}
\end{figure}

Figure~\ref{fig:atk:sim-know-supp} gives the results.
The experimental parameters
are described in Section~\ref{sec:add-eval-info}.
The code for this attack is 
in the subdirectory \mytt{simpleKnowledgeBasedSuppress}.

Both PI and PR increase as both the privacy settings and the value frequency
increase.
Though not apparent from Figure~\ref{fig:atk:sim-know-supp}, different
suppression parameters take affect depending on whether the attacker is
optimizing for PI or for PR.
When optimizing for PR (the cluster of points on the lower right),
it is the increase in suppression standard deviation
\mytt{supp\_sd} that leads to increased PI. Increase in the \mytt{low\_mean\_gap}
does not affect PR.
By contrast, when PI is optimized (data points at the top where $PI=1$),
it is \mytt{low\_mean\_gap} that leads to a higher PR: \mytt{supp\_sd} has no effect.

From the data, we can see that attacks with
privacy setting P and XP fall within our risk area.
For setting P, this includes Value Frequencies where 10\% or more
of protected entities have the unknown value, and for setting XP,
where 50\% or more of protected entities have the unknown value.
The XXP setting has no attacks that fall in the risk area.

\paragraph{Discussion}

As with the knowledge-based attack using noise (Section~\ref{atk:sim-know-noise}),
this attack requires Class C prior knowledge, and is therefore extremely
unlikely to be possible in practice. If the DPA or DPO is nevertheless concerned
with this possibility, then an XP privacy setting leads to a high-precision
prediction in roughly 1/1000 predictions for values that are 90\% common,
and roughly 1/20000 predictions for values that are 50\% common. Note
that such common values are rarely sensitive.

\subsection{Averaging: naïve}
\label{atk:avg-naive}

\paragraph{Prior Knowledge: None}

\paragraph{Additional conditions: None}

\paragraph{Goal}
Eliminate the noise from counts. While a successful attack
wouldn't break anonymity in and of itself,
the resulting noise-free counts could then be used in other attacks, for instance
the Linear program reconstruction: aggregate combinations attack
(Section \ref{atk:lpr-aggr}).

\paragraph{Attack}
Repeat the query multiple times and take the average of the noise samples.

\paragraph{Evaluation}
Because counts are sticky, the same query always produces the same noise.
No averaging is possible with this attack.

\paragraph{Discussion}
In prior versions of Diffix, considerable effort went into ensuring that
the stickiness couldn't be fooled, for instance by composing the same query
in different formats. Those efforts are not required in Diffix Elm because
the SQL constraints don't offer opportunities for generating the same query
in different ways.

\subsection{Averaging: different semantics, same result}
\label{atk:avg-sem}

\paragraph{Prior Knowledge: None}

\paragraph{Additional conditions: Common}
This attack requires specific conditions in the data: it must
be the case that multiple different bucket conditions generate the same data.
In the case of text columns, this could occur when specific characters
in fixed positions are the same for a given value and not for other values.
For instance, suppose that a text column had three values, ``Married'',
``Single'', and ``Divorced''. The following bucket conditions would all
produce outputs consisting of the same protected entities in the same buckets:

\begin{lstlisting}
  SELECT substring(col FOR 1),count()...
  SELECT substring(col FOR 2),count()...
  SELECT substring(col FOR 3),count()...
   ...
\end{lstlisting}

In UA-Mode, only substrings starting at offset 1 may be formed.

A similar effect is possible with numeric and datetime columns, but far less
likely to occur. It would require for instance that
all protected entities in the bucket 0-100 also exist in the bucket 0-50. As a
result, \mytt{floor(col/100)*100} and \mytt{floor(col/50)*50} would produce the same bucket.

\paragraph{Goal}
Eliminate the noise from counts. While a successful attack wouldn't break
anonymity in and of itself,
the resulting noise-free counts could then be used in other attacks, for instance
the Linear program reconstruction: aggregate combinations attack
(Section \ref{atk:lpr-aggr}).

\paragraph{Attack}
The attack is to form multiple buckets using different bucket conditions as
described above, and then to average out the resulting multiple noise values.

\paragraph{Evaluation}
The attack fails because of the aid-layer noise. Although the
sql-layer changes with each query, and can therefore
be averaged out, the aid-layer remains the same.

Nevertheless, this attack effectively reduces the total amount of noise
applied to counts, and so could be used in combination with other attacks
to weaken the anonymity of those attacks.
For example, in the case where $SD=1.5$, the effective noise (if the
sql-layer noise could be completely removed) would be just over 1.0.
It is therefore useful to know under what conditions this attack can
succeed.

\begin{figure}[tp]
\begin{center}
\includegraphics[width=\linewidth]{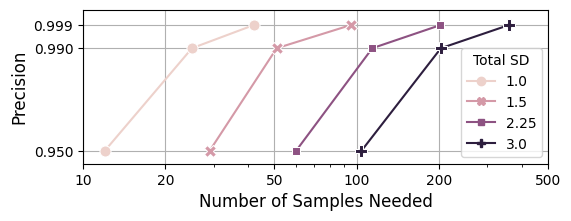}
\caption{The number of noise samples needed to reduce the effective noise
to less than $\pm$0.5 for different levels of precision and noise. Noise less
than $\pm$0.5 exposes the true count for \mytt{count(DISTINCT aid)} queries.
Data points generated through simulation.
}
\label{fig:samples-needed}
\end{center}
\end{figure}

Figure~\ref{fig:samples-needed} shows how many attack queries are required to
result in noise less than $\pm$0.5 for different levels of precision and
amounts of noise. From this we see, for instance, that to achieve 99\%
precision, and a standard deviation of 1.0 (which is roughly that of one
noise layer when \mytt{base\_sd=1.5}), around 25 queries are needed.

The simulation used to derive these numbers can be found in the code \mytt{attack.py}
in the subdirectory \mytt{avgDiffSyntaxSameSemantic}. The number of required
samples represents the number of characters that would be needed to generate
the samples in UA-Mode using \mytt{substring()}.

\paragraph{Discussion}

This attack would not accidentally be executed by a non-malicious analyst.
It therefore only applies to UA-mode operation.

This attack increases in likelihood as the number of columns with
few distinct values and lengthy text strings increases (relative to
the per-layer \mytt{SD}). In data where this is a concern, the
columns can be pre-processed so that the strings are reduced in size
or replaced with digits.

\subsection{Linear program reconstruction: randomness in column}
\label{atk:lpr-random}

\paragraph{Prior Knowledge: Class B}

In this attack, there are one or more \emph{identifying} columns, and
an \emph{unknown} column. The identifying columns, taken together,
uniquely identify each protected entity being attacked. The unknown column is
what is being learned.
The attacker must know all values of identifying columns.

\paragraph{Additional conditions: Common}
The identifying columns must be text columns (the attack uses the
\mytt{substring()} function which only works on text columns). The
identifying columns must have substantial internal randomness.
A substantial number of characters in the text string must be randomly assigned,
and as such have no correlation with other random characters.

Most commonly this would
be a column that serves as an identifier, and whose values are
randomly assigned (for instance a UUID value).

\paragraph{Goal}

The goal is to reconstruct the identifying and unknown column values. If this can
be done, then each protected entity can be singled out because of the identifying columns.

\paragraph{Attack}

The attack is patterned after the original 2003 reconstruction attack of Dinur and
Nissim~\cite{dinur2003revealing}, and a later variant successfully executed against
Diffix Cedar by Cohen and Nissim~\cite{cohen2019linear}. As with Diffix Elm, the attacker can request
the count of protected entities that have a given value in the unknown column. Also like
Diffix Elm, noise is added to the counts (though there is no suppression).
Critically, in the Dinur attack, the attacker has the ability
to specify which protected entities are included in each count. This allows the attacker to
select counts composed of random but known protected entities.

The corresponding SQL for the Dinur attack could for instance be:

\begin{lstlisting}
   SELECT count(DISTINCT aid) FROM table
   WHERE unknown_col = X AND
         identifying_col IN
             (i1,i6,i11,i12,...,i142)
\end{lstlisting}

For each count with a set of selected protected entities, a pair of equations are formed:

$i_{1} + i_{6} + i_{11} + i_{12} + ... + i_{142} > count - \delta$

$i_{1} + i_{6} + i_{11} + i_{12} + ... + i_{142} < count + \delta$

Each variable $i_X$ represents one protected entity, and can take the values
1 or 0 corresponding to whether the protected entity has or does not have the
unknown value. $\pm \delta$ is the range of noise that can be added to the
count.

The attacker makes multiple queries, each with a randomly selected subset of
protected entities. This results in a set of equations that can be solved for the
values of $i_X$. If there are enough equations relative to the amount of noise,
then there is a single correct solution to the equations and the attacker can
determine the correct value of each $i_X$, and therefore the unknown column
value of each protected entity. As a result, all protected entities are correctly singled
out.

Unlike the Dinur setup, Diffix Elm does not allow the attacker to specify
which protected entities can be included in an answer. Therefore, the attacker must
rely on randomness in the identifying column itself, combined with prior
knowledge of the values of the identifying column, to build the equations.

In T-mode, the attacker can make a set of queries of the form:

\begin{lstlisting}
   SELECT substring(identifying_col
                    FROM X FOR Y),
          unknown_col, count(DISTINCT aid)
   FROM table
   GROUP BY 1,2
\end{lstlisting}

By varying X and Y, the attacker creates different sets of protected entities.
In the case of UA-Mode, X is always 1, so the attacker can form only a
relatively small number of equations (limited by the length of the
identifying column or columns themselves). In TA-Mode, however, the attacker
(if the analysts indeed turned out to be malicious) can both generate more
random groups (vary X), and can better control the size of the groups
(vary Y).

\paragraph{Evaluation}

The code for this attack may be found in subdirectory
\mytt{linearReconstructionRandom}. The constraint builder and
solver is in file \mytt{lrAttack.py}, routine \mytt{makeProblem()}.
There is a Jupyter notebook at file \mytt{basic.ipynb} that explores
the results.

In our experiments, we assumed a single identifying column, and assumed
only two values in the unknown column. We tested the attack for both
UA-Mode and TA-Mode across a range of parameters:

\begin{description}
  \item[Number of protected entities being attacked:] From 10 to 100 protected entities
  for untrusted, 10 to 800 for trusted
  \item[Length of ID string:] 120 characters for untrusted, and
  from 15 to 240 characters for trusted
  \item[Number of symbols per ID character:] 2, 8, and 32 symbols
  \item[The frequency of the unknown value:] 10\% and 50\% frequency
\end{description}

\begin{figure}[tp]
\begin{center}
\includegraphics[width=\linewidth]{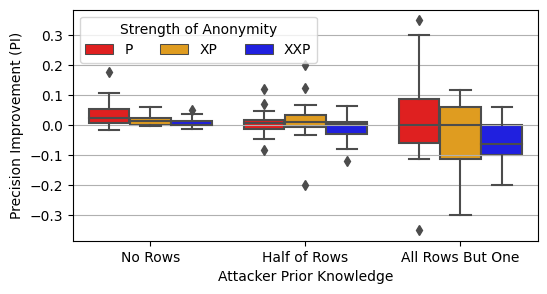}
\caption{\textbf{Untrusted Analyst Mode:}
Precision Improvement for different strengths of anonymity
and amounts of prior knowledge. Prediction Rate (PR) is always 1.0.
The different points comprising the box plots represent different experimental
parameter settings.
The parameters associated with P, XP, and XXP are
given in Table~\ref{tab:eval-params}.
}
\label{fig:lr-ran-plot-untrusted}
\end{center}
\end{figure}

The main result for UA-Mode is shown in Figure~\ref{fig:lr-ran-plot-untrusted},
which shows the Precision Improvement (PI) for different anonymity strengths and
prior knowledge. Prediction Rate (PR) is always 1.0 for this attack.
The different points on the box plots represent the
different combinations of the above experimental parameter settings (see
the Jupyter notebook for more detail). The core result is that, even when
the attacker knows the unknown values for half of the protected entities, PI is never
more that 0.2.

Even where the attacker knows all data except for one protected entity, the
attacker never achieved better PI than 0.5. (The high point for the 'P'
anonymization strength is for an attack on 10 protected entities and 8 symbols per ID character.)

In short, the attack for UA-Mode is not effective.

\begin{figure}[tp]
\begin{center}
\includegraphics[width=\linewidth]{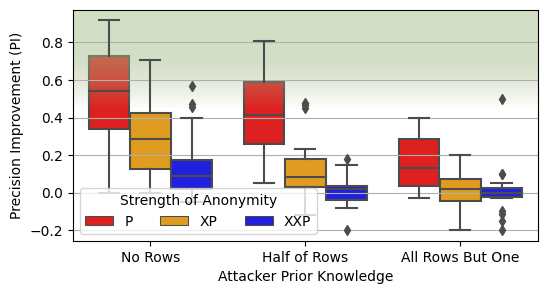}
\caption{\textbf{Trusted Analyst Mode (but malicious analyst):}
Precision Improvement for different strengths of anonymity
and amounts of prior knowledge. Prediction Rate (PR) is always 1.0.
The different points comprising the box plots represent different experimental
parameter settings.
The parameters associated with P, XP, and XXP are
given in Table~\ref{tab:eval-params}.
}
\label{fig:lr-ran-plot-trusted}
\end{center}
\end{figure}

In spite of the fact that a trusted analyst would not accidentally
run the attack, we should
understand the extent to which the attack is effective in TA-Mode.
The main result for TA-Mode is shown in Figure~\ref{fig:lr-ran-plot-trusted},
which shows the Precision Improvement (PI) for different anonymity strengths and
prior knowledge. Prediction Rate (PR) is always 1.0 for this attack.
The different points on the box plots represent the
different combinations of the above experimental parameter settings (see
the Jupyter notebook for more detail). The core result is that some reconstruction
attacks are very effective in TA-Mode.

\begin{figure}[tp]
\begin{center}
\includegraphics[width=\linewidth]{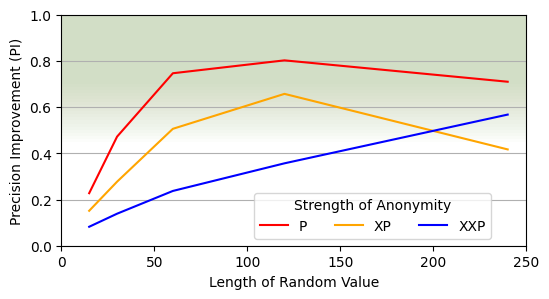}
\caption{\textbf{Trusted Analyst Mode (but malicious analyst):}
Precision Improvement for different amounts of randomness in the
identifying columns (based on length of ID).  Prediction Rate (PR) is always 1.0.
200 protected entities were attacked,
there were 8 symbols per ID character, and the unknown value appeared with 50\%
The parameters associated with P, XP, and XXP are
given in Table~\ref{tab:eval-params}.
}
\label{fig:lr-ran-plot-aid-len-trusted}
\end{center}
\end{figure}

The experimental variable that has the strongest effect is the amount of randomness
in the identifying columns: more randomness leads to more effective attacks because
the attacker can make more equations and so reduce the possible set of correct
answers. Figure~\ref{fig:lr-ran-plot-aid-len-trusted} shows the effect of
the length of the ID value on PI. For this graph, 200 protected entities were attacked,
there were 8 symbols per ID character, and the unknown value appeared with 50\%
probability. These are conditions favorable for the attacker.

From Figure~\ref{fig:lr-ran-plot-aid-len-trusted}, we see that anonymization
strength of 'P' hits 50\% PI at around 30 3-bit symbols (90 bits of randomness)
'XP' at around 60 3-bit symbols (180 bits of randomness), and
'XXP' at around 200 3-bit symbols (600 bits of randomness).
For comparison, a typical UUID has around 96 bits of randomness (24 4-bit random symbols).

Note that PI eventually starts decreasing with still more random bits. We speculate
that this is because, once diminishing returns in the amount of randomness is reached,
more symbols only leads to more possible solutions,
and so the solver has a larger chance of producing an incorrect solution.

\paragraph{Discussion}

It seems virtually impossible for a trusted analyst to accidentally execute
the queries necessary to run this reconstruction attack. The analyst would have
to run a sequence of \mytt{substring()} over a column with no particular
analytics value (because of the randomness).
Note also that the attack leaves a very distinctive signature. If query activity
is monitored, this could help to dissuade an attack in TA-Mode.

Note finally that it may be possible to pre-process the data so that
excess randomness is removed, especially given that randomness has little
analytic value.

\subsection{Linear program reconstruction: aggregate combinations}
\label{atk:lpr-aggr}

\paragraph{Prior Knowledge: None}
For the columns being attacked, the attacker must know the set of distinct
column values (i.e. if the column is \mytt{account status}), the attacker
would need to know that the possible values are '\mytt{active}' and
'\mytt{inactive}'). This is typically public knowledge.

While the attacker does not need to know any data about protected entities to
run the attack, some knowledge of protected entities may help reconstruct the
data of unknown protected entities.

\paragraph{Additional conditions: None}

\paragraph{Goal}
The goal of this attack is to reconstruct the column values in the
data. Individual rows with a distinct set of values are effectively
singled out.

\paragraph{Attack}
This attack uses a constraint solver to try to compute what the original
table values must be. It makes a set of queries that cover every combination
of the columns that are being attacked. For instance, if three columns,
C1, C2, and C3 are being attacked, then the attacker queries for
each column separately, each of three combinations of two columns, and
all three columns.

Based on the answers, the attacker can then define a set of constraints:

\begin{enumerate}
  \item There are N protected entities, where N is the noisy answer to
      the count of all rows.
  \item For any given histogram of one or more columns, a protected entity
      appears in exactly one bucket.
  \item The count of protected entities in any reported bucket is constrained by
      the noisy count plus or minus some range (typically between 1 and
      3 standard deviations of the noise), but no fewer than \mytt{low\_thresh}.
  \item The count of protected entities in a suppressed bucket (which
      is known to be suppressed because all possible column values
      are known a priori), is constrained by zero and a range above the
      mean (\mytt{low\_thresh+low\_mean\_gap}, typically between 1 and 3 standard
      deviations of the suppression noise \mytt{supp\_sd}).
  \item Each protected entity in a sub-bucket also appears in the associated
      parent buckets. For example, if a protected entity appears in the bucket
      defined by the two column values \mytt{age=20,zip=12345}, then the
      protected entity also appears in the one-column buckets \mytt{age=20} and
      \mytt{zip=12345}.
\end{enumerate}

A solution for these constraints results in a reconstructed table
containing zero or more distinct protected entities, where a protected entity is
distinct if it has a unique set of column values. A singling-out
prediction is made for each distinct protected entity. No singling-out predictions are
made for non-distinct protected entities (leading to a lower PR).

Note that column values can themselves be generalizations. For instance,
an age group of 25-years may serve as a column value (leading to four
distinct "values" to attack). This is important
because the attack scales exponentially with the number of columns and column
values. Generalization effectively reduces the number of column values
that need to be solved for, though at the expense of the attacker obtaining less
precise information about the data.

\paragraph{Evaluation}
The code for this attack may be found in subdirectory
\mytt{linearReconstructionAggregate}. The constraint builder and
solver is in file \mytt{lrAttack.py}, routine \mytt{makeProblem()}.
There is a Jupyter notebook at file \mytt{basic.ipynb} that explores
the results.

\begin{figure}[tp]
\begin{center}
\includegraphics[width=\linewidth]{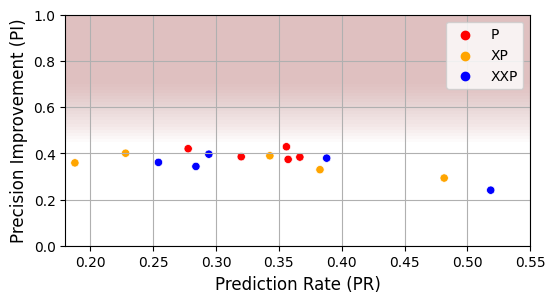}
\caption{Scatterplot of PI and PR for aggregate-based linear reconstruction
attacks using different
anonymization parameters against tables of different sizes and shapes.
This is for attacks where the attacker has no prior knowledge.
Each point is the average of 30 attacks over tables generated with
different random number seeds.
The parameters associated with P, XP, and XXP are
given in Table~\ref{tab:eval-params}.
}
\label{fig:lr-agg-plot}
\end{center}
\end{figure}

Figure~\ref{fig:lr-agg-plot} gives the main result.
Each point on the graph is the average PI and average PR over 30
attack runs for tables with different numbers of columns and distinct
values per column. The number of columns and values is relatively
small, ranging from 3 to 5 columns and from 3 to 5 distinct values
per column. The reason we tested with relatively small tables is because
the solver scales with the product of the number of column/value
combinations and the number of rows. The number of rows assigned to
each table is equivalent to the number of column/value combinations.
Each row is assigned a value from each column randomly with uniform
probability.

As a validation, we also ran the attack with no anonymization at all.
These attacks show perfect reconstruction, and obtained $PI=1.0$
and PR in the range between 35\% and 45\%. This PR range is because only this
fraction of entries in the table had distinct column values, and so
predictions were made only on these entries.

The key result is that the attack is unable to achieve PI greater
than 50\% for even the lowest privacy setting of P. This attack
is not effective. Note as well that stronger anonymization does not make
a huge difference in the effectiveness of the attack. Even a small
amount of noise leads to incorrect solutions.

\begin{figure}[tp]
\begin{center}
\includegraphics[width=\linewidth]{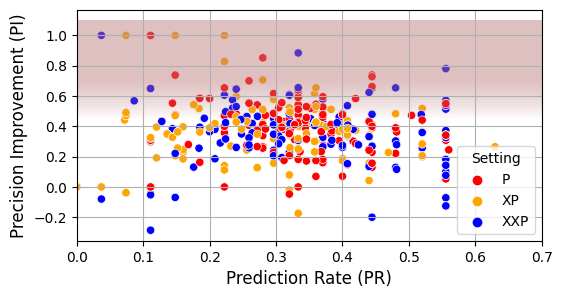}
\caption{Scatterplot of PI and PR for the individual runs of the
aggregate-based linear reconstruction attacks. Figure~\ref{fig:lr-agg-plot}
shows the average of these individual runs.
}
\label{fig:lr-agg-plot-no-avg}
\end{center}
\end{figure}

Note that, while the date points shown in Figure~\ref{fig:lr-agg-plot}
each represent the average of 30 runs of the attack, the difference
between individual attacks is quite large. Figure~\ref{fig:lr-agg-plot-no-avg}
gives the data for the individual attacks. Here we can see that individual
runs range from perfect reconstruction to far worse than a statistical guess
(negative PI). The attacker has no way of knowing where on this spectrum
any given attack lies, and so the average from Figure~\ref{fig:lr-agg-plot}
approximates the actual PI and PR overall.

\begin{figure}[tp]
\begin{center}
\includegraphics[width=\linewidth]{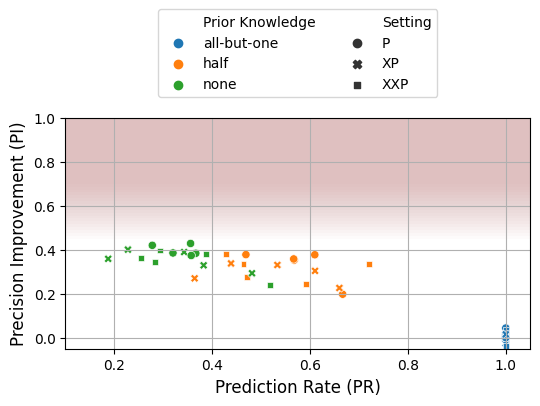}
\caption{Scatterplot of PI and PR for different amounts of prior knowledge.
}
\label{fig:lr-agg-plot-all-pri}
\end{center}
\end{figure}

Figure~\ref{fig:lr-agg-plot-all-pri} gives the results for different amounts
of prior knowledge. Contrary to what one might expect, prior knowledge does not
improve PI. Indeed it lowers PI while increasing PR, although this effect is
an artifact of how we measure PI and PR in this case.

Specifically, what we do
is to remove the prior known protected entities from both the original and reconstructed
tables, and then measure PI and PR on the resulting tables. More prior knowledge
leads to smaller measured tables. At the extreme, when the attacker knows all of
the date except for one row, then the measured original and reconstructed tables
have only one entry each.

As the number of rows in the measured tables shrinks, the proportion of unique rows
increases. Indeed a table with one row only by definition has a unique row. This
causes PR to increase, and at the same time lowers the accuracy of predictions simply
because the decision to make a prediction is based relatively less on the selectivity
of the solution, and more on the probability of there being a unique entry by
chance.

Note, however, that the PI difference between no prior knowledge and half prior knowledge is
not that much. We believe that the solver simply often finds the wrong solution,
and so there are always incorrect predictions. We could not test this, however, because
larger tables take very long to solve because of the exponential increase in
variables.

\paragraph{Discussion}

This version of the reconstruction attack is not effective. With noise
and suppression, there are too many solutions that are correct in that
they satisfy the constraints, and we don't know how to detect which
solutions might be better. We have not aggressively explored how one might
do this: we don't have any good ideas and our intuition is that this simply
isn't a fruitful avenue of attack.

\subsection{Difference: positive AND, single victim}
\label{atk:diff-first}

Note that this attack does not work because of suppressed bucket
merging (see~\ref{sec:qh-step-4}).
However, we describe it here to motivate the need for suppressed bucket merging.

\paragraph{Prior Knowledge: Class C}
This attack requires that a certain condition holds in the data, and that the attacker
knows of the condition. Specifically, it must be the case that a single user has
a different value from all other protected entities in a given column (the \emph{isolating column})
for some subset of the data, and that
the attacker knows this. For example,
everyone in the subset computer science department has isolating column
value \mytt{gender='male'} except one person.

\paragraph{Additional conditions: Rare}
There must be enough protected entities with the common value (i.e. males) in the
subset (i.e. CS department) that very few if any of the
buckets corresponding to the unknown values are suppressed.

\paragraph{Goal}
The goal of the attacker is to single out the protected entity with the unique value in the
given subset of the data (i.e. the female among males).

\paragraph{Attack}
The attacker creates two queries, one that creates a bucket that excludes the victim (i.e.
the female in the CS department), and another that creates a bucket that may or may not include the victim
depending on whether the victim has the unknown attribute. For example, in the following
two queries, the unknown attribute is \mytt{title}.

\begin{lstlisting}
   SELECT dept, gender, title, count()
   FROM table GROUP BY 1,2,3
\end{lstlisting}

\begin{lstlisting}
   SELECT dept, title, count()
   FROM table GROUP BY 1,2
\end{lstlisting}

The attacker is interested only in the buckets with \mytt{dept='CS'}.
The victim is never in the bucket of the first query (where \mytt{gender='male'}).

If not for suppressed bucket merging,
the victim would be in the bucket of the second query where the title matches that of
the victim (V In), and not in the other buckets of the second query (V Not In).
Therefore, the underlying true count between the second and first query would differ
by 1 for the V In pair, and wouldn't differ for the V Not In pair.

The AIDV set would always the same for the V Not In pairs,
and different for the V In pair. Correspondingly, the seed material related
to the aid-layer would be the same in the V Not In pairs, and would differ only for the
V In pair.

The seed material for the sql-layer differs for every bucket of both queries.

If we take the difference between the first and second noisy count for matching
buckets (again, assuming no suppressed bucket merging), we find that:
\begin{itemize}
   \item For V Not In pairs, there is no difference in the underlying count, and
       the difference in noise is that of one layer.
   \item For V In pair, the underlying count differs by 1, and the
       difference in noise is that of two layers.
\end{itemize}

In other words, there would be two signals that the attacker could use to try
to deduce the victim's bucket.

Given these two signals, the attacker has two strategies. The first is to
make a prediction with every attack (\mytt{PR=1}) by assuming that the bucket where
the difference in count between the second and first queries
is largest is the one that holds the victim. The second is to only make a prediction if
the magnitude of the difference exceeds some threshold. This improves PI at
the expense of a lower PR.

\paragraph{Evaluation of likelihood that table conditions exist}
The file \mytt{findConditions.py} in subdirectory \mytt{findConditions} contains code
that measures the extent to which the conditions for this attack exist in the data.
Using this code, we evaluated the number of times the conditions exist in three
real datasets:
\begin{description}
   \item[Census:] 15 columns and 3.8 million protected entities
   \item[Banking:] 15 columns and 5369 protected entities
   \item[Taxi:] 21 columns and 12995 protected entities
\end{description}

\mytt{findConditions.py} operates in two phases. First, it examines pairs of
columns, an isolating column and an subset column, looking for the attack
condition whereby there are two values in the isolating column, and only one
protected entity has one of the values. When discovered, it then examines the
remaining columns as unknown columns to ensure that no suppression takes place.
When all these conditions are met, then we have a working attack.

The conditions for the attack occur whenever the isolating column has a small
number of distinct values, and:
\begin{enumerate}
   \item There is a strong negative correlation between an isolating column
   value and a subset column value (as in the CS department example above), or

   \item One of the values in the isolating column has a very high occurrence.
\end{enumerate}

In our measures, we never found a case where the attack conditions exist 
for the first reason. In all there datasets, however, there are columns were
one value dominates.

For instance, in the census dataset used in our measure,
the second existed for four columns:
\begin{description}
   \item[citizen:] Four values, dominant value 93\%
   \item[race:] Five values, dominant value 88\%
   \item[school:] Two values, dominant value 83\%
   \item[speaks\_english:] Three values, dominant value 73\%
\end{description}
(Note that the census dataset has a \mytt{gender} column with only two distinct values,
but they are
roughly evenly split and don't correlate with other columns, and so no attack
conditions were found using \mytt{gender} as the isolating column.)

The number of protected entities for which the attack conditions existed at
least once are:
\begin{description}
   \item[Census:] 70 of 3.8 million protected entities (1/54000)
   \item[Banking:] 14 of 5369 protected entities (1/380)
   \item[Taxi:] 5236 of 12995 protected entities (1/2.5)
\end{description}

The reason that the taxi dataset has a high occurrence relative to the other two
datasets is because it is a time-series dataset with 440K rows (average 33 rows
per protected entity, where protected entities are taxi drivers, and each row corresponds to
a trip). Each trip creates a scenario where the conditions might hold with respect
to that trip. Usually the subset in the taxi measure was a column like trip
start time or trip start latitude, which effectively isolated a taxi ride
for some isolating column.

\paragraph{Evaluation of effectiveness (assuming conditions exist and no
suppressed bucket merging)}
Suppressed bucket merging prevents this attack from working.
It detects the condition, and places
the rows from the victim into the corresponding bucket.

Nevertheless, we ran the attack on the assumption of no suppressed
bucket merging.
The code for this attack may be found in subdirectory
\mytt{diffAttack} in file \mytt{diffAttackClass.py}, where the configuration
\mytt{attackType = 'diffAttack'} is set.

We found that if the attacker
uses $PR=1$, then PI is always below 50\%. If, however, the attacker
lowers PR, then the attacker can achieve $PR > 0.95$ with prediction rates
that fall within the designated risk area. In the worst case, with the
minimum noise amount of \mytt{base\_sd=1.5}, 95\% PI is obtained for
1/50 protected entities when 2 unknown values are being attacked. The attack
is less effective as the noise or the number of attacked
unknown values grows.

\paragraph{Discussion}
We believe that the likelihood of this attack occurring in practice
(if suppressed bucket merging did not exist) would be extremely small. The data conditions
are rare, and the required prior knowledge is substantial. Further,
when the conditions
do occur, the attacker can learn only one of a small number of unknown
values, usually just 2. Normally values that are shared by a substantial
portion of the population are not as sensitive.

On the other hand, not all users may agree with this assessment, and
it is the case that a trusted analyst could inadvertently formulate
the attack (at least, more likely than other attacks like linear
reconstruction or range creep). Therefore, from an abundance of caution,
we implement suppressed bucket merging and prevent this attack.

\subsection{Difference: positive AND, group of victims}
\label{atk:diff-group}

\paragraph{Prior Knowledge: Class C}
As with the difference attack exploiting positive AND against a single victim
(Section~\ref{atk:diff-first}),
this attack requires that certain conditions exist in the data, and that the
attacker knows of the conditions. The difference is that here the conditions
apply to multiple protected entities instead of a single one.
It must be the case that a group of protected entities (victims)
a different value from all other protected entities in a given column (the \emph{isolating column})
for some subset of the data, and that
the attacker knows this.

\paragraph{Additional conditions: Rare}
The protected entities should all have the same value for the unknown attribute.
To the extent that they do not, the attack is less effective.
The number of protected entities must be enough that suppressed
bucket merging is not triggered.

\paragraph{Goal}
The goal is to infer the unknown value shared by the group of victims. By the
strict GDA score definition of inference, the goal fails if a single victim
does not share the unknown value. An alternative goal would be to guess the
unknown value of a single one of the victims based on prior knowledge that that
specific protected entity is one of the victims. In this case, the attack may
succeed even if the group of victims does not share the same unknown value. 
In this latter case, singling out would occur if the attacker has prior
knowledge that distinguishes the specific victim from the other victims.

\paragraph{Attack}
The attack mechanism is the same as that of Section~\ref{atk:diff-first}.

\paragraph{Evaluation}
The code for this attack may be found in subdirectory
\mytt{diffAttack} in file \mytt{diffAttackClass.py}, where the configuration
\mytt{attackType = 'diffAttackLed'} is set. The attack run by this code
is the singling-out goal, not the inference goal. The attacker assumes that
the victim is in the bucket that exhibits the greatest noisy count difference.
We used the \mytt{count(DISTINCT AID)} aggregate.

To run the attack, we label \emph{Num Isolated} protected entities as the group of victims. We
randomly assign an unknown value to each victim (they do not necessarily
all have the same value). We randomly select one from the group of
victims as the singled-out victim. The attack succeeds if we correctly guess the
unknown value of this specific victim.

\begin{figure}[tp]
\begin{center}
\includegraphics[width=\linewidth]{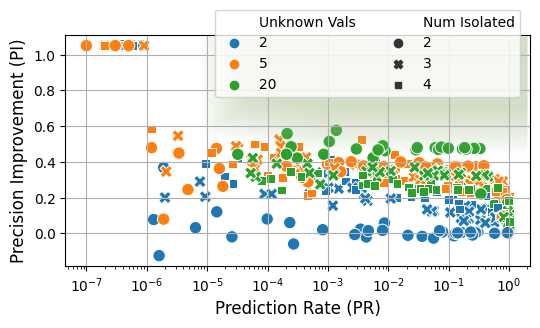}
\caption{Scatterplot for \emph{difference attack with a group
of victims} (Section~\ref{atk:diff-group}). \emph{Unknown Vals} is the
number of distinct values for the unknown attribute. \emph{Num Isolated}
is the number of victims in the isolated group. \emph{PI} values above 1.0 denote
cases where there were insufficient correct predictions to derive a statistically
meaningful value (fewer than 10).
}
\label{fig:diff-attack-led}
\end{center}
\end{figure}

The results of the attack are shown in Figure~\ref{fig:diff-attack-led}. As can
be seen, the attack is ineffective. The fact that PI is higher for for a larger
number of unknown values seems counter-intuitive (it should be easier to guess
among fewer values than more values). This is an artifact of the definition of
precision improvement: the absolute precision is much less for more unknown
values.

\paragraph{Discussion}
Suppressed bucket merging only merges when
all AIDVs in the otherwise-suppressed bucket share a potential unknown value.
When this happens, then the first and second query answers are identical, and
the attacker learns nothing.

When this is not the case (and suppressed bucket merging is not
invoked), then the attack fails
because the isolated AIDVs are spread around, leading to 1) ambiguity as to
which bucket may contain the victim, and 2) less of a difference between
the bucket pair, which in turn means that the noise is more likely to obscure
what is happening. Due to both of these effects, the attack is not effective.

Note that if this were not the case, then we would have to modify 
suppressed bucket merging so as to
merge even in this case, which would create additional distortion.

\subsection{Range creep with averaging (TA-Mode only)}
\label{atk:range-creep}

Note that this attack only works in TA-Mode. In UA-Mode the precision of X
in the expression \mytt{floor(col/X)*X} is not enough to generate the attack
conditions.

\paragraph{Prior Knowledge: Class A}
The attacker must have prior knowledge of the victim's value in a numeric
column, and must either have explicit knowledge of the next lower and higher values
in the column, or be able to deduce with high probability how far away the next
lower and higher values are (e.g. through knowledge of the precision on column values).

\paragraph{Additional conditions: Common}
This attack requires that the victim has a value in a numeric column that is
distinct from all other users. It also requires that there are enough other
protected entities that have values higher than the victim's next lower
value, and lower than the victim's
next higher value, to avoid suppression (see attack description below).

\paragraph{Goal}
The goal is to learn the victim's value for an unknown column by
singling out the victim by averaging out the noise using slight
increments of the \mytt{floor()} bucketizing function.

\paragraph{Attack}
By way of example, suppose
that the victim has the value 1000 in some integer column, that no other user has this
value, and that the attacker knows it.

The attacker makes the following query:

\begin{lstlisting}
   SELECT floor(int_col/999.9)*999.9,
          unknown_col, count(*)
   FROM table GROUP BY 1,2
\end{lstlisting}

The buckets with \mytt{int\_col} in the range 0-999.9 exclude the victim's row.
The victim's row is in one of the buckets in the range 999.9-1999.8 (the one matching
the victim's unknown value).

The attacker then makes a series of queries with very slight increments of the
\mytt{int\_col} width, for instance 999.901, 999.902, 999.903 etc. Each of these
queries results in the same set of protected entities in the buckets. However, the seed
material for the sql-layer noise \mytt{sql\_noise} changes with each query, leading to different
noise values for each query, which in turn allows the attacker to average out the
noise. As a result, the attacker establishes a
noisy count for the buckets from the lower range as being composed
of \mytt{true\_count + aid\_noise}.

Next, the attacker makes a query as follows:

\begin{lstlisting}
   SELECT floor(int_col/1000.01)*1000.01,
          unknown_col, count(DISTINCT aid)
   FROM table GROUP BY 1,2
\end{lstlisting}

followed by queries that slightly increment the bucket width (1000.02, 1000.03, etc.).
In each of these queries, the victim's row will be added to the lower-range bucket corresponding
to the victim's value in \mytt{unknown\_col}, and likewise removed from the upper-range bucket.
This series of queries also averages out the \mytt{sql\_noise}. As a result, the
\mytt{true\_count} changes for only the two buckets that
match the victim's value in \mytt{unknown\_col}. Further, if there is one row per
protected entity, the \mytt{aid\_noise} will also change only for the bucket with the
victim's value.

\paragraph{Evaluation}
Assuming the attack conditions and prior knowledge exist, this attack certainly works
with high probability (TA-Mode). Figure~\ref{fig:samples-needed} shows how many samples are
needed to overcome the noise for a single layer in this attack. In any event, given
arbitrary precision in choosing bucket boundaries, an attacker can certainly generate
enough queries.

\paragraph{Discussion}
The probability that a trusted analyst would accidentally run this attack is virtually
zero. There is no reason that an analyst would make such small increments to the
bucket size, given that nothing is learned from doing so.

The attack leaves a distinctive fingerprint, and so a system that logs queries would
act as a deterrent for a trusted analyst that nevertheless wishes to run the attack.

The conditions required for this attack to work in UA-Mode almost certainly cannot
exist. Referring to Figure~\ref{fig:samples-needed}, let's assume that the attacker
requires 95\% precision, and that \mytt{base\_sd=1.5}, which puts the per-layer
standard deviation at roughly 1. The attacker therefore needs to make 12 queries
on each side. Now suppose that the value prior to that of the victim is 1000.
To exclude the victim, the attacker would need to generate queries with bucket
sizes of 2000, 5000, 10000, and so on. The 12th such query would have a bucket
size of 10M. In other words, there would have to be a gap between the protected entity
with value 1000 and the victim of nearly 10M, and the victim's value must
be greater than 10M. If there is a protected entity with a value higher than that
of the victim's, then to form the 12 queries that place the victim in the
lower bucket, the next protected entity's value would need to be greater than
$10^{11}$. Note that if the victim has the maximum value, then the attack
doesn't work because the upper bucket would be suppressed.

\subsection{Salt: Dictionary attack on table}
\label{atk:salt-tab}

\paragraph{Prior Knowledge: Class X (assumed not possible)}
The attacker must have near-complete knowledge of the contents of the
table. In addition, the attacker must know the possible values of
the remaining unknown values. (Almost by definition attackers cannot have
this much prior knowledge. It is the moral equivalent of knowing the
first 10 characters of an 11-character password. As such, it falls
outside of the A/B/C classification of prior knowledge.)

\paragraph{Additional conditions: Common}
The number of possible values of the unknown values must be small
enough that a brute-force attack on these values is feasible.

\paragraph{Goal}
To determine the value of the remaining unknown contents of the
table by determining the salt and validating that the salt is correct.

\paragraph{Attack}
In this brute-force dictionary attack, the attacker tries every
combination of potential values that the
unknown values can take. Each such combination produces a
proposed replication of the table. Given each replica,
the attacker can duplicate the behavior of Diffix Elm, first
by computing a proposed salt, and then replicating noise and
suppression.

The attacker tries a number of queries, and compares the duplicate
results with the results from the Diffix Elm system. If the proposed salt
is incorrect, then even with the
minimum noise of \mytt{SD=1.5} and counting distinct AIDs, the
probability that any given noisy count differs between the
Diffix Elm system and the attacker's replicate is roughly 0.74.
With only a few 10s of buckets that all match, the attacker can determine
with very high probability the correct table has been replicated.

\paragraph{Evaluation}
This brute-force attack on the table is analogous to a brute-force
dictionary attack on a password. So long as the number of unknown values
and corresponding possible values that they can take
is small enough, this attack works.

\paragraph{Discussion}
A scenario where this attack is possible seems very unlikely. It requires
a situation whereby an attacker on one hand has almost complete knowledge
of the table, but on the other hand should not have knowledge of the
remaining small portion of the table. Further, it only makes sense if the
size of the unknown knowledge is less than the size of the salt. Otherwise a brute-force attack on the salt would be easier.

In any event, Diffix Elm should not be used in scenarios where this attack
is feasible.

\subsection{Salt: Knowledge attack}
\label{atk:salt-know}

\paragraph{Prior Knowledge: Class X (assumed not possible)}
The attacker knows the salt value and the AID values.

\paragraph{Additional conditions: None}
Note that if the table has one row per AID, and the Diffix Elm implementation
uses the row index as the AID value, then the attacker knows that the AID
values are simply the sequential values from 0 to the table size (which the
attacker knows approximately from a simple count query).

\paragraph{Goal}
Reconstruct the table.

\paragraph{Attack}
Here we provide a sketch of the attack.

The attacker generates a set of queries that produce relatively low counts.
This can be done by selecting a large number of columns, or by selecting
columns with a large number of distinct values.

By way of example, suppose that the noisy count for one such bucket is 2.
Knowing the salt, the attacker can replicate the sql noise layer. Given this,
the attacker knows the possible number of distinct AIDs in the bucket with
high probability. They can then
try different combinations of AID values, compute the resulting aid noise, and
compare with the noisy count. When the attacker's noisy count does not match
the system's noisy count, then the attacker knows that at least one of the
AID values does not match. Given this, the attacker builds a set of constraints
and uses a solver to determine which AID values belong to which buckets.

In this way, the attacker learns some of the column values associated with AIDs.
Given this knowledge, the attacker can then build larger buckets composed partially
of known columns, and again solve for the unknown parts. Eventually the attacker
can reconstruct the entire table.

\paragraph{Evaluation}
We do not know if this attack is feasible, because we have not tried it.

\paragraph{Discussion}
In TA-Mode, this attack appears impossible because the attacker would not have
access to the set of queries required to build up the reconstruction.

Protecting the salt in Diffix Elm is somewhat analogous to protecting an encryption key
or a password. A system deploying Diffix Elm must protect the salt just as
encryption keys or passwords must be protected. Note, however, that it is far
easier to protect the salt because, unlike a key or a password, it never needs
to exist external to the system. 

\subsection{Access to multiple (incorrect) instances}
\label{atk:mult-instance}

\paragraph{Prior Knowledge: None}

\paragraph{Additional conditions: (should never happen)}
The data is deployed on multiple instances of Diffix Elm (for instance
for scalability). The attacker has access to the multiple instances.
Critically, initialization of the salt is \emph{incorrectly implemented}, such that
each instance has a different salt.

\paragraph{Goal}
Remove the noise, and from there launch a linear reconstruction attack.

\paragraph{Attack}
Replicate the same query on each of the instances, and compute the average count
to eliminate the noise.

\paragraph{Evaluation}
As long as there are enough instances (see Figure~\ref{fig:samples-needed})
this attack will work (noting that the condition of incorrect implementation exists).

\paragraph{Discussion}
This attack isn't possible on a correctly implemented system. We describe it here
primarily to document the need for correctly implementing seed initialization.

\subsection{Incremental data update: difference}
\label{atk:change-diff}

\paragraph{Prior Knowledge: Class A}
The attacker must have knowledge that only one protected entity's data changes
in the data update (see Additional conditions below), or must be able to
infer this with high probability (based on knowledge of the general rate of
change and specific knowledge of the victim).

\paragraph{Additional conditions: Common}
The table is an update table (the salt is changed with modifications to the table).
Among the subset of data being queried, the update must pertain to only a
single protected entity. 

\paragraph{Goal}
Detect that the change has taken place, therefore inferring information about a
single protected entity.

\paragraph{Attack}
An example of this attack would be one where a person in a given
department has been promoted, and only that person. The attacker knows that
the promotion may come with a salary raise, and that no other protected entity in
the department has a salary raise at the same time. The attacker makes the
following query both before and after the promotion:

\begin{lstlisting}
   SELECT floor(salary/10000)*10000,
          dept, count(DISTINCT aid)
   FROM table GROUP BY 1
\end{lstlisting}

The underlying true count for the two queries will change for two buckets,
that of the prior salary, and that of the new salary (assuming that the
salary change enough to move from one bucket to the other). The salary bucket
with the largest increase from before to after the change is that of the victim.
The attacker can improve PI at the expense of PR
by requiring that the change exceed a threshold.

\paragraph{Evaluation}
The file \mytt{diffAttackClass.py} in subdirectory \mytt{diffAttack} contains code
that measures this attack, setting \mytt{attackType=changeDiffAttack}.

\begin{figure}[tp]
\begin{center}
\includegraphics[width=\linewidth]{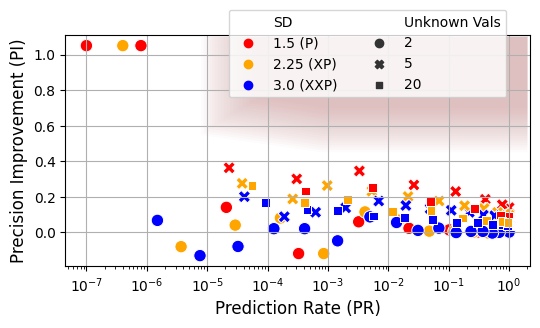}
\caption{Difference attack where the underlying data has changed.
\emph{Unknown Vals} is the number of distinct unknown data values
in the unknown column. \emph{SD} is the standard deviation of the
total amount of noise.
}
\label{fig:change-attack}
\end{center}
\end{figure}

Figure~\ref{fig:change-attack} shows that the attack is not effective.
The attack fails because the \mytt{salt} will have changed after the
table update, and so every salary bucket will have different noise from
both noise layers.

\paragraph{Discussion}
This attack would be effective if the \mytt{salt} were not changed (i.e. as
with append table salt management). In this case, the noisy count for every
bucket except those of the victim would remain the same. This underscores
the importance of managing the salt and data changes appropriately.

\subsection{Incremental data update: averaging}
\label{atk:change-avg}

\paragraph{Prior Knowledge: Class A}
The attacker must have knowledge that only one protected entity's data changes
during the set of data updates (see Additional conditions below).

\paragraph{Additional conditions: Very rare}
The table must be updated multiple times (with a change of salt each time).
For a given subset of the data, the data for only one protected entity changes once over
the course of the multiple table updates. In addition, there are multiple updates
both before and after the change.

\paragraph{Goal}
The goal is to reduce or eliminate the noise from counts both before and after the
change through averaging, and in this way learn the change for a specific protected entity.

\paragraph{Attack}
This attack is the same as the incremental data update attack of
Section~\ref{atk:change-diff}, except that here the attacker gets multiple
samples and averages them.

\paragraph{Evaluation}
The file \mytt{diffAttackClass.py} in subdirectory \mytt{diffAttack} contains code
that measures this attack, setting \mytt{attackType=changeAvgAttack}.

\begin{figure}[tp]
\begin{center}
\includegraphics[width=\linewidth]{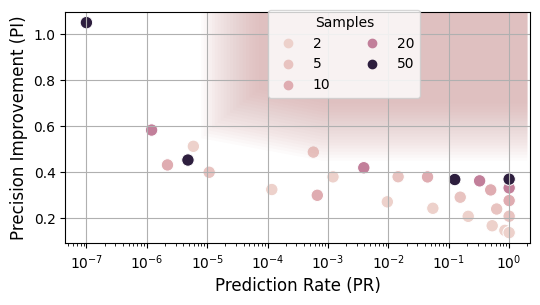}
\caption{Difference attack where the underlying data has changed
multiple times.
This data is for \mytt{SD=2.25} and 5 distinct unknown column values.
PI values above 1.0 represent cases where there are zero predictions or too few predictions
to produce a statistically meaningful PI value.
}
\label{fig:change-avg-attack}
\end{center}
\end{figure}

Figure~\ref{fig:change-avg-attack} shows that the attack is not effective
even with up to 50 table changes before and after the modified data.

\paragraph{Discussion}

Figure~\ref{fig:change-avg-attack} shows that more samples does lead to
better PI (see data points for PR=1.0). Since the effective standard deviation
grows with the square root of the number of samples, however, it takes a
large number of samples to get meaningful increases in PI. This means that,
while the administrator should take care not to update the dataset too
frequently, the system can tolerate a substantial number of updates with
few changes to a given subset of the dataset.

\subsection{Detect outlier bucket}
\label{atk:outlier-bucket}

\paragraph{Prior Knowledge: Class C}
The attacker knows of one or more of a small number of protected entities that have substantially more rows than all other protected entities. These protected entities are here called \emph{outliers}.

\paragraph{Additional conditions: Very rare}
The number of outliers must be more than the minimum \mytt{outlier\_range}
so that at least one outlier is sometimes assigned to the \mytt{top\_group}. The number of
outliers must also be somewhat less than the sum of the max \mytt{outlier\_range}
and max \mytt{top\_range}. This is so that the amount of noise is not dominated
by the outliers themselves.

The unknown column that is being inferred in the attack must be one whereby all
rows of a given protected entity are assigned the same value, and therefore the
victim appears only in a single bucket.

\paragraph{Goal}
The goal is infer an unknown value of a single protected entity outlier by detecting
when a given bucket has a substantially higher count than expected.
This higher count is due to the fact that one or two of the few outliers is in the
\mytt{top\_group}, and therefore isn't flattened, thus pushing up the bucket's
count.

\paragraph{Attack}
The attack comes in two phases. In the first phase, the attacker
determines the following:

\begin{enumerate}
   \item The total number of protected entities (noisy count)
   \item The total number of rows (noisy count)
   \item The total number of protected entities per bucket
\end{enumerate}

From this, the attacker computes the average number of rows per protected entity,
and a baseline expected number of rows per bucket
by multiplying the number of protected entities by the average number of rows
per protected entity.

In phase two, the attacker queries for the actual (noisy) number of rows per
bucket, and assumes that the victim is in the bucket where the actual number
of rows exceeds the expected number of rows the most.

\paragraph{Evaluation (for the normal case):}

The file \mytt{betaAttackClass.py} in subdirectory \mytt{outlierAttack} contains code
that measures this attack.

To evaluate the effectiveness of this attack, we first consider a
"normal" case where the distribution of individual contributions
is quite skewed towards a few extreme contributors (using a Beta distribution),
but otherwise randomly assigned to buckets.

\begin{figure*}[tp]
  \captionsetup[subfigure]{aboveskip=-5pt,belowskip=-5pt}
\centering
  \hspace{0.8cm}
  \begin{subfigure}[b]{0.25\textwidth}
     \centering
     \includegraphics[width=\textwidth]{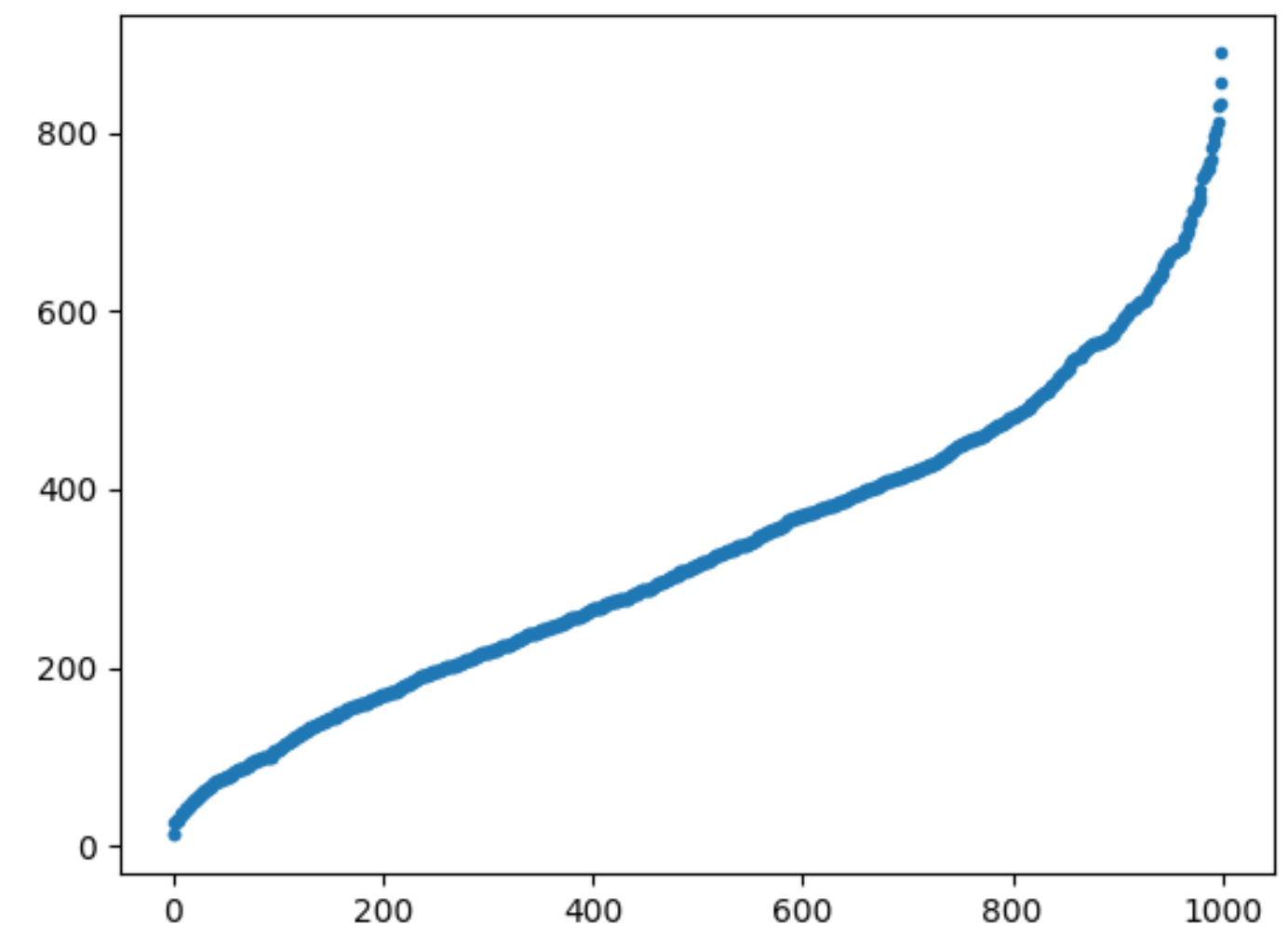}
     \label{fig:a2b4}
     \caption{Alpha = 2, Beta = 4}
  \end{subfigure}
  \hfill
  \begin{subfigure}[b]{0.25\textwidth}
     \centering
     \includegraphics[width=\textwidth]{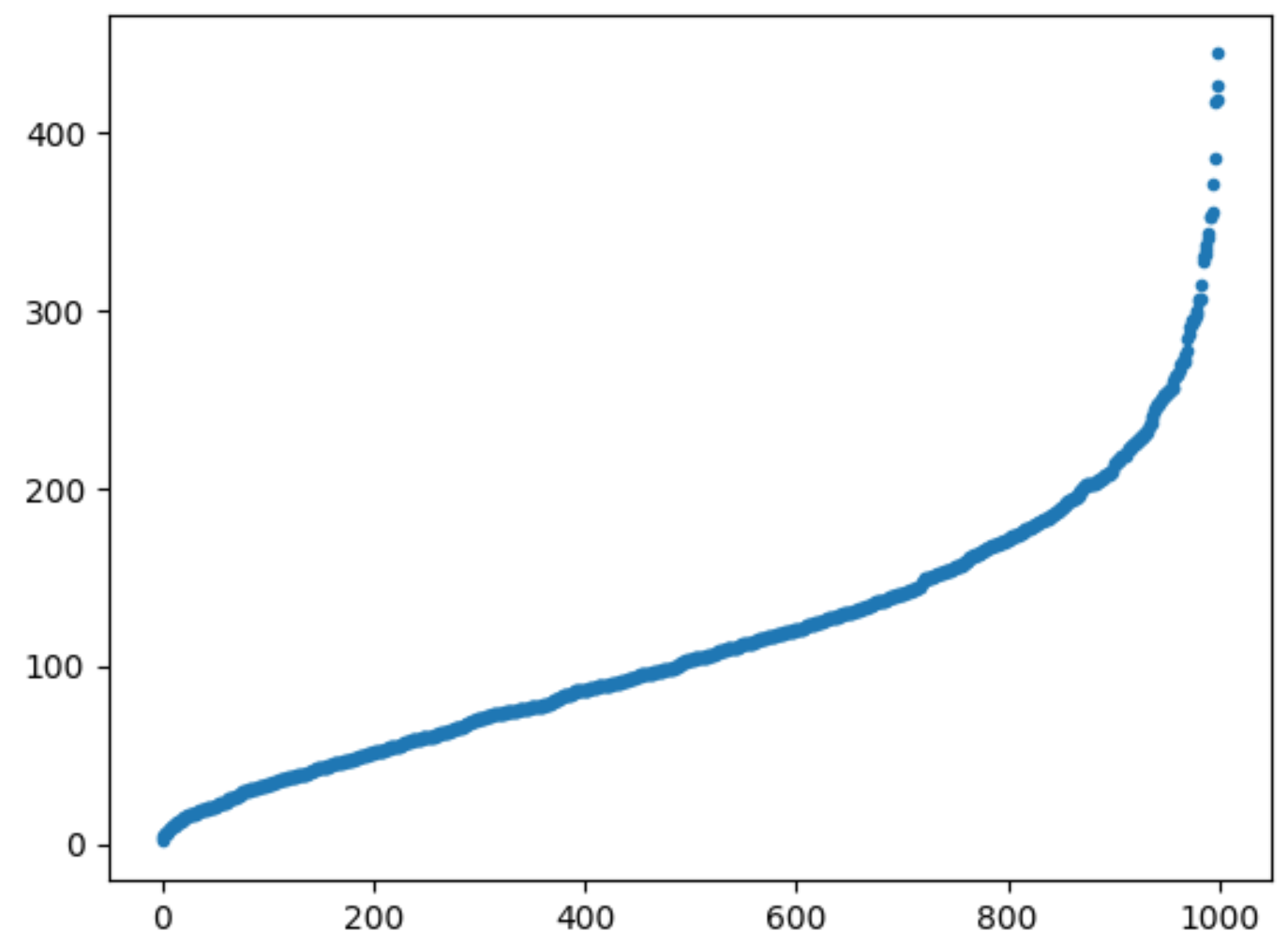}
     \label{fig:a2b16}
     \caption{Alpha = 2, Beta = 16}
  \end{subfigure}
  \hfill
  \begin{subfigure}[b]{0.25\textwidth}
     \centering
     \includegraphics[width=\textwidth]{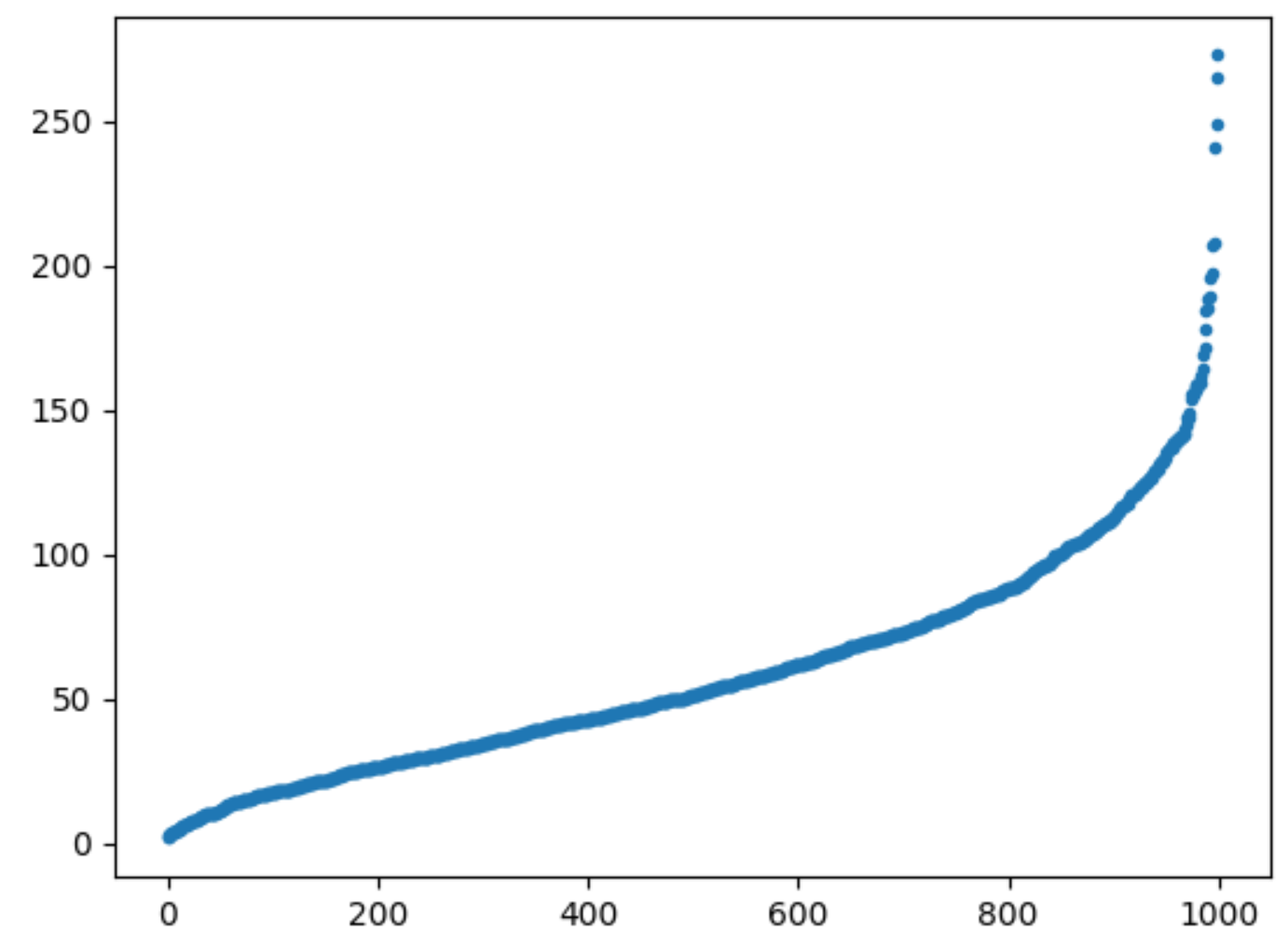}
     \label{fig:a2b32}
     \caption{Alpha = 2, Beta = 32}
  \end{subfigure}
  \hspace{0.8cm}
  \caption{Examples of the Beta distributions used in the outlier detection
  attack. The Y axis represents the number of rows for the corresponding
  AID.}
\label{fig:alphabetadist}
\end{figure*}

Specifically, we generate 1000 protected entities. We assign the protected entities evenly to a varying number of buckets (2, 5, and 20). We assign a number of rows to each protected entity according to a Beta distribution, variably using alpha:beta of 2:4, 2:16, and 2:32. Examples of these three distributions are shown in Figure~\ref{fig:alphabetadist}. As beta increases, these distributions generate increasingly extreme outliers.

We assign the protected entity with the most rows to be the victim. We
assume that
the bucket with the most rows contains the victim. We also vary the value
of a threshold, requiring that the bucket with the most rows must exceed
the average bucket size multiplied by the threshold. This increases
PI at the expense of lower PR.

The code for this attack is in the \mytt{outlierBucket} subdirectory,
file \mytt{betaAttackClass.py}.

\begin{figure}[tp]
\begin{center}
\includegraphics[width=\linewidth]{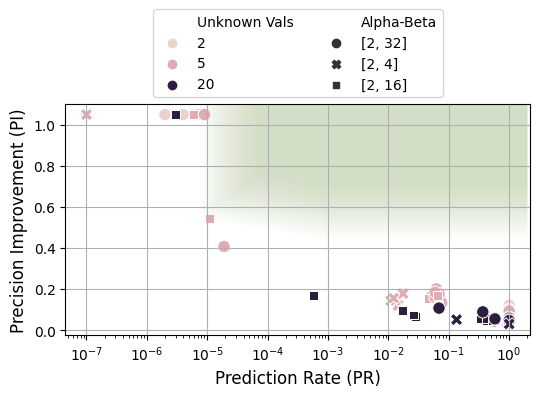}
\caption{Results of an attack trying to detect the value of the
extreme contributor. \emph{Unknown Vals} is the number of distinct values
in the unknown column. \emph{Alpha-Beta} are the parameters of the beta
distribution of individual contributions (see Figure~\ref{fig:alphabetadist}).
PI values above 1.0 represent cases where there are zero predictions or too few predictions
to produce a statistically meaningful PI value.
}
\label{fig:beta-outlier}
\end{center}
\end{figure}

The results are shown in Figure~\ref{fig:beta-outlier}. This shows that,
for the distributions in the attack, the flattening and proportional
noise mechanism of Diffix is very effective, even for skewed distributions that
generate extreme contributors.

\paragraph{Evaluation (for the worst case, very rare):}

The file \mytt{attack.py} in subdirectory \mytt{outlierAttack} contains code
that measures this attack.

The above evaluated skewed distributions, but did not evaluate worst-case
distribution. 
Here we evaluate row counts that represent the worst
case for the Diffix flattening and proportional noise mechanism. Specifically,
we generate row counts that perfectly match the conditions required for the
attack to work. In practice, we expect this scenario to be extremely rare.

\begin{figure}[tp]
\begin{center}
\includegraphics[width=\linewidth]{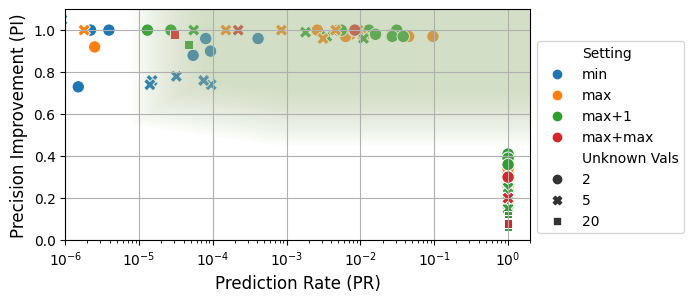}
\caption{Theoretical worst-case results of an attack trying to detect
the value of the extreme contributor. The dataset conditions with Settings
\mytt{max} and \mytt{max+1} are
pessimal for the attack, and presumed to be extremely rare.
Data points are for SD values between 1.5 and 3.0, and for
a mean extreme contribution of roughly 37 rows.
}
\label{fig:worst-outlier}
\end{center}
\end{figure}

\begin{figure}[tp]
\begin{center}
\includegraphics[width=\linewidth]{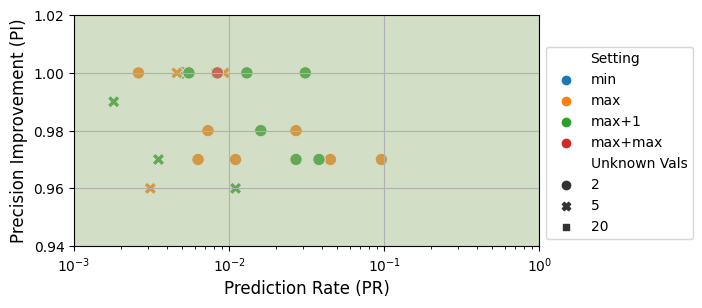}
\caption{This is a zoom-in of the green risk area of
    Figure~\ref{fig:worst-outlier}.
}
\label{fig:worst-outlier-close}
\end{center}
\end{figure}

The worst-case distribution used in our test is a combination of two distributions.
One distribution generates a set of protected entities called \emph{normal contributors}.
The number of rows for each normal contributor is uniformly distributed between
1 row and 10 rows. The second distribution generates a set of protected entities
called \emph{extreme contributors}. The number of rows for each extreme contributor
in Figures~\ref{fig:worst-outlier} and ~\ref{fig:worst-outlier-close}
comes from a uniform distribution between roughly 35 and 40 rows.

We vary the number of extreme contributors relative to how many extreme
contributors are in the noise/flattening
groups \mytt{outlier\_group} and \mytt{top\_group} as:

\begin{description}
   \item[min:] The number of
   extreme contributors equal to minimum \mytt{outlier\_range}, leading to
   extreme contributors only in \mytt{outlier\_group}.
   \item[max:] The number of
   extreme contributors equal to maximum \mytt{outlier\_range}, leading to
   one or two extreme contributors often in the \mytt{top\_group}.
   \item[max+1:] The number of
   extreme contributors equal to maximum \mytt{outlier\_range} plus one,
   leading to at least one
   extreme contributor in the \mytt{top\_group}.
   \item[max+max:] The number of
   extreme contributors equal to maximum \mytt{outlier\_range} plus 
   maximum \mytt{top\_range}, leading to both the \mytt{outler\_group} and
   \mytt{top\_group} being filled with extreme contributors.
\end{description}

The results is shown in Figure~\ref{fig:worst-outlier}, with Figure
\ref{fig:worst-outlier-close} zooming in on the green risk area.
At PR=1, the attack is ineffective for all data distributions.
The attack is also ineffective for the \mytt{min} setting, where all
of the extreme contributors are in the \mytt{outlier\_group}, and
are all flattened to match the average contribution of the normal
contributors in the \mytt{top\_group}. 

For the \mytt{max+max} setting, where the extreme contributors
make up both the \mytt{top\_group} and \mytt{outlier\_group},
there is no flattening. In this case, the attack is effective
when all or most extreme contributors have the same unknown value,
including the victim, and the noise value is large enough to
exceed the attack threshold. When there are only two unknown values, this
happens roughly once every 200 attacks. It happens less often with
more distinct unknown values.

The \mytt{max} and \mytt{max+1} settings are the worst case. In these
cases, there is typically one or two extreme contributors in the
\mytt{top\_group} that are on one hand not flattened, but on the
other don't contribute enough to the noise amount to become hidden.
From the zoom-in of Figure~\ref{fig:worst-outlier-close}, we see that
this pessimal data distribution can easily lead to high PI with
PR between 1/10 and 1/100 when there are only two distinct unknown values. 

\begin{figure}[tp]
\begin{center}
\includegraphics[width=\linewidth]{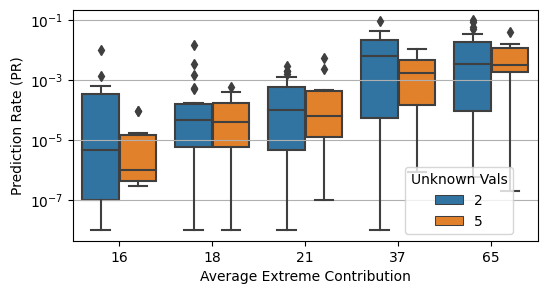}
\caption{Theoretical worst-case results of an attack trying to detect
the value of the extreme contributor, where PI $>$ 0.95.  
The extreme contributors distribution is uniform between plus and
minus 10\% of the average.
The normal contributors distribution is uniform between 1 and 10.
Data points are for SD values between 1.5 and 3.0, and all
four Setting types.
}
\label{fig:worst-by-out-factor}
\end{center}
\end{figure}

Figure~\ref{fig:worst-by-out-factor} illustrates the effect of the size of the gap
between the extreme and normal contributors in the bimodal distribution.
This shows that as the gap in the bimodal distribution grows, PR shrinks.

\paragraph{Discussion}
\label{sec:discuss:outlier}

For almost any real dataset, the proportional noise and flattening
mechanism is very effective. Nevertheless, we demonstrate that for
worst-case scenarios, high PI can be obtained along with relatively high
PR.
 
We argue that these worst-case scenarios are extremely rare and therefore
should not be a concern in the general case. There are several reasons:

\begin{enumerate}
   \item The data conditions themselves are rare (bimodal distribution with
   just the right number of extreme contributors),
   \item The attacker knows that these data conditions exist,
   \item The attacker is interested in learning from unknown columns
       with a very small number of distinct values.
\end{enumerate}

Nevertheless, if this case is a concern, then the underlying data can
be measured to determine if the data conditions exist. If they do, then
the extreme contributors may be completely removed from the data prior to
anonymization. Figures~\ref{fig:worst-outlier-close} and
\ref{fig:worst-by-out-factor} suggest that
data with a bimodal distribution of row counts, where the gap between
the upper and lower modes is roughly 2x or more, and where the number
of extreme contributors is around the maximum \mytt{outlier\_range}
value plus one or two, may require removal of the extreme contributors
or adjustment of the \mytt{outlier\_range}/\mytt{top\_range} parameters.

\subsubsection{Outlier based on distinct values}
\label{atk:outlier-bucket-distinct}

In a variation of this attack, the attacker knows that a small number of protected entities have an unusually large effect on the number of distinct values in a given column. This would most normally occur if a column consisted primarily of unique values, and a small number of protected entities simply have substantially more rows than other protected entities.

When this is the case, then the algorithm for counting distinct values (Section~\ref{sec:distinct}) detects the contributions of outlier protected entities, and uses flattening and proportional noise to hide them.

When the column values associated with \mytt{count(DISTINCT column)} are all unique, then the sorted list traversals in DQH step 3.3 (Section~\ref{sec:dqh3}) cause all of the unique column values associated with a given protected entity to be assigned to that protected entity as its contribution. In this case, the contribution of each protected entity is identical to the number of rows, and so the results of Figures~\ref{fig:beta-outlier} and \ref{fig:worst-outlier} apply directly.

At the other extreme, when each column value has a large number of protected entities associated with it (enough to prevent suppression), then the count of distinct values is exact, and there is no observable effect to do large contributors.

When each column value is shared by a small number of protected entities, only one of the protected entities is assigned a contribution for the value. The total magnitude of contributions is less, but protected entities with more rows will still have higher contributions because they persist longer in the sorted list traversals in DQH step 3.3. Flattening hides these higher contributions.

The setup for this attack is similar to that of the Beta distribution attack above (Section~\ref{atk:outlier-bucket}). There are 1000 protected entities, randomly assigned to a varying number of buckets (2, 5, and 20), and with a random number of rows taken from a Beta distribution (2:4, 2:16, and 2:32). In addition, values are assigned such that each value is represented by from 2 to 4 AIDVs. The attacker then tries to determine which bucket the most extreme outlier is in by selecting the bucket with the highest number of distinct values.

\begin{figure}[tp]
\begin{center}
\includegraphics[width=\linewidth]{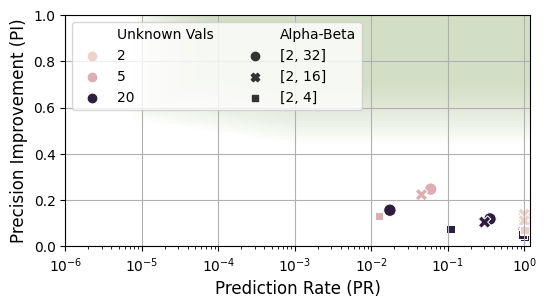}
\caption{Results of an attack trying to detect the value of the
extreme contributor to \mytt{count(distinct col)}.
\emph{Unknown Vals} is the number of distinct values
in the unknown column. \emph{Alpha-Beta} are the parameters of the beta
distribution of individual contributions (see Figure~\ref{fig:alphabetadist}).
}
\label{fig:distinct-beta-outlier}
\end{center}
\end{figure}

Figure~\ref{fig:distinct-beta-outlier} shows the results. We see that the attack is not at all effective: the extreme outlier is hidden through flattening.

%
%
%
%
%
%


\subsection{Attack Summary}
\label{sec:attk-summ}

Table~\ref{tab:attacks-summary} summarizes the attacks according to the
three main risk criteria, PI and PR measures, required prior knowledge,
and necessary conditions.

Green shading denotes very strong protection,
either because the PI or PR measures are very good, the required prior
knowledge is very unlikely to exist, or the necessary conditions are very
rare in practice. Yellow shading denotes strong protection.

\begin{table*}
\begin{center}
\begin{tabular}{lp{0.6\columnwidth}|cccp{0.75\columnwidth}}
  \toprule
    & Attack & 
    \rot{PI / PR} &
    \rot{PK Class} &
    \rot{Conditions} &
    Comments \\
  \midrule
   \ref{atk:attr} & Attribute value inspection &
                   \colorbox{green}{X} &  &    &   
                   Must ensure that the issues described in Sections
                   \ref{sec:relate-ind-pe} and \ref{sec:small-group}
                   are addressed.  \\ \hline
   \ref{atk:unique-infer} & Unique Inference &
                   \colorbox{green}{VS} &  & Com &   
                   May wish to inspect unique inference output bins with 
                   high AIDV counts that deviate from table-wide
                   distribution (\ref{sec:inspect-unique}).  \\ \hline
   \ref{atk:sim-know-noise} & Simple knowledge-based: Noise &
                   W-\colorbox{green}{VS} & \colorbox{green}{C}  &  Com  &   \\ \hline
   \ref{atk:sim-know-supp} & Simple knowledge-based: Suppression & 
                   W-\colorbox{green}{VS} & \colorbox{green}{C} & Com & 
                   May require XP or XXP level suppression \\ \hline
   \ref{atk:avg-naive} & Averaging: naïve & 
                     \colorbox{green}{X} &   &    &   \\ \hline
   \ref{atk:avg-sem} & Averaging: different semantics, same result & 
                     \colorbox{green}{X(T)} &   & Com &
                     Not an attack per se, but could partially reduce noise amount.
                     Would not accidentally happen with trusted analyst. \\ \hline
   \ref{atk:lpr-random} & LPR: randomness in column & 
                     &   &    &   \\
                        & \multicolumn{1}{r|}{(UA-mode)} &
                     W-\colorbox{green}{VS} & \colorbox{yellow}{B} & Com &   
                     May want higher noise levels for untrusted analyst. \\
                        & \multicolumn{1}{r|}{(TA-mode)} &
                     \colorbox{green}{X(T)} & \colorbox{yellow}{B} & Com &
                     Would not accidentally happen with trusted analyst. \\ \hline
   \ref{atk:lpr-aggr} & LPR: aggregate combinations & 
                     \colorbox{green}{VS} &  &    &   \\ \hline
   \ref{atk:diff-first} & Difference: positive AND, single victim & 
                     \colorbox{green}{X} & \colorbox{green}{C} & \colorbox{yellow}{R} &   \\ \hline
   \ref{atk:diff-group} & Difference: positive AND, group of victims & 
                     \colorbox{green}{VS} & \colorbox{green}{C} & \colorbox{yellow}{R}  &   \\ \hline
   \ref{atk:range-creep} & Range creep with averaging & 
                     &   &   &   \\
                        & \multicolumn{1}{r|}{(UA-mode)} &
                     \colorbox{green}{X} & A & Com  &   \\
                        & \multicolumn{1}{r|}{(TA-mode)} &
                     \colorbox{green}{X(T)} & A & Com &
                     Would not accidentally happen with trusted analyst. \\ \hline
   \ref{atk:salt-tab} & Salt: Dictionary attack on table & 
                     \colorbox{green}{X(T)} & \colorbox{green}{X} & Com &  
                     Morally equivalent to a password dictionary attack.
                     Would not accidentally happen with trusted analyst. \\ \hline
   \ref{atk:salt-know} & Salt: Knowledge attack & 
                     \colorbox{green}{X(T)} & \colorbox{green}{X} &   &  
                     Requires knowledge of the secret salt.
                     Would not accidentally happen with trusted analyst. \\ \hline
   \ref{atk:mult-instance} & Access to multiple instances & 
                      &  & \colorbox{green}{X} &  
                     Requires \textbf{incorrect} implementation of salt. \\ \hline
   \ref{atk:change-diff} & Incremental data update: difference & 
                     \colorbox{green}{VS} & A & Com &   \\ \hline
   \ref{atk:change-avg} & Incremental data update: averaging & 
                     \colorbox{green}{VS} & A & \colorbox{green}{VR} &  
                     Depends on poor administration of data. \\ \hline
   \ref{atk:outlier-bucket} & Detect outlier bucket & 
                     W & \colorbox{green}{C} & \colorbox{green}{X} &  
                     Only effective if learning one of a few distinct values.
                     Data conditions can be detected and prevented in advance. \\ \hline
   \ref{atk:outlier-bucket-distinct} & Detect outlier bucket (distinct) & 
                     \colorbox{green}{VS} & \colorbox{green}{C} & \colorbox{green}{X} &  \\
  \bottomrule
\end{tabular}
\end{center}
  \caption{\textbf{Attack Summary:} The \emph{PI/PR} column indicates strength
  of protection from PI and PR measures. The \emph{PK Class} columns indicates
  the class of prior knowledge needed for the attack. The \emph{Condition}
  column indicates the likelihood that the required conditions exist in the
  data. \colorbox{green}{Green} shading denotes very strong protection, while
  \colorbox{yellow}{yellow} shading denotes strong protection. See
  Section~\ref{sec:attk-summ} for descriptions of the codes.
  }
  \label{tab:attacks-summary}
\end{table*}

\paragraph{PI/PR column:}
The \emph{PI/PR} column summarizes the strength of anonymization against the
given attack as measured by PI and PR. The codes are:
\begin{description}
   \item[\colorbox{green}{X}] The attack simply doesn't work: nothing to measure.
   \item[\colorbox{green}{X(T)}] The attack would not accidentally take place with normal
       trusted analyst behavior.
   \item[\colorbox{green}{VS}] Very Strong: $PI < 0.5$ or $PR < 1/100000$.
   \item[\colorbox{yellow}{S}] Strong: $PI < 0.5$ or $PR < 1/1000$.
   \item[W] Weak: $PI < 0.5$ or $PR < 1/10$.
   \item[W-\colorbox{green}{VS}] Protection ranges from Weak to Very Strong depending
       on the privacy settings.
\end{description}

\paragraph{PK Class column:}
The \emph{PK Class} column summarizes the class of prior knowledge needed by
the attacker for the given attack. The codes are:
\begin{description}
   \item[\colorbox{green}{X}] The attacker cannot have the necessary prior
       knowledge (for instance because protected by administrator)
   \item[\colorbox{green}{C}] Class C (prior knowledge of multiple protected
   entities and uniqueness in data).
   \item[\colorbox{yellow}{C}] Class B (prior knowledge of multiple protected
   entities).
   \item[A] Class A (prior knowledge of a single protected entity).
   \item[- blank -] No prior knowledge is required.
\end{description}

\paragraph{Conditions column:}
The \emph{Conditions} column summarizes the likelihood of the
conditions necessary for the attack.
\begin{description}
   \item[\colorbox{green}{X}] The conditions can be detected and eliminated,
       or can only occur through deployment errors.
   \item[\colorbox{green}{VR}] Very Rare: The conditions are so rare as to
       never occur for all practical purposes.
   \item[\colorbox{yellow}{R}] Rare: The conditions sometimes occur for a
   small fraction of protected entities.
   \item[Com] Common: The conditions commonly occur.
   \item[- blank -] There are no special conditions (i.e. all datasets
       can be attacked).
\end{description}

Table~\ref{tab:params-atk} summarizes the attacks that are affected by
each anonymization configuration parameter.

\begin{table}
\begin{center}
\begin{small}
\begin{tabular}{r|p{0.6\columnwidth}}
   \toprule
   Parameter & Associated attacks \\
   \midrule
    \multicolumn{2}{l}{\kern -0.8em Suppression:}\\
   \mytt{low\_thresh} & \ref{atk:attr} Attribute value inspection \\
    & (\ref{atk:lpr-random} Linear program reconstruction: randomness in column) \\
   \mytt{low\_mean\_gap} & \ref{atk:sim-know-supp} Simple knowledge-based: Suppression \\
    & (\ref{atk:lpr-random}) \\
   \mytt{supp\_sd} & \ref{atk:sim-know-supp}, (\ref{atk:lpr-random}) \\
   \midrule
    \multicolumn{2}{l}{\kern -0.8em Noise:}\\
   \mytt{base\_sd} & \ref{atk:sim-know-noise} Simple knowledge-based: Noise \\
    & \ref{atk:avg-sem} Averaging: different semantics, same result \\
    & \ref{atk:lpr-random} Linear program reconstruction: randomness in column \\
   \midrule
    \multicolumn{2}{l}{\kern -0.8em Flattening:}\\
   \mytt{outlier\_range} & \ref{atk:outlier-bucket}, \ref{atk:outlier-bucket-distinct} Detect outlier bucket \\
   \mytt{top\_range} & \ref{atk:outlier-bucket}, \ref{atk:outlier-bucket-distinct} \\
   \bottomrule
\end{tabular}
\end{small}
\end{center}
  \caption{Table summarizing which attacks are affected by which
  anonymization parameters. Attacks in parenthesis are affected less.
  }
\label{tab:params-atk}
\end{table}

While it is of course up to the DPA or DPO to determine the thresholds and
criteria for anonymity and the associated configuration parameters, we
regard the settings reflected in the Table~\ref{tab:attacks-summary} as
quite conservative.

The picture that emerges from this analysis, and Table~\ref{tab:attacks-summary}
in particular, is that the protection afforded by Diffix Elm is very strong
and can certainly be regarded as anonymous.

Of the PI/PR scores, only one attack does not achieve a Very Strong
(VS) score, namely the Detect outlier
bucket~\ref{atk:outlier-bucket} attack. In this case,
the data conditions can be detected and prevented in
advance, thus leading to a Very Strong PI/PR score.

In TA-mode, two attacks can have a PI/PR score below Very Strong depending on
the anonymization parameters (the two Simple knowledge-based
attacks~\ref{atk:sim-know-noise} and~\ref{atk:sim-know-supp}).
In both cases, the prior knowledge is Class C, and so a lower
PI/PR score may be perfectly reasonable, especially in a non-public
data sharing scenario.

All other attacks have a Very Strong PI/PR score (and in some cases also
Class B or Class C prior knowledge).

\section{DPA, DPO, and data controller guidance}
\label{sec:dpa-guidance}

This section provides guidance to DPAs and DPOs
and data controllers or data processors for evaluating the privacy risk
associated with any given deployment of Diffix Elm.
In this section, we refer to all of these four entities as simply the DPx.

Tables~\ref{tab:non-attacks} and~\ref{tab:attacks-summary} taken together
list all of the known attacks on various versions of Diffix compiled over the
last five years or so. These are the result of extensive and repeated analysis
from researchers at MPI-SWS, employees of Aircloak GmbH, and external researchers
responding to our publications and two bounty programs.

Table~\ref{tab:non-attacks} lists the attacks from prior versions of Diffix
that can't be executed on Diffix Elm simply because the query syntax does
not exist. Table~\ref{tab:attacks-summary} is a summary of the attacks from
Section~\ref{sec:evaluation}. Table~\ref{tab:params-atk} lists
which anonymization parameters affect which attacks.

This section focuses specifically on issues related to the correct configuration
of Diffix Elm, and on the preparation of data prior to use
with Diffix Elm. The broader issue of how to do a risk evaluation in light
of how to set the PR/PI thresholds relative to risks of prior knowledge and
data conditions is out of scope.

\subsection{Protected entities}
\label{sec:pro-entity}

The DPx must ensure that the privacy of
individuals (natural persons) in the original dataset is protected.
To do this, it must be clear how the individual is identified in
the dataset.

Strictly speaking, Diffix Elm protects the privacy of \emph{protected entities}.
A protected entity may literally be an individual, for instance as defined
by a social security number. A protected entity may also be something that
is closely associated with an individual, like a mobile phone, a car, or a
credit card. In these cases, the correlation between protected entity and
individual may not be perfect: more than one individual may use a given
phone or drive a given car.

Furthermore, the protected entity may refer to a small group of strongly related
individuals, for instance two individuals sharing a bank account, or
the members of a household.

Datasets may have one row per protected entity, or multiple rows per protected entity.
Survey data, demographic data, and census data typically are \emph{one-row}
datasets (see Table~\ref{tab:one-row} for example). Time-series data is
\emph{multi-row} (see Table~\ref{tab:multi-row} for example).

\begin{table*}[tp]
    \centering
    \begin{tabular}{|l|l|l|l|l|l|}
    \hline
        Gender & Zip Code & Age & Education & Job & ... \\ \hline
        M & 12345 & 46 & High School & Plumber & ... \\ \hline
        O & 54321 & 23 & Bachelor & None & ... \\ \hline
        F & 48572 & 32 & PhD & Professor & ... \\ \hline
    \end{tabular}
    \caption{This is an example of a \emph{one-row} dataset, where each
    protected entity occupies exactly one row of the dataset. One-row datasets
    do not require an AID column (Diffix Elm automatically inserts an
    AID column). Typical examples of one-row datasets are survey data,
    census data, and demographic data.}
    \label{tab:one-row}
\end{table*}
\begin{table*}[tp]
    \centering
    \begin{tabular}{|l|l|l|l|}
    \hline
        IMEI & Time & Latitude & Longitude \\ \hline
        123 & 2021-10-01 21:34:19 & 43.27366 & 81.36623 \\ \hline
        123 & 2021-10-01 21:36:21 & 43.43884 & 81.39229 \\ \hline
        123 & 2021-10-01 22:02:51 & 43.81922 & 81.40221 \\ \hline
        ... & ... & ... & ... \\ \hline
        456 & 2021-02-13 17:34:19 & -17.27366 & 67.36623 \\ \hline
        456 & 2021-02-13 17:36:21 & -17.43884 & 67.39229 \\ \hline
        456 & 2021-02-13 17:02:51 & -17.67883 & 81.40221 \\ \hline
        ... & ... & ... & ... \\ \hline
    \end{tabular}
    \caption{This is an example of a \emph{multi-row} dataset, where each
    protected entity may occupy more than one row in the dataset. Multi-row datasets
    must have at least one column that identifies the protected entity (here the
    \mytt{IMEI} column). Time-series data is multi-row.}
    \label{tab:multi-row}
\end{table*}

Datasets may have multiple protected entities. There are primarily three scenarios where this can occur:
\begin{enumerate}
    \item There is a relationship or interaction between two individuals, for instance send and receive email, sender and receiver in a bank transaction, doctor and patient, or friends on a social network.
    \item There are strongly related groups of individuals, for instance a household or a joint bank account, where the group is deemed to have privacy concerns similar to the individual.
    \item The data owner wants to protect some other kind of information (perhaps to maintain business secrets), such as a branch office or store.
\end{enumerate}

\subsection{Relationship between individual and protected entity}
\label{sec:relate-ind-pe}

Ideally, a given individual (natural person) is associated with one and only one protected entity (as identified by the AID column). This association can be one-to-one (each AID value is associated with one individual), or many-to-one (each AID value is associated with multiple individuals).  An example of one-to-one is a national identity number like a social security number as the AID. An example of many-to-one would be the address of a single dwelling is the AID (where the residents of the address are the individuals).

Real datasets may deviate from this ideal to a greater or lesser extent.  For instance, if the AID is a mobile phone identifier, and a person uses multiple mobile phones (at a single time or over time), then that person appears in the dataset as different persons. This is an example of a one-to-many relationship between individual and protected entity.  If an attacker can link the AIDs related to the person, then the anonymity of that person is weakened.

For example, suppose that the addresses of mobile phone owners is in the dataset, and the individual with multiple phones has the same address each time. Then an attack could select that address in the query and learn information about a single individual.

One way to remedy this is to assign two AIDs, one for the mobile phone identifier, and one for the address. Another way to remedy this would be to remove the address information from the dataset (though this might unnecessarily degrade data quality).

Note finally that an individual may be associated with multiple protected entities simply because the data is dirty. For instance, the user id of an individual may have been typed incorrectly, thus leading to two entries for the same individual.

It is therefore important that the DPx understands to what extent individuals may appear as multiple protected entities, and ensure either that there are no columns in the dataset that can link the protected entities, or that multiple AIDs have been assigned to protect against the linking.

\subsection{Small groups of strongly related individuals}
\label{sec:small-group}

Often it can happen that small groups of individuals are strongly correlated in a dataset. This can easily happen for instance with family units or married couples.

As an example, suppose a hospital dataset has a column for insurance number, but that the insurance number is shared by the whole family. If the protected entity is individual persons, and the insurance number remains in the dataset, then information about the entire family can be viewed. For instance, the following query would give the family's total health care expenditure so long as the suppression threshold for the associated bucket is lower than the number of family members:

\begin{lstlisting}
   SELECT insurance_num, sum(paid)
   FROM hospital_dataset
   GROUP BY insurance_num
\end{lstlisting}

There are several remedies to this problem.

First, the offending column (here \mytt{insurance\_num}) can be removed.

Second, the group itself (i.e. \mytt{insurance\_num}) can be defined as a second protected entity.

A third approach is to set the suppression threshold \mytt{low\_thresh} to be larger than the maximum number of individuals that share an insurance number (while keeping only the individual as the protected entity).

\subsection{Trust Mode}
\label{sec:trust-mode}

Diffix Elm has two modes, Trusted Analyst mode (TA-mode), and
Untrusted Analyst mode (UA-mode). TA-mode has more generalization
capabilities (any numeric bucket size instead of snapped, and any
substring offset instead of only first-character offset).

If Diffix Elm is operated in UA-mode, then the attacks
\emph{LPR: randomness in column} (\ref{atk:lpr-random}) and
\emph{Range creep with averaging} (\ref{atk:range-creep}) are ineffective.

In TA-mode, the attacks would be effective if an analyst executed them.
There is no reason that a trusted analyst
would accidentally run these attacks. In addition, Diffix for Desktop
gives the analyst access to the original data, and so if an analyst were
malicious, then they could simply exploit the original data directly rather
than run an attack.

The DPx must verify that safeguards are in place to ensure that
the set of queries necessary to exploit the above-mentioned attacks are not 
released to untrusted individuals or to the public. Simply ensuring
that the analysts are trusted may be adequate protection, since the
queries would not be accidentally released in a normal analytic task.

Additionally, trusted analysts could be informed about the possibility
of the above attacks. Finally, the DPx may require that multiple
parties approve any data release to ensure that the queries necessary
for the attacks are not released.

\subsection{Worst-case extreme contributors}
\label{sec:extr-cont}

The attack \emph{Detect outlier bucket} (\ref{atk:outlier-bucket}) has
a possible worst-case PI/PR measure that falls well within the designated risk
area.  Although the prior knowledge requirement is Class C for this attack,
the DPx should either:
\begin{itemize}
   \item Verify that the data conditions do not exist, and if they do:
   \item Verify that the prior knowledge is not viable, or remove
   extreme contributors until the data conditions no longer exist.
\end{itemize}

\subsection{Optionally inspect unique inferences}
\label{sec:inspect-unique}

Diffix Elm does not explicitly prevent output buckets that allow unique
inferences (see~\ref{atk:unique-infer}). A unique inference occurs when,
in an output bucket with $N$ columns, the values for
$N-k$ of the columns are unique to this bucket. In this case, the
values for the remaining $k$ columns may be inferred.
While PI is always zero for unique inference buckets, in cases where
the number of AIDVs in a unique inference buckets substantially exceeds
the suppression threshold, the absolute precision of an inference
is high.

The PDx may require that such high-precision unique inferences are
inspected to ensure that the inferences are not surprising or
sensitive (see~\ref{sec:atk:unique-inference-discuss}).



\bibliographystyle{abbrv}
\bibliography{../../masterBib/master}

\onecolumn
\appendix

\section{PDx questionnaire}
\label{sec:questions}

\begin{enumerate}
   \setlength\itemsep{2em}
   \setlength\parskip{1em}
   \item What is the protected entity?

   \item Do individuals correspond one-to-many or many-to-many
   with the protected entity (\ref{sec:relate-ind-pe})?

   \item If yes, do any columns link an individual across
   multiple protected entities?

   If yes, then either the linking columns must be removed, or the
   suppression threshold \mytt{low\_thresh} must be set to the
   maximum number of protected entities to which a given individual
   is linked.
   
   \item Can closely-related groups of individuals (like married couples or
   families) be linked by some column or columns in the
   dataset (\ref{sec:small-group})?

   If yes, then either the group must be the protected entity, or the
   ability to link the group must be removed (i.e. by removing the
   columns), or the suppression threshold \mytt{low\_thresh} must be set
   to a value higher than the largest group.

   \item Does the data set have one row per protected entity
   or multiple rows per protected entity?

   If one row, then no AID column is explicitly selected, and we may regard
   the row index number as an implicit AID column.
   
   \item If multiple rows, does the selected AID column correctly identify the protected entity or entities?
   
   \item If TA-mode (Trusted Analyst mode) is deployed, are proper
   procedures in place
   to ensure that queries conforming to the attack conditions listed in
   Section~\ref{sec:trust-mode} are prevented?

   \item Do the data conditions exist for the worst-case \emph{Detect outlier
   bucket} attack (\ref{sec:extr-cont})?

   \item If so, has it been determined that the prior knowledge requirements
   are not viable (\ref{sec:extr-cont})?

   \item Is it necessary to inspect output buckets for privacy-leaking unique
   inferences (\ref{sec:inspect-unique})?

\end{enumerate}

\end{document}